\documentclass{aa}
\usepackage[varg]{txfonts}
\usepackage{graphicx}
\usepackage{natbib}
\usepackage{booktabs}
\usepackage{morefloats}
\bibpunct{(}{)}{;}{a}{}{,}

\usepackage{xcolor}
\begin{document}

\title{Spectral modelling of Type IIb Supernovae.}
\subtitle{Comparison to SN 2011dh and the effect of macroscopic mixing.}

\author{Mattias~Ergon\inst{\ref{inst1}} \and Claes~Fransson\inst{\ref{inst1}}}

\institute{The Oskar Klein Centre, Department of Astronomy, AlbaNova, Stockholm University, 106 91 Stockholm, Sweden 
\label{inst1}}

\date{Accepted for publication by Astronomy and Astrophysics.}

\abstract{We use the new non-local-thermodynamical-equilibrium (NLTE) lightcurve and spectral synthesis code JEKYLL to evolve a macroscopically mixed ejecta model of a type IIb Supernova (SN) originating from a star with an initial mass of 12 M$_\odot$ through the photospheric and nebular phase. The ejecta model is adopted from earlier work and has a mass of 1.7 M$_\odot$, a kinetic energy of 0.7 foe and contains 0.075 M$_\odot$ of $^{56}$Ni. The macroscopic mixing is simulated through a statistical representation of ejecta fragmented into small clumps, but spherically symmetric on average. We compare to SN 2011dh, and find that \emph{both} the spectra and the lightcurves are well reproduced in \emph{both} the photospheric and nebular phase, although there are also some differences. Our work further strengthens the evidence that this SN originated from a star with an initial mass of $\sim$12 M$_\odot$ that had lost all but tiny (<0.1 M$_\odot$) fraction of its hydrogen envelope, strongly suggesting a binary origin. We also investigate the effects of the macroscopic mixing by comparing macroscopically and microscopically mixed models, and by varying the clumping geometry. In the photospheric phase, we find strong effects on the effective opacity in the macroscopically mixed regions, which affects the model lightcurves. The diffusion peak is considerably narrower (rise time decreased by 11 percent) in the macroscopically mixed case, and differs strongly (rise time decreased by 29 percent) if the radioactive material in the helium envelope is allowed to expand more than in our standard model. The effect is mainly geometrical, and is driven by the expansion of the clumps containing radioactive material, which tend to decrease the effective opacity. In the limit of optically thick clumps, the decrease is roughly given by the product of the (volume) expansion and filling factors for the radioactive material, and in our models values up to $\sim$8 are explored. These findings has implications for lightcurve modelling of stripped-envelope SNe in general, and the effect would increase the estimated ejecta masses. In the nebular phase, we find strong effects on the collisional cooling rates in the macroscopically mixed regions, which affects lines driven by collisional cooling, in particular the [\ion{Ca}{ii}] 7291, 7323 \AA~and [\ion{O}{i}] 6300, 6364 \AA~lines. The effect is mainly related to differences in composition between macroscopically and microscopically mixed ejecta. As these lines are often used for mass determinations, it highlights the importance of how, and to what extent the calcium- and oxygen-rich material is mixed. As shown in this and earlier work, \textit{both} NLTE and macroscopic mixing are essential ingredients to accurately model the lightcurves and spectra of Type IIb SNe throughout their evolution.}


\titlerunning{Spectral modelling of Type IIb Supernovae.}
\authorrunning{M. Ergon and C. Fransson.}
\maketitle

\defcitealias{Maz93}{ML93}
\defcitealias{Luc02}{L02}
\defcitealias{Luc03}{L03}
\defcitealias{Luc05}{L05}
\defcitealias{Koz92}{KF92}
\defcitealias{Ker14}{K14}
\defcitealias{Jer15}{J15}
\defcitealias{Jer11}{J11}
\defcitealias{Jer12}{J12}
\defcitealias{Erg14}{E14}
\defcitealias{Erg15}{E15}
\defcitealias{Erg18}{Paper I}
\defcitealias{Kro09}{K09}

\section{Introduction}
\label{s_intruduction}

For a quantitative understanding of supernovae (SNe) and their progenitor stars an accurate modelling of the radiative transfer and the microphysics based on a realistic structure of the ejecta is necessary. In this respect Type IIb SNe offer an interesting case. Their evolution from a hydrogen dominated early phase to a helium and metal dominated late phase links them observationally to Type II as well as Type Ib SNe. Several detections of the progenitor stars in pre-explosion imaging provide insights about their origin, and the Type IIb supernova remnant Cas A provides insights about the structure of the ejecta. Type IIb SNe likely originate from stars that have lost most, but not all of their hydrogen envelopes, leaving behind an (almost) bare but still intact helium core. This provides a more solid connection between the SN ejecta and the initial mass of the progenitor star.

Except for the prototypical Type IIb SN 1993J, the best observed Type IIb SN is 2011dh, for which we presented observations and modelling of the lightcurves in \citet[hereafter \citetalias{Erg14}, \citetalias{Erg15}]{Erg14,Erg15}, as well as modelling of nebular spectra in \citet[hereafter \citetalias{Jer15}]{Jer15}. The modelling suggest an initial mass of $\sim$12 M$_\odot$ for the progenitor, a conclusion supported by observations of the star in pre-explosion images \citep{Mau11}. Modelling of the early lightcurve \citep{Ber12} and photospheric spectra \citepalias{Erg14} show that the yellow supergiant visible in these images had lost all but a tiny fraction of its hydrogen envelope. As stellar winds for stars of this mass seem too weak to expel the hydrogen envelope this strongly suggests a binary origin, where most of the hydrogen envelope was lost through interaction with a companion star. A similar conclusion applies to SN 1993J, based on modelling of the SN \citep{Nom93,Shi94,Woo94}, pre-explosion observations of the progenitor star \citep{Ald94}, as well as a likely post-explosion detection of the companion star \citep{Mau04,Fox14}. We also studied a large sample of Type IIb SNe using a grid of hydrodynamical models in \citet{ErgPHD}, where we found most of those to originate from relatively low-mass stars. However, the simplified treatment of the opacity in the hydrodynamical modelling, makes this result somewhat uncertain. Further evidence for a binary origin comes from lightcurve modelling \citep[e.g.][]{Ben12} as well as spectral modelling \citep{Des15,Des16,Des18b} based on binary evolutionary stellar models.

Mixing of the SN ejecta occurs in the explosion due to hydrodynamical instabilities \citep[e.g.][]{Mul91}. Hydrodynamical modelling suggests that the mixing is extensive in Type IIb SNe \citep{Wong17}, a conclusion that is supported by observations of Type IIb SNe  (e.g.~\citetalias{Erg15}) and the distribution of O- and Si-burning products in Cas A \citep[e.g.][]{Wil02}. In addition, both theoretical arguments \citep[e.g.][]{Fry91}, and observations of SNe \citep[e.g.][]{Fra89} suggest that it occurs on macroscopic scales only. Evidence for this also comes from the spatial correlation of nuclear burning products in Cas A \citep[e.g][]{Enn06}. Such \textit{macroscopic mixing} is caused by advection, where the ejecta are fragmented into clumps in which the composition remains intact. This in contrast to \textit{microscopic mixing}, which is caused by diffusion, and where the composition is altered. In traditional 1-D codes, macroscopic mixing can not be self-consistently simulated, although important aspects of it have been incorporated by \citet{Koz98a,Koz98b} and in CMFGEN by \citet{Des18} and \citet{Des20}. Therefore, in works based on traditional 1-D codes, the mixing has typically been assumed to be microscopic. On the other hand, in Monte-Carlo (MC) based codes it is possible to relax the 1-D constraint in a statistical sense while keeping it in an average sense, JEKYLL uses the Virtual Grid method \citep[hereafter \citetalias{Jer11}]{Jer11}, which allows for a statistical representation of ejecta consisting of small clumps of different composition, but spherically symmetric on average. We can therefore take the macroscopic mixing of the ejecta into account, although the treatment is idealized and does not account for large-scale asymmetries. Note, that the clumping geometry is not only determined by the fragmentation and mixing that occur in the explosion, but also by the subsequent expansion of clumps containing radioactive material due to heating from radioactive decays \citep[e.g][]{Her91}. Evidence of such "Ni bubbles" is also seen in Cas A, whose interior seems to consist of large cavities surrounded by O-burning material \citep{Mil15}.

In \citet[hereafter \citetalias{Erg18}]{Erg18} we presented, described and tested JEKYLL, and investigated the effect of non-local-thermodynamical-equilibrium (NLTE) on the spectra and lightcurves of a microscopically mixed Type IIb model. In this paper we use JEKYLL to evolve a macroscopically mixed Type IIb model through the photospheric and nebular phase, and investigate the effect of the macroscopic mixing on the spectra and lightcurves. This model belongs to a set of models that was earlier evolved through the nebular phase using SUMO \citepalias{Jer11}, and compared to the observed nebular spectra and lightcurves of SN 2011dh in \citetalias{Jer15} and \citetalias{Erg15}, respectively. Out of those, the strongly mixed 12 M$_\odot$ model explored here, showed the best agreement with the nebular spectra and lightcurves of SN 2011dh. It is therefore of great interest to investigate how well this model compares to the spectra and lightcurves of SN 2011dh in the photospheric phase, which is the first objective of this paper. The second objective is to investigate the effects of the macroscopic mixing, and we have therefore constructed a microscopically mixed version of the model, and a set of macroscopically mixed models differing in the clumping geometry. In our treatment, the clumping geometry is determined by the sizes of the clumps and the total volume they occupy (filling factors).

It should be emphasized that the SN models explored here are not only of interest for SN 2011dh, but have broader implications for Type IIb SNe, stripped-envelope (SE) SNe and core-collapse (CC) SNe in general. This is particularly true for the macroscopic mixing and its effects, which have not been self-consistently investigated before throughout the SN evolution. Due to the excellent data obtained for SN 2011dh, as well as its progenitor star, this SN makes an excellent starting point for such investigations.

The paper is organized as follows. In Sect.~\ref{s_ejecta_model_and_config} we describe the methods used and our set of models, in Sect.~\ref{s_application} we discuss the modelling results for our standard model, and compare to observations of SN 2011dh, and in Sect.~\ref{s_effect_macro} we investigate the effect of macroscopic mixing on the modelling results. Finally, in Sect.~\ref{s_conclusions} we conclude and summarize the paper.

\section{Methods and models}
\label{s_ejecta_model_and_config}

All SN models presented in this work were calculated with the JEKYLL code, which was described in detail in \citetalias{Erg18}. Here, we briefly repeat the general methods used in JEKYLL, and discuss the treatment of the macroscopic mixing in some more detail. The configuration of JEKYLL and the atomic data used are described in Appendix \ref{a_configuration} and \ref{a_atomic_data}, respectively.

The ejecta models are based on model 12C from the set of Type IIb models presented by \citetalias{Jer15}, which corresponds to a progenitor star with an initial mass of 12 M$_\odot$ and strong macroscopic mixing of the ejecta. Based on this ejecta model, we construct a standard model, which differs slightly from the original one, and a set of models that differs in the type of mixing (macroscopic or microscopic) and the clumping geometry. 

\subsection{General methods}
\label{s_method_general}

JEKYLL is a lightcurve and spectral-synthesis code based on a MC method for the time-dependent 3-D radiative transfer developed by \citet{Luc02,Luc03,Luc05}, and extended as described in \citetalias{Erg18}. To calculate the radiation field and the state of matter\footnote{With state of matter we refer to the temperature and the populations of ionized and excited states.} an iterative procedure is used, which is similar to an accelerated $\Lambda$-iteration (see discussion in \citetalias{Erg18}). JEKYLL has several solvers to calculate the state of matter, but here we use the NLTE solver, where the statistical and thermal equilibrium equations are solved for taking into account all relevant processes. In particular, this includes heating, excitation and ionization by non-thermal electrons calculated using the method by \citet{Koz92}. In the inner region, where the matter and radiation field are assumed to be coupled, we use a diffusion solver to calculate the temperature. The main limitations in JEKYLL are the assumptions of homologous expansion, thermal and statistical equilibrium, and a spherically symmetric distribution of the matter. The latter is, however, only assumed on large scales and on average, as we discuss in Sect.~\ref{s_method_macro}. Another important limitation is the lack of a treatment of the ejecta chemistry (i.e. molecules and dust).

Homologous expansion is often well justified after a few doublings of the SN radius, but some hydrodynamical effects like circumstellar interaction and expansion of regions containing radioactive material may last longer. Large-scale departures from spherical symmetry, as probably exist to some extent in most SNe, may have a considerable impact on the spectra and lightcurves \citep[e.g.][]{Kro09}. At early times, ionization freeze-out may affect the strength of lines originating from the outer layers \citep{Des10}, whereas at late times it may actually dominate the energy budget \citep{Koz98a}. Molecules may form in the C/O and Si/O rich regions of the ejecta, and affect both the IR emission and the cooling of these regions \citep{Liu94,Lil20}. Dust may also form in these regions and affect both spectra and lightcurves (see \citetalias{Jer15} and \citetalias{Erg15} with respect to SN 2011dh). All of this should be kept in mind while analysing our models.

\subsection{Treatment of the macroscopic mixing}
\label{s_method_macro}

To simulate the macroscopic mixing JEKYLL uses the Virtual Grid method \citepalias{Jer11}. The ejecta are assumed to be spherically symmetric on average, and the macroscopic mixing is represented by distinct types of small spherical clumps, characterized by their composition, density, size and filling factor. In the MC radiative transfer, the clumps are drawn based on their geometrical cross-section as the MC packets propagate through a macroscopically mixed region. Note, that the clumps are virtual in the sense that they only exist as long as a MC packet propagates through them. In Appendix~\ref{a_effective_opacity} we compare to radiative transfer calculations on actual 3D-grids and show that the method is well justified over the parameter space of interest. The state of matter is calculated separately for each type of clumps using a MC radiation field constructed from MC packets passing through these clumps. The internal stratification of the state of matter in the clumps is ignored. In the inner region, where we use the diffusion solver, we assume a uniform temperature in the clumps and use an effective Rosseland mean opacity (see Sect.~\ref{s_effect_macro_rad} and Appendix~\ref{a_effective_opacity}). This is justified if the diffusion time in the clumps is small compared to the expansion and decay timescales. Given the temperature the state of matter is calculated separately for each type of clumps. Note, that the MC radiative transfer for the $\gamma$-rays emitted in the radioactive decays uses the Virtual Grid method both in the inner and outer region.

As mentioned, methods to treat different aspects of the macroscopic mixing have been incorporated also in traditional 1-D codes. The method of alternating spherical shells introduced by \citet{Koz98a,Koz98b} and incorporated in CMFGEN by \cite{Des20} takes into account the effect of a different density and composition in the clumps, whereas the method introduced by \citet{Des18} takes into account the effect of a different density in the clumps. However, as the 1-D constraint is strictly enforced, any 3-D effects on the radiative transfer can not be simulated. The advantage of the Virtual Grid method is that the 1-D constraint is relaxed in a statistical sense, and therefore the effects of a small-scale 3-D clumping geometry on the radiative transfer can be taken into account. As we discuss thoroughly in Sect.~\ref{s_effect_macro_rad} and Appendix.~\ref{a_effective_opacity} these effects become important in the regime of optically thick clumps.

However, the Virtual Grid method is still a simplification, and might be considered as a convenient statistical parametrization of a more general 3-D problem, where small-scale asymmetries are idealized and large-scale asymmetries ignored. Although it would be possible to map a 3-D hydrodynamical simulation to this parametrization (and we may explore that path in the future), the models explored here are based on 1-D hydrodynamical simulations, where the compositional layers have been artificially mixed with each-other. In this work we do not explore the extent of the mixing, but focus on the type of mixing (macroscopic or microscopic) and the clumping geometry. Given the mass-fractions of the compositional layers in a macroscopically mixed region, the clumping geometry is determined by the sizes and the filling factors of the clumps. These are in turn determined by the original sizes of the clumps and the expansion of the clumps containing radioactive material, which results in a corresponding compression of the other clumps. The clumping geometry in our models is discussed in Sect.~\ref{s_ejecta_model}. 

As the dynamics in JEKYLL is limited to homologous expansion, we assume that all important hydrodynamical effects occurred before the start of the simulation. This assumption is safe with respect to the fragmentation of the ejecta into clumps, which occurs in the explosion, but less so for the subsequent expansion of the clumps containing radioactive material. At early times, the decay energy is deposited locally, and if the diffusion time in the clumps is long compared to the decay and expansion timescales, this creates a temperature difference. If, in addition, the hydrodynamical timescale is small compared to the diffusion timescale, the corresponding pressure difference drives an expansion of the clumps. This means that we are on the safe side if the diffusion time in the clumps is small compared to the decay and expansion timescales. As we demonstrate in Appendix \ref{a_tdiff_clump}, this condition, which also applies to the diffusion solver (see above), is increasingly well fulfilled in all our models after a few days. Nevertheless, the homologous assumption introduces an uncertainty during the first week.

\subsection{Ejecta models}
\label{s_ejecta_model}

A full description of the Type IIb model 12C is given in \citetalias{Jer15}, but we repeat the basic properties here. It is based on a SN model by \citet{Woo07} for a star with an initial mass of 12 M$_\odot$, from which the masses and abundances for the 0.6 M$_\odot$ carbon-oxygen core and the 1.0 M$_\odot$ helium envelope have been adopted. The carbon-oxygen core is assumed to have a constant (average) density, and the helium envelope to have the same (average) density profile as the best-fit model for SN 2011dh by \citet{Ber12}. In addition, a 0.1 M$_\odot$ hydrogen envelope based on models by \citet{Woo94} is attached. The velocities of the interfaces between the carbon-oxygen core, the helium envelope and the hydrogen envelope are set to 3500 and 11000 km s$^{-1}$, respectively, based on observations of SN 2011dh. It should be emphasized that model 12C is not a self-consistent hydrodynamical model, but rather a phenomenological model based on results from hydrodynamical simulations.

\begin{figure*}[tbp!]
\includegraphics[width=1.0\textwidth,angle=0]{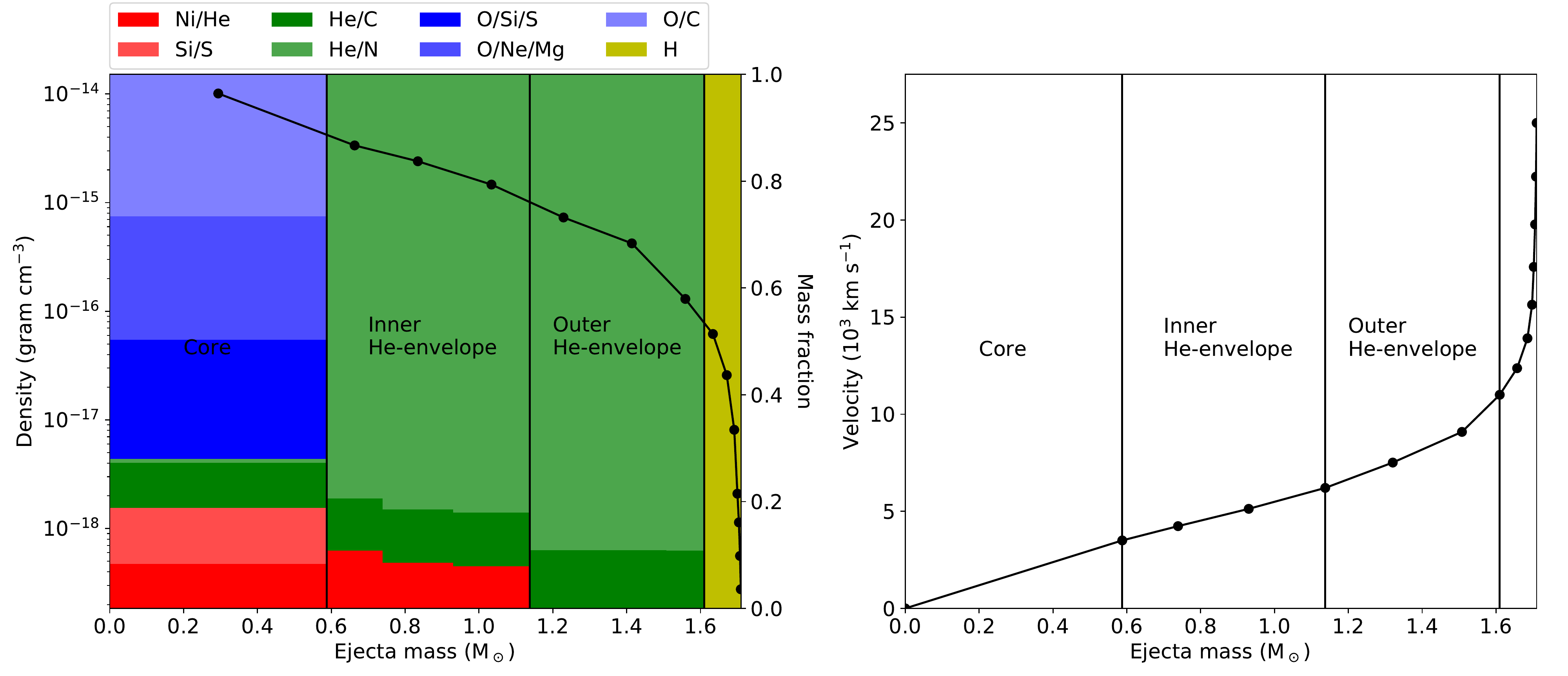}
\caption{Left panel: Density profile at 100 days (black dots) and the mass fractions of the compositional zones (coloured as shown at the top) for model 12C. Right panel: Velocity profile for model 12 C.}
\label{f_ejecta_structure}
\end{figure*}

Based on the original onion-like compositional structure, \citetalias{Jer15} identified five compositional zones (O/C, O/Ne/Mg, O/Si/S, Si/S and Ni/He\footnote{Referred to as the Fe/Co/He zone in \citetalias{Jer15}.}) in the carbon-oxygen core and two compositional zones (He/N and He/C) in the helium envelope. To mimic the mixing of the compositional zones in the explosion, two scenarios with different degrees of mixing (medium and strong) were explored in \citetalias{Jer15}. Model 12C has strong mixing, corresponding to a fully mixed carbon-oxygen core and about half of the radioactive Ni/He material mixed into the inner part of the helium envelope. Note, that the other material in the core is \emph{not} mixed into the helium envelope, which is a simplification. The density profile, the masses of the compositional zones and the velocity profile for model 12C are shown in Fig~\ref{f_ejecta_structure}, and given in tabulated form in Table~\ref{t_ejecta_structure}. The composition of each zone is given in \citetalias{Jer15}. The ejecta has a total mass of 1.7 M$_\odot$, a kinetic energy of 0.7 foe and contains 0.075 M$_\odot$ of $^{56}$Ni.

The constraints on the clumping geometry are weak, but as discussed in \citetalias{Jer15}, there are some observational constraints on the clumping geometry of the core, which were used as guidelines for the original model. In \citet{Jer12}, a density contrast of $\sim$30 between the Ni/He and oxygen-rich clumps was derived for the Type IIP SN 2004et, and based on that a density contrast of 30 and 210 was explored in \citetalias{Jer15}. Model 12C has a density contrast of 210, corresponding to an expansion factor\footnote{Here and in the following this refers to the volume expansion factor.} of 10 and a filling factor of 0.85 for the Ni/He clumps, and an compression factor of 21 and a filling factor of 0.035 for the oxygen-rich clumps. We emphasize that the amount of expansion/compression in Type IIb SNe is uncertain and might differ from that in Type IIP SNe, but the chosen values are consistent with constraints on the filling factor of the oxygen-rich material ($0.02<\Phi<0.07$) derived for SN 2011dh in \citetalias{Erg15} and \citetalias{Jer15}. In \citetalias{Erg15}, a lower limit on the number of clumps in the O/Ne/Mg zone of $\sim$900 was derived from small-scale variations in the [\ion{O}{i}] 6300, 6364 \AA~and \ion{Mg}{i}] 4571 \AA~line profiles of SN 2011dh, and based on that \citetalias{Jer15} assumed that each compositional zone in the core consisted of 10000 clumps.

For the comparison with SN 2011dh we have constructed a standard model, which is similar to model 12C. However, we have chosen to deviate in two aspects. First, we assume that all clumps have the same mass (instead of the same number per compositional zone), and second, we assume that the Ni/He clumps expand also in the helium envelope. The cavities observed in Cas A extend well beyond $\sim$5000 km s$^{-1}$ \citep{Mil13b,Mil15}, which gives support to the latter assumption. The clump mass was chosen to give 15000 clumps in the O/Ne/Mg zone, and the expansion of the Ni/He clumps in the helium envelope was assumed to be half of that in the core. These assumptions seem more physically sound than the original ones, but the choices for the mass and expansion of the Ni/He clumps in the helium envelope are somewhat arbitrary.

To investigate the effect of the type of mixing, we have constructed a microscopically mixed version of the standard model by averaging the density and the abundances over the compositional zones.  Similarly, to investigate the effect of the clumping geometry, we have constructed a set of models that differs in the expansion of the clumps containing radioactive material and the sizes of the clumps. These are: (1), a model where the Ni/He and Si/S clumps have not been expanded at all, (2), a model where the Ni/He clumps in the helium envelope have been expanded with a larger factor (8.5), and (3), a model where the number of clumps have been increased by a factor of 100. All models and their parameters are listed in Table \ref{t_ejecta_models}, and in Table~\ref{t_fillingfactor_density} we list the filling factors and the densities at 100 days for the compositional zones in the macroscopically mixed models.

Note, that in real SN ejecta there will be a distribution of clump properties, whereas in our models each compositional zone in each spatial shell is represented by a single type of clumps. Although JEKYLL allows for a distribution of densities and sizes, this would require more processing power, and as the clumping geometry is nevertheless badly constrained, we leave such an exercise to future works. 

\begin{table*}[tb]
\caption{Velocity ranges, mean density at 100 days and the masses of the compositional zones for model 12C.}
\begin{center}
\begin{tabular}{llllllllll}
\toprule
Velocity range & Mean density & Ni/He & Si/S & O/Si/S & O/Ne/Mg & O/C & He/C & He/N & H \\
(km s$^{-1}$) & (10$^{-14}$ g cm$^{-3}$) & (M$_\odot$) & (M$_\odot$) & (M$_\odot$) & (M$_\odot$) & (M$_\odot$) & (M$_\odot$) & (M$_\odot$) & (M$_\odot$) \\
\midrule
0 - 3.50e3 & 1.01 & 4.91e-2 & 6.13e-2 & 1.31e-1 & 1.36e-1 & 1.56e-1 & 4.99e-2 & 4.36e-3 & -\\
3.50e3 - 4.24e3 & 3.36e-1 & 1.64e-2 & - & - & - & - & 1.47e-2 & 1.20e-1 & -\\
4.24e3 - 5.13e3 & 2.40e-1 & 1.64e-2 & - & - & - & - & 1.91e-2 & 1.56e-1 & -\\
5.13e3 - 6.20e3 & 1.47e-1 & 1.64e-2 & - & - & - & - & 2.08e-2 & 1.70e-1 & -\\
6.20e3 - 7.51e3 & 7.29e-2 & - & - & - & - & - & 1.99e-2 & 1.63e-1 & -\\
7.51e3 - 9.01e3 & 4.22e-2 & - & - & - & - & - & 2.04-e2 & 1.67e-1 & -\\
9.01e3 - 1.10e4 & 1.30e-2 & - & - & - & - & - & 1.11e-2 & 9.12e-2 & -\\
1.10e4 - 1.24e4 & 6.18e-3 & - & - & - & - & - & - & - & 4.71e-2\\
1.24e4 - 1.39e4 & 2.58e-3 & - & - & - & - & - & - & - & 2.80e-2\\
1.39e4 - 1.56e4 & 8.11e-4 & - & - & - & - & - & - & - & 1.25e-2\\
1.56e4 - 1.76e4 & 2.10e-4 & - & - & - & - & - & - & - & 4.59e-3\\
1.76e4 - 1.98e4 & 1.14e-4 & - & - & - & - & - & - & - & 3.55e-3\\
1.98e4 - 2.22e4 & 5.60e-5 & - & - & - & - & - & - & - & 2.48e-3\\
2.22e4 - 2.50e4 & 3.76e-5 & - & - & - & - & - & - & - & 1.74e-3\\
\bottomrule
\end{tabular}
\end{center}
\label{t_ejecta_structure}
\end{table*}

\begin{table*}[tb]
\caption{Ejecta models derived from model 12C. For each model we list the type of mixing, the (volume) expansion (blue) and compression (red) factors for the compositional zones in the core and the inner helium envelope, and the clump mass.}
\begin{center}
\begin{tabular}{llllllll}
\toprule
Model & Type of mixing & Ni/He & Si/S & O and He & Ni/He & He & Clump mass (M$_\odot$)\\
& & core & core & core & envelope & envelope &\\
\midrule
Standard model & macroscopic & \textcolor{blue}{10} & \textcolor{blue}{1.1} & \textcolor{red}{21} & \textcolor{blue}{5.0} & \textcolor{red}{1.8} & 8.4e-6\\
Microscopic mixing & microscopic & - & - & - & - & - & -\\
No expansion & macroscopic & 1 & 1 & 1 & 1 & 1 & 8.4e-6\\
Strong expansion & macroscopic & \textcolor{blue}{10} & \textcolor{blue}{1.1} & \textcolor{red}{21} & \textcolor{blue}{8.5} & \textcolor{red}{6.0} & 8.4e-6\\
More clumps & macroscopic & \textcolor{blue}{10} & \textcolor{blue}{1.1} & \textcolor{red}{21} & \textcolor{blue}{8.5} & \textcolor{red}{6.0} & 8.4e-6\\
\bottomrule
\end{tabular}
\end{center}
\label{t_ejecta_models}
\end{table*}

\begin{table*}[tb]
\caption{Filling factors (upper part) and density at 100 days (lower part; 10$^{-14}$ g cm$^{-1}$) for the compositional zones in the core and the inner helium envelope for the macroscopically mixed models.}
\begin{center}
\resizebox{\linewidth}{!}{%
\begin{tabular}{lllllllllll}
\toprule
Model & Ni/He & Si/S & O/Si/S & O/Ne/Mg & O/C & He/C & He/N & Ni/He & He/C & He/N \\
& core & core & core & core & core & core & core & envelope & envelope & envelope\\
\midrule
No expansion & 8.36e-2  & 1.04e-1 & 2.23e-1 & 2.31e-1 & 2.65e-1 & 8.49e-2 & 7.41e-3 & 1.00e-1 & 9.80e-2 & 8.02e-1\\
Standard model & 8.50e-1 & 1.10e-1 & 1.10e-2 & 1.10e-2 & 1.30e-2 & 4.10e-3 & 3.60e-4 & 5.00e-1 & 5.40e-2 & 4.46e-1 \\
Strong expansion & 8.50e-1 & 1.10e-1 & 1.10e-2 & 1.10e-2 & 1.30e-2 & 4.10e-3 & 3.60e-4 & 8.50e-1 & 1.60e-2 & 1.34e-1 \\
\midrule
No expansion & 1.01 & 1.01 & 1.01 & 1.01 & 1.01 & 1.01 & 1.01 & 2.07e-1 & 2.07e-1 & 2.07e-1\\
Standard model & 9.88e-2 & 9.88e-1 & 2.07e+1 & 2.07e+1 & 2.07e+1 & 2.07e+1 & 2.07e+1 & 4.14e-2 & 3.76e-1 & 3.76e-1 \\
Strong expansion & 9.88e-2 & 9.88e-1 & 2.07e+1 & 2.07e+1 & 2.07e+1 & 2.07e+1 & 2.07e+1 & 2.44e-2 & 1.27 & 1.27 \\
\bottomrule
\end{tabular}}
\end{center}
\label{t_fillingfactor_density}
\end{table*}

\subsection{SN models} 
\label{s_sn_models}

The ejecta models described in Sect.~\ref{s_ejecta_model} were resampled to a finer spatial grid (29 radial cells) and (homologously) rescaled to one day. Based on an initial temperature profile, the SN models were then evolved with JEKYLL using 130 logarithmically spaced timesteps to 398 days. The SN models were calculated using an frequency grid of 5350 logarithmically spaced intervals between 10 \AA~and 40 $\mu$m, and each model required $\sim$9000 CPU hours, which using 128 CPUs resulted in a computing time of $\sim$3 days. The initial temperature profile was taken from the best-fit hydrodynamical model for SN 2011dh from \citetalias{Erg15}, with the mixing of the $^{56}$Ni adjusted to match that of model 12C. This SN model was based on a bare helium core stellar model, and therefore the cooling of the thermal explosion energy, lasting for a few days in a model with a hydrogen envelope, is ignored. The subsequent evolution is powered by the continuous injection of radioactive decay energy, and the choice of initial temperature profile is not critical, although it may have some effect on the early evolution.

\section{Comparisons to observations}
\label{s_application}

Among the Type IIb models presented in \citetalias{Jer15}, the strongly mixed 12 M$_\odot$ model 12C (see Sect~\ref{s_ejecta_model}) was found to give the best match to the observed optical and NIR nebular spectra \citepalias{Jer15} and the broad-band and pseudo-bolometric lightcurves\footnote{The best match was found for model 12F, which differs from model 12C only in the optical depth of the dust.} \citepalias{Erg15} of SN 2011dh. It is therefore of great interest to explore how well this model reproduces the early spectra and lightcurves, something which is now possible using JEKYLL. In addition, it is interesting in itself to run a self-consistent model from the photospheric to the nebular phase. However, as the nebular phase was discussed in detail in \citetalias{Jer15}, we mainly focus on the first 150 days, which gives some overlap without being repetitive. Note, that in the comparison we use our standard version of model 12C, which differs slightly from the original model (see Sect.\ref{s_ejecta_model}).

The comparison also serves the purpose to discuss the behaviour of Type IIb in general and in several ways it is applicable to SE SNe as well. This is particularly true for Type Ib SNe, as they differ from Type IIb SNe only by the absence of a low-mass hydrogen envelope.

In Sects.~\ref{s_spec_evo}, \ref{s_phot_evo}, \ref{s_colour_evo} and \ref{s_bol_evo} we discuss the spectral, photometric, colour and bolometric evolution of the standard model and compare this to observations of SN 2011dh. First, however, we briefly discuss the physical conditions in the model, to provide some background for the discussion of the observed quantities that follows.

\subsection{Physical conditions}
\label{s_matter_state}

\begin{figure*}[tbp!]
\includegraphics[width=1.0\textwidth,angle=0]{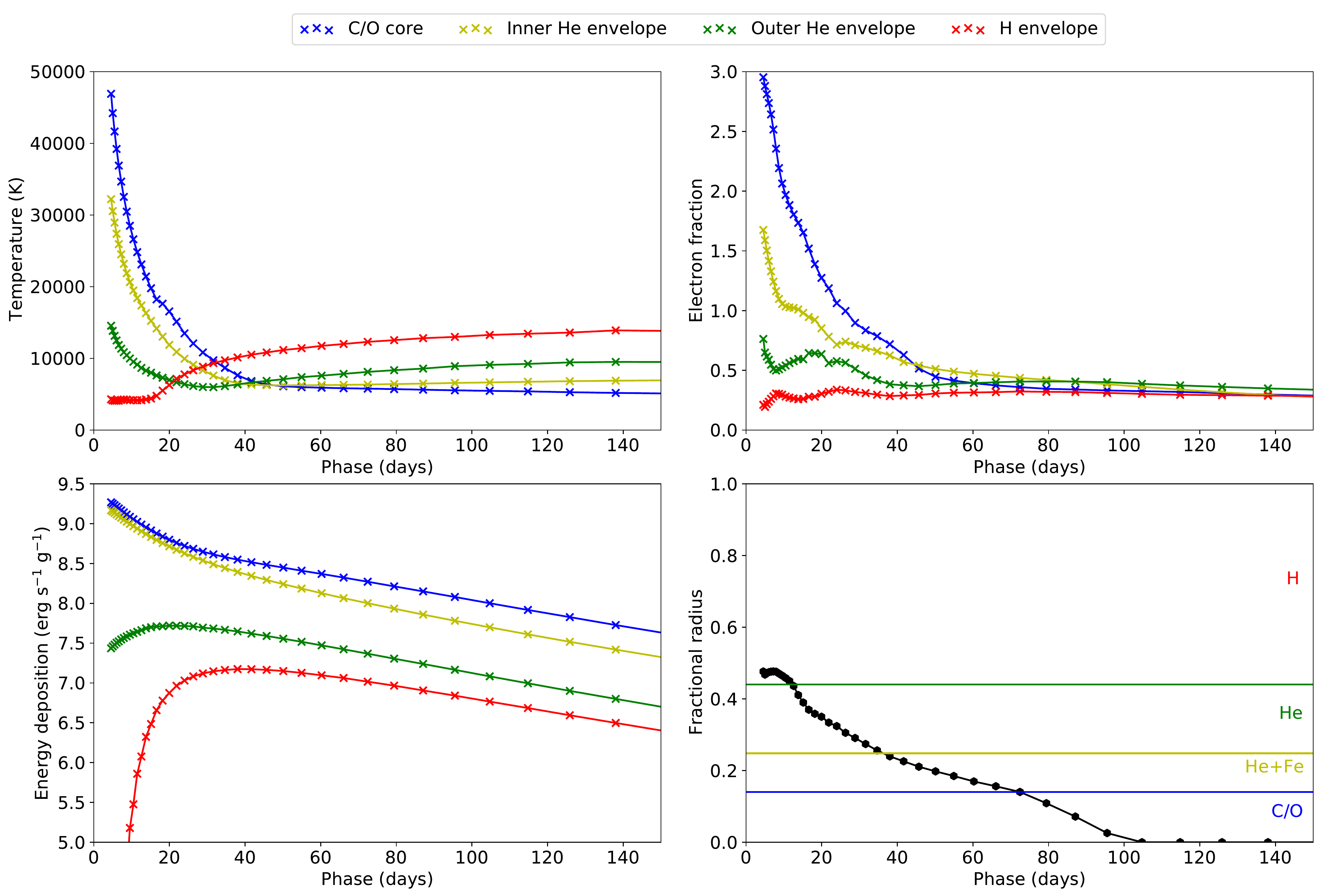}
\caption{Evolution of the temperature (upper left panel), electron fraction (upper right panel) and radioactive energy deposition (lower left panel) in the oxygen core (blue), inner/outer (yellow/green) helium envelope and the hydrogen envelope (red) for the standard model. In the lower right panel we show the evolution of the (Rosseland mean) continuum photosphere (black) as well as the outer borders of the carbon-oxygen core (blue) and inner/outer (green/yellow) helium envelope.}
\label{f_12C_matter_evo}
\end{figure*}

Figure~\ref{f_12C_matter_evo} shows the evolution of the temperature, electron fraction and radioactive energy deposition in the core, the inner/outer helium envelope and the hydrogen envelope for the standard model. These are averages over the spatial cells and the compositional zones, but in Sects.~\ref{s_effect_macro_matter} and \ref{s_effect_macro_energy} we get back to this and discuss the variation of these quantities between the compositional zones. In Figure~\ref{f_12C_matter_evo} we also show the evolution of the (Rosseland mean) continuum photosphere. 

Initially, the photosphere is at the border of the hydrogen envelope, which is relatively cool ($T\sim5000$ K) and mainly recombined ($x_e\sim 0.2$) whereas the core is hot ($T\sim50000$ K) and highly ionized ($x_e>3$). Towards $\sim$40 days, when the photosphere reaches the inner helium envelope, the temperature and the electron fraction in the inner parts decrease quickly, whereas the temperature begins to rise in the outer parts. After this the evolution slows down, the temperature continues to rise in the outer parts and exceeds 10,000 K in the hydrogen envelope, whereas the electron fraction drops below 0.5 towards 100 days, roughly when the ejecta become optically thin in the continuum. The radioactive energy is initially deposited in the inner parts where the radioactive material resides, but at $\sim$10 days the deposition begins to rise quickly in the outer parts, and we will get back to this issue below.

\subsection{Spectral evolution}
\label{s_spec_evo}

\begin{figure*}[tbp!]
\includegraphics[width=0.97\textwidth,angle=0]{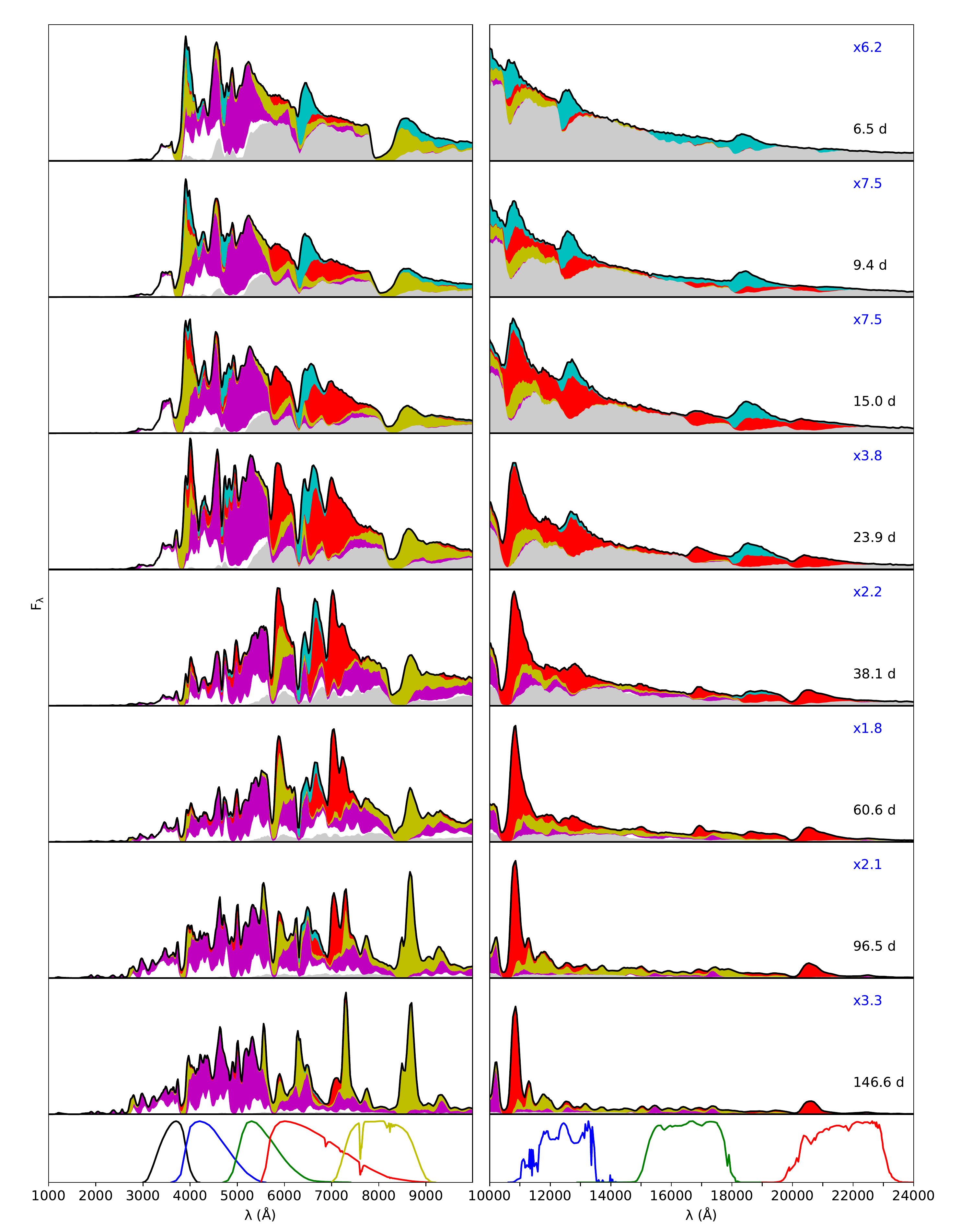}
\caption{Spectral evolution in the optical (left panel) and NIR (right panel) for the standard model, where the NIR flux has been scaled as indicated in blue. In the spectra we show the contributions (last scattering/emission event excluding electron scattering) to the flux from bound-bound transitions of hydrogen (cyan), helium (red), carbon to calcium (yellow), scandium to manganese (white) and iron to nickel (magenta) as well as continuum processes (grey). At the bottom we show the transmission profiles of the optical Johnson-Cousins $U$ (black), $B$ (blue), $V$ (green), $R$ (red) and $I$ (yellow) bands and the NIR 2MASS $J$ (blue), $H$ (green) and $K$ (red) bands.}
\label{f_12C_spec_trans_evo}
\end{figure*}

\begin{figure*}[tbp!]
\includegraphics[width=0.97\textwidth,angle=0]{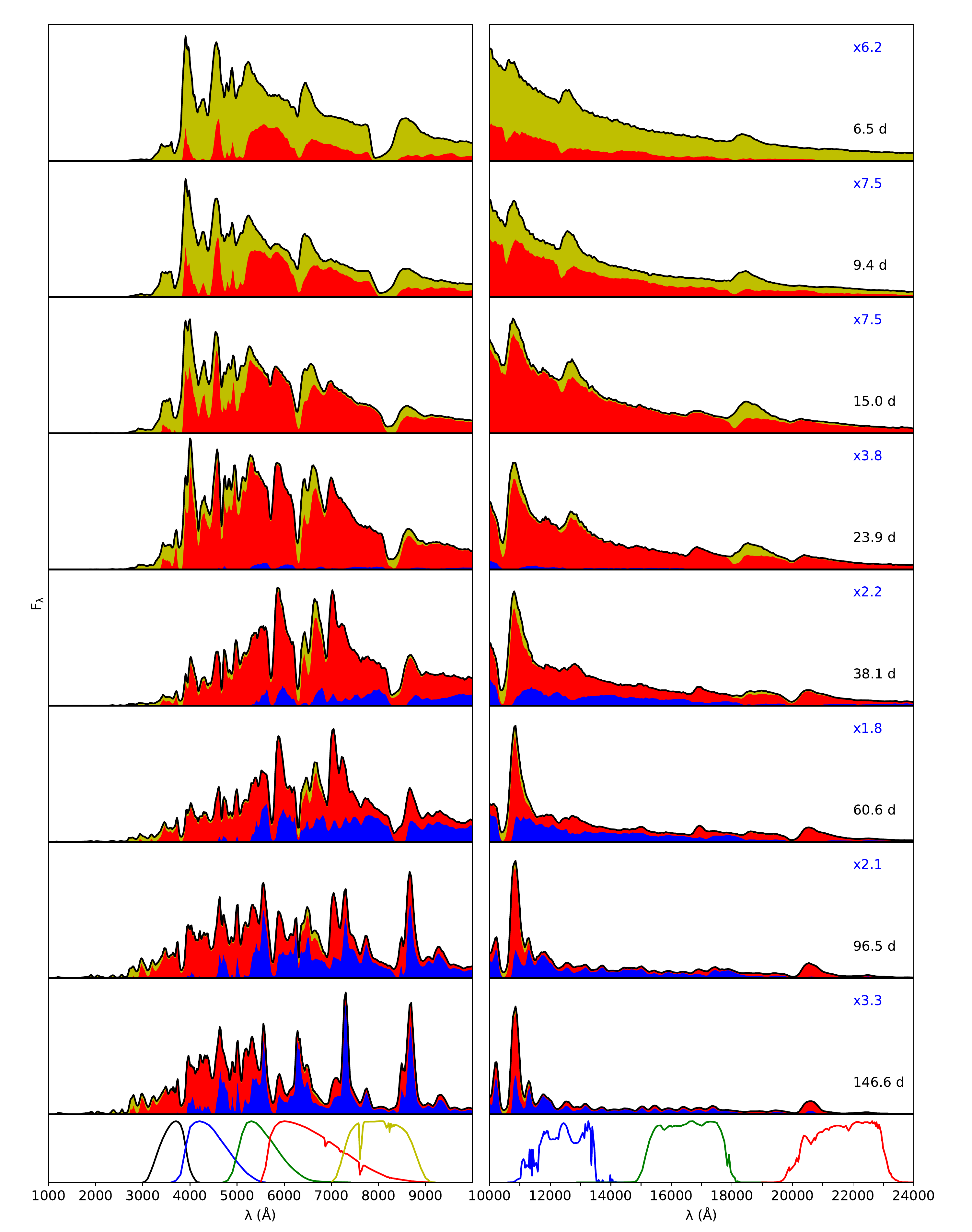}
\caption{Spectral evolution in the optical (left panel) and NIR (right panel) for the standard model, where the NIR flux has been scaled as indicated in blue. In the spectra we show the contributions (last scattering/emission event excluding electron scattering) to the flux from the carbon-oxygen core (blue), and the helium (red) and hydrogen (yellow) envelopes. At the bottom we show the transmission profiles of the optical Johnson-Cousins $U$ (black), $B$ (blue), $V$ (green), $R$ (red) and $I$ (yellow) bands and the NIR 2MASS $J$ (blue), $H$ (green) and $K$ (red) bands.}
\label{f_12C_spec_cell_evo}
\end{figure*}

\begin{figure*}[tbp!]
\includegraphics[width=0.97\textwidth,angle=0]{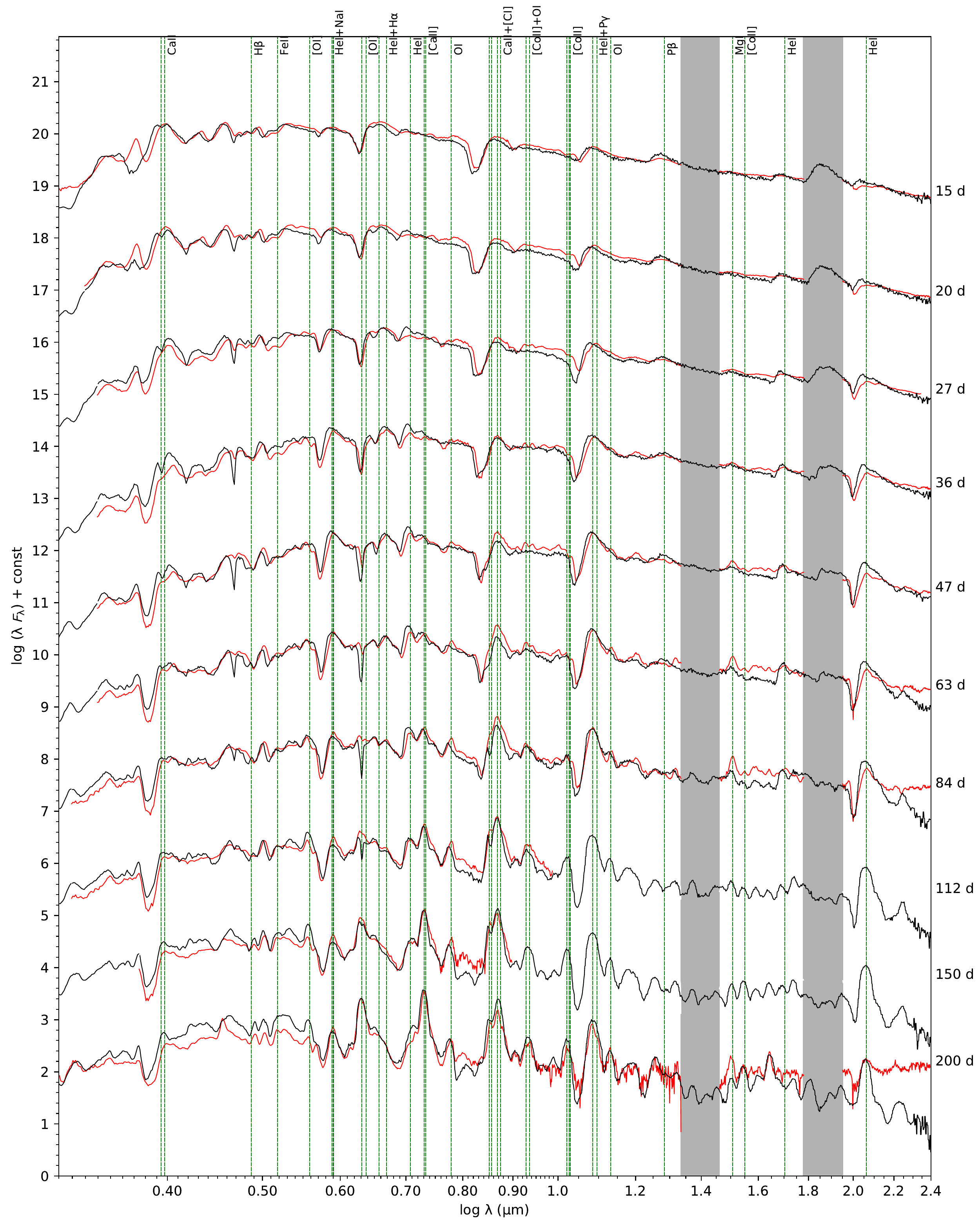}
\caption{Spectral evolution for the standard model (black) compared to the observations of SN 2011dh (red). Spectra from 10 logarithmically spaced epochs between 15 and 200 days are shown, where the model and observed spectra at each epoch have been shifted by the same, but otherwise arbitrary amount. Here, and in the following figures the spectra of SN 2011dh have been interpolated as described in \citetalias{Erg14}, but we only show interpolations that have an observational counterpart close (less than 20 percent) in time. In addition, the rest-wavelengths of the most important lines are shown as red dashed lines and the NIR telluric absorption bands as grey vertical bars.} 
\label{f_dh_spec_evo_comp}
\end{figure*}

Figures \ref{f_12C_spec_trans_evo} and \ref{f_12C_spec_cell_evo} show the spectral evolution for the standard model, where the former figure displays the radiative process, and the latter figure the location giving rise to the emission. In both cases this refers to the last emission/scattering events for the MC packets excluding electron scattering. Furthermore, in Fig.~\ref{f_dh_spec_evo_comp} we compare the spectral evolution in the optical and near-infrared (NIR) to observations of SN 2011dh. In this and other observational comparison figures, all spectra have been re-binned to 10 \AA. In all other spectral figures, the model spectra have been gently smoothed with a gaussian with $\sigma$=1.0 frequency bins. In Appendix \ref{a_additional_figures} we provide additional figures showing the bound-bound contributions from all ions and compositional zones, as well as a version of Fig.~\ref{f_dh_spec_evo_comp} on a linear flux-scale.

As seen in Fig.~\ref{f_dh_spec_evo_comp}, there is a good qualitative, and in most aspects also quantitative agreement, between the standard model and the observations of SN 2011dh. Before $\sim$10 days, when the emission comes mainly from the hydrogen envelope the agreement is a bit worse (not shown). This is possibly an effect of the choice of initial conditions for the model (see Sect.~\ref{s_ejecta_model}), where the initial cooling of the thermal explosion energy, lasting a few days, has been ignored. Models for the spectral evolution of Type IIb SNe during this phase, which depends critically on the radius and the mass of the hydrogen envelope, has been presented by \citet{Des18b}. Another difference is the evolution redwards $\sim$2 $\mu$m where a strong excess develops in the observed spectrum, beginning already at $\sim$60 days. This excess was discussed in \citetalias{Erg15}, and was attributed to dust formation in the ejecta, although CO first overtone emission was detected at 206 days, and therefore contributes to the excess between $\sim$2.3 and $\sim$2.5 $\mu$m.

The main signature of a Type IIb SN is the transition from a hydrogen to a helium dominated spectrum, and this is well reproduced by the model. Initially, the hydrogen lines are strong and emission from the hydrogen envelope is dominating. Already at $\sim$10 days emission from the helium envelope starts to dominate redwards $\sim$5000 \AA, and between 10 and 15 days the helium lines appear, grow stronger, and eventually dominate the spectrum at $\sim$40 days. Hydrogen line emission disappears on a similar time-scale, completing the transition, although the Balmer lines remain considerably longer in absorption. The first 40 days is also the period over which the contribution from continuum processes fades away. Initially, this contribution is substantial redwards $\sim$5000 \AA~and dominating in the NIR, but then quickly fades away, although it remains important in the $H$-band and redwards 2.3 $\mu$m until $\sim$40 days. 
 
After $\sim$40 days, emission from the carbon-oxygen core becomes increasingly important and at $\sim$100 days it dominates redwards $\sim$5000 \AA. As a consequence, emission from heavier elements abundant in the core increases, in particular after $\sim$100 days, when the characteristic [\ion{O}{i}] 6300,6364 \AA~and [\ion{Ca}{ii}] 7291,7323 \AA~lines appear. This is also the moment when the carbon-oxygen core becomes fully transparent in the continuum (see Fig.~\ref{f_12C_matter_evo}), and therefore marks the transition into the nebular phase. This transition, in itself a demanding test of the code, is nicely reproduced by the model. As discussed in Sect.~\ref{s_effect_macro}, this is partly due to our treatment of the macroscopic mixing. 

Below, we discuss the most important lines originating from the different elements in some detail (Sects.~\ref{s_spec_evo_H}-\ref{s_spec_evo_Fe_Ni}), as well as the line velocities measured from their absorption maxima (Sect.~\ref{s_spec_evo_line_vel}), and again compare to observations of SN 2011dh. In addition, with JWST in mind, we also discuss the model infrared (IR) spectra in Sect.~\ref{sec_IR}.

\subsubsection{Hydrogen}
\label{s_spec_evo_H}

The contribution from hydrogen lines is shown in Fig.~\ref{f_12C_spec_trans_evo}, and as mentioned it is initially strong, but fades away after $\sim$10 days, when the photosphere retreats into the increasingly transparent helium envelope (see Fig.~\ref{f_12C_matter_evo}). This trend is most pronounced for the recombination driven Paschen lines, which disappear towards $\sim$40 days. Balmer line emission fades on a similar time-scale, whereas absorption remains for a longer time, and even increases before $\sim$40 days. Contrary to the other Balmer lines, H$\alpha$ initially shows a clear P-Cygni profile, but after $\sim$10 days it becomes increasingly blended with the \ion{He}{i} 6678 \AA~line and attains the double-peaked shape so characteristic in Type IIb SNe. Note, that Paschen $\gamma$ is blended with the \ion{He}{i} 1.083 $\mu$m line during the transition period.

\begin{figure}[tbp!]
\includegraphics[width=0.49\textwidth,angle=0]{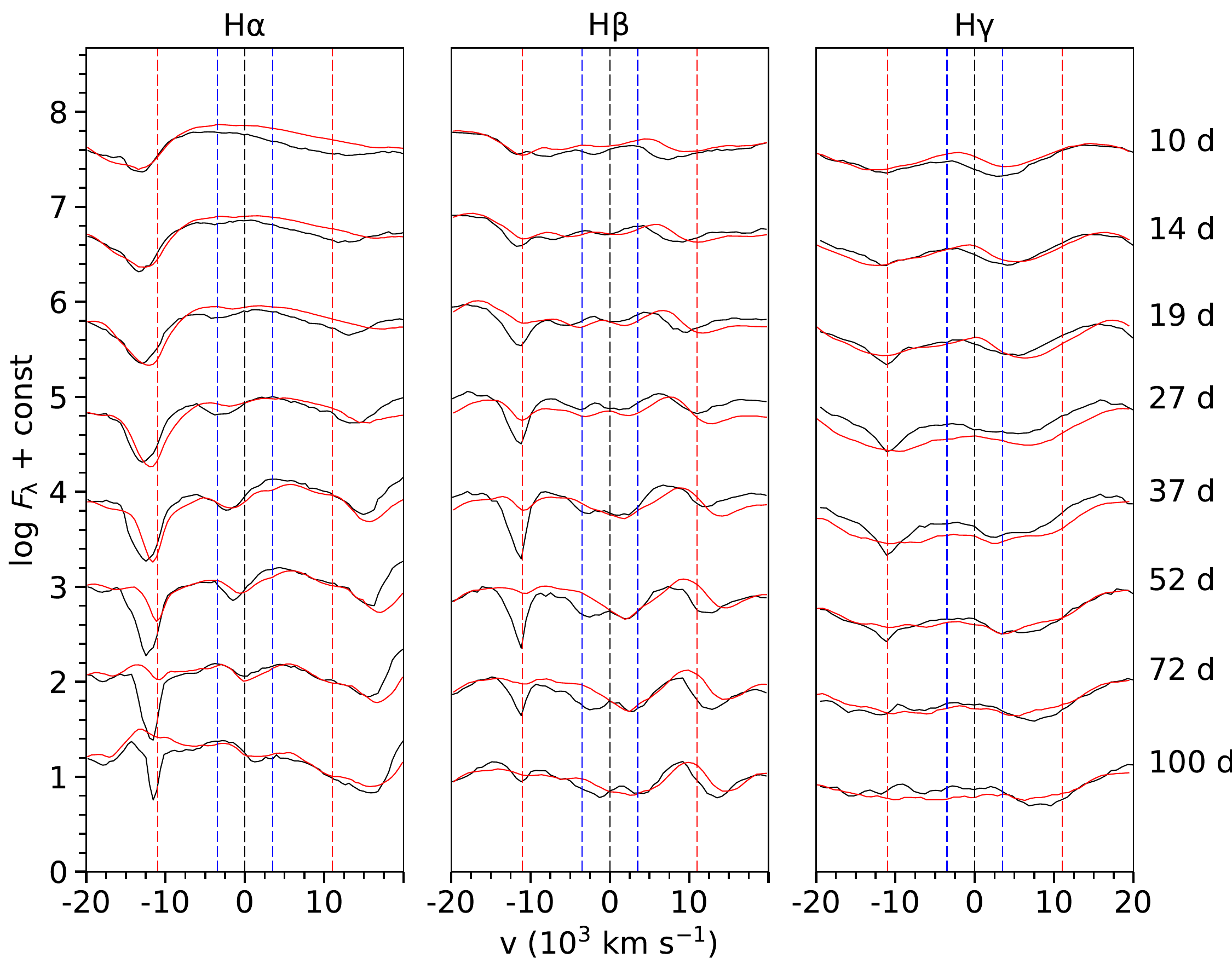}
\caption{Evolution of hydrogen lines for the standard model (black) compared to the observations of SN 2011dh (red). Spectra from 8 logarithmically spaced epochs between 10 and 100 days are shown, where the model and observed spectra at each epoch have been shifted by the same, but otherwise arbitrary amount. Here, as well as in Figs.~\ref{f_dh_spec_evo_He_opt_NIR}-\ref{f_dh_spec_evo_Fe_Co}, we also show the velocity extent of the helium envelope (red lines) and the carbon-oxygen core (blue lines).}
\label{f_dh_spec_evo_H}
\end{figure}

Figure \ref{f_dh_spec_evo_H} shows the evolution of the Balmer lines compared to SN 2011dh. The evolution is qualitatively similar, but the absorption is significantly stronger and remains longer in the model, suggesting that the $\sim$0.05 M$_\odot$ of hydrogen in the model is too high. This is in line with the 0.02-0.04 and 0.024 M$_\odot$ of hydrogen estimated through spectral modelling of SN 2011dh by \citetalias{Erg14}\footnote{Using an early version of JEKYLL assuming steady-state, LTE and (electron and line) scattering only.} and \citet{Mar13}, respectively. Note, that this implies either a less massive hydrogen envelope or a lower mass-fraction of hydrogen than in the model, an issue that can not be resolved without further modelling. In agreement with the spectral modelling in \citetalias{Erg14}, we also find that the absorption minimum of the Balmer lines asymptotically approaches the velocity of the helium/hydrogen envelope interface. The Type IIb models by \citet{Des15,Des16,Des18b} behave in a similar way, and a stagnation of the absorption velocity for the Balmer lines is observed in most Type IIb SNe \citep[see e.g.][]{Liu16}. The stagnation velocity varies among different SNe, and for the most well-observed Type IIb SNe 1993J, 2011dh and 2008ax it is $\sim$9000, $\sim$11000 and $\sim$13000 km s$^{-1}$ respectively, suggesting  progressively lower hydrogen masses for these SNe \citepalias{Erg14}.

\subsubsection{Helium}
\label{s_spec_evo_He}

The contribution from helium lines is shown in Fig.~\ref{f_12C_spec_trans_evo}, and as mentioned, it increases strongly between 10 and 15 days, dominates the spectrum between 20 and 60 days and thereafter fades away in the optical, but remains important in the NIR. As demonstrated by \citet{Luc91}, non-thermal excitation and ionization are essential to populate the excited levels of \ion{He}{i}, in turn required to produce the lines observed. This was confirmed by \citet{Des12}, and in \citetalias{Erg18} we showed that if the non-thermal excitation and ionization were turned off in a microscopically mixed version of the model explored here, the helium lines disappeared.

\begin{figure}[tbp!]
\includegraphics[width=0.49\textwidth,angle=0]{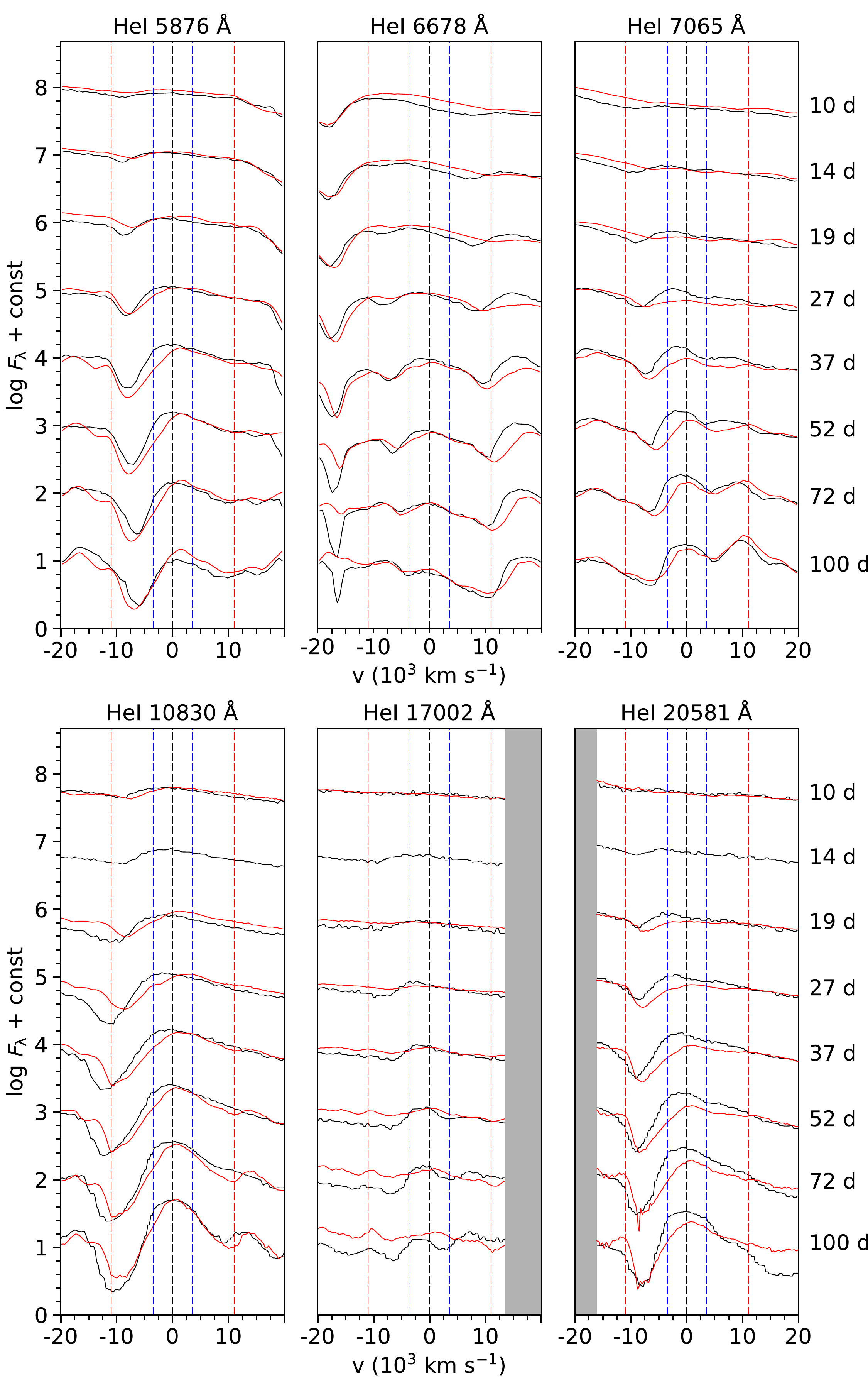}
\caption{Evolution of optical (upper panel) and NIR (lower panel) helium lines for the standard model (black) compared to the observations of SN 2011dh (red). Otherwise as in Fig.~\ref{f_dh_spec_evo_H}.}
\label{f_dh_spec_evo_He_opt_NIR}
\end{figure}

Figure \ref{f_dh_spec_evo_He_opt_NIR} shows the evolution of the \ion{He}{i} 5876 \AA, 6678 \AA, 7065 \AA, 1.083 $\mu$m, 1.700 $\mu$m and 2.058 $\mu$m lines compared to SN 2011dh. These are the lines that stand out most clearly in the model, but several other weaker and blended lines like \ion{He}{i} 3889 \AA, 4471 \AA, 5016 \AA, 7281 \AA~and 1.278 $\mu$m are also present. The helium lines almost exclusively originate from the helium-rich material in the envelope, but after $\sim$150 days an increasing contribution to the \ion{He}{i} 1.083 $\mu$m line comes from the Ni/He clumps in the core. The \ion{He}{i} 5876 \AA, 6678 \AA~and 1.083 $\mu$m lines are quite well reproduced by the model, whereas the \ion{He}{i} 7065 \AA~and 2.058 $\mu$m lines are overproduced in emission, and the weaker \ion{He}{i} 1.700 $\mu$m line (mainly seen in emission) is considerably stronger than in the observed spectra. The P-Cygni profiles for most of the helium lines are shifted towards higher velocities than observed, most pronounced at early times. Given the importance of non-thermal processes for the helium lines, we may speculate that these differences are related to the distribution of the radioactive material, and that too much of this have been mixed into the helium envelope in the model. Note, that after $\sim$40 days most of the \ion{He}{i} 5876 \AA~line emission is scattered by the \ion{Na}{i} 5890, 5896 \AA~lines, and before $\sim$25 days the \ion{He}{i} 1.083 $\mu$m line is blended with Paschen $\gamma$ and the \ion{Mg}{ii} 1.091, 1.095 $\mu$m lines.

\begin{figure}[tbp!]
\includegraphics[width=0.49\textwidth,angle=0]{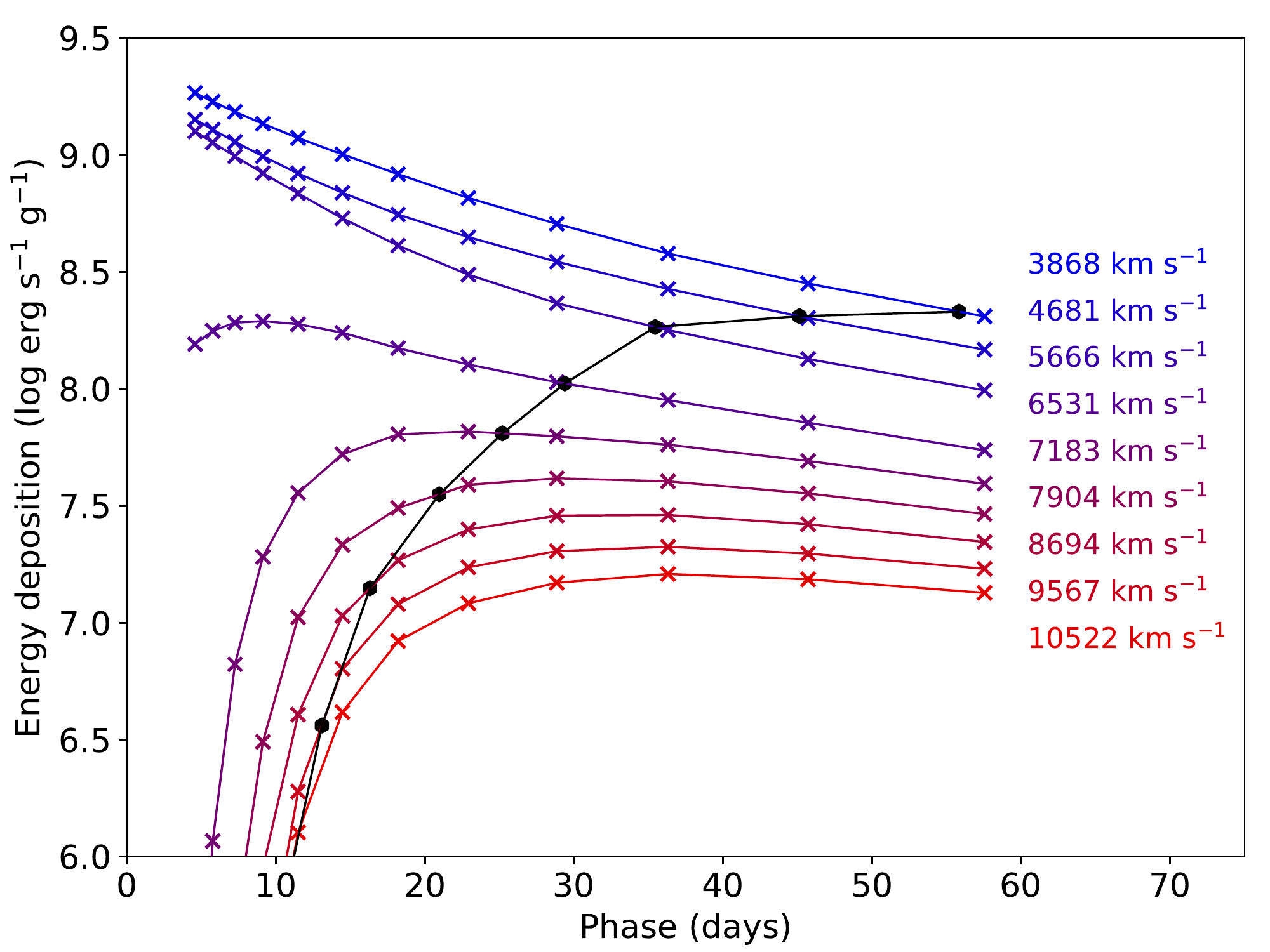}
\caption{Evolution of the radioactive energy deposition in the helium envelope (red to blue crosses) and the position of the (Rosseland mean) continuum photosphere (black circles) for the standard model.}
\label{f_12C_eRadio_evo}
\end{figure}

As seen in Fig.~\ref{f_dh_spec_evo_He_opt_NIR}, the \ion{He}{i} 1.083 $\mu$m absorption migrates outward in velocity until ~$\sim$40 days. This behaviour is also observed in SN 2011dh, although in that case the other helium lines showed a similar, but less pronounced trend. In \citetalias{Erg14} we suggested that the evolution of the helium lines is driven mainly by the ejecta becoming optically thin to the $\gamma$-rays. This idea is supported by Fig.~\ref{f_12C_eRadio_evo}, which shows the evolution of the radioactive energy deposition in the helium envelope. Between 10 and 20 days there is a strong increase in the energy deposition outside the photosphere, corresponding well to the period when the helium lines grow in strength. Note, however, that there is a weak helium signature also before $\sim$10 days, mainly visible in the \ion{He}{i} 5876 \AA~line. We also see that the energy deposition in the outermost helium layers continues to increase to $\sim$40 days, which may explain the evolution of the \ion{He}{i} 1.083 $\mu$m line. The outward migration of the \ion{He}{i} 1.083 $\mu$m absorption is also present in the Type I/IIb models (e.g.~model 3p65Ax1) by \citet{Des15,Des16}, and was noted and discussed by the authors, who also provide a similar explanation.

\subsubsection{Carbon to calcium}
\label{s_spec_evo_C_Ca}

\begin{figure}[tbp!]
\includegraphics[width=0.49\textwidth,angle=0]{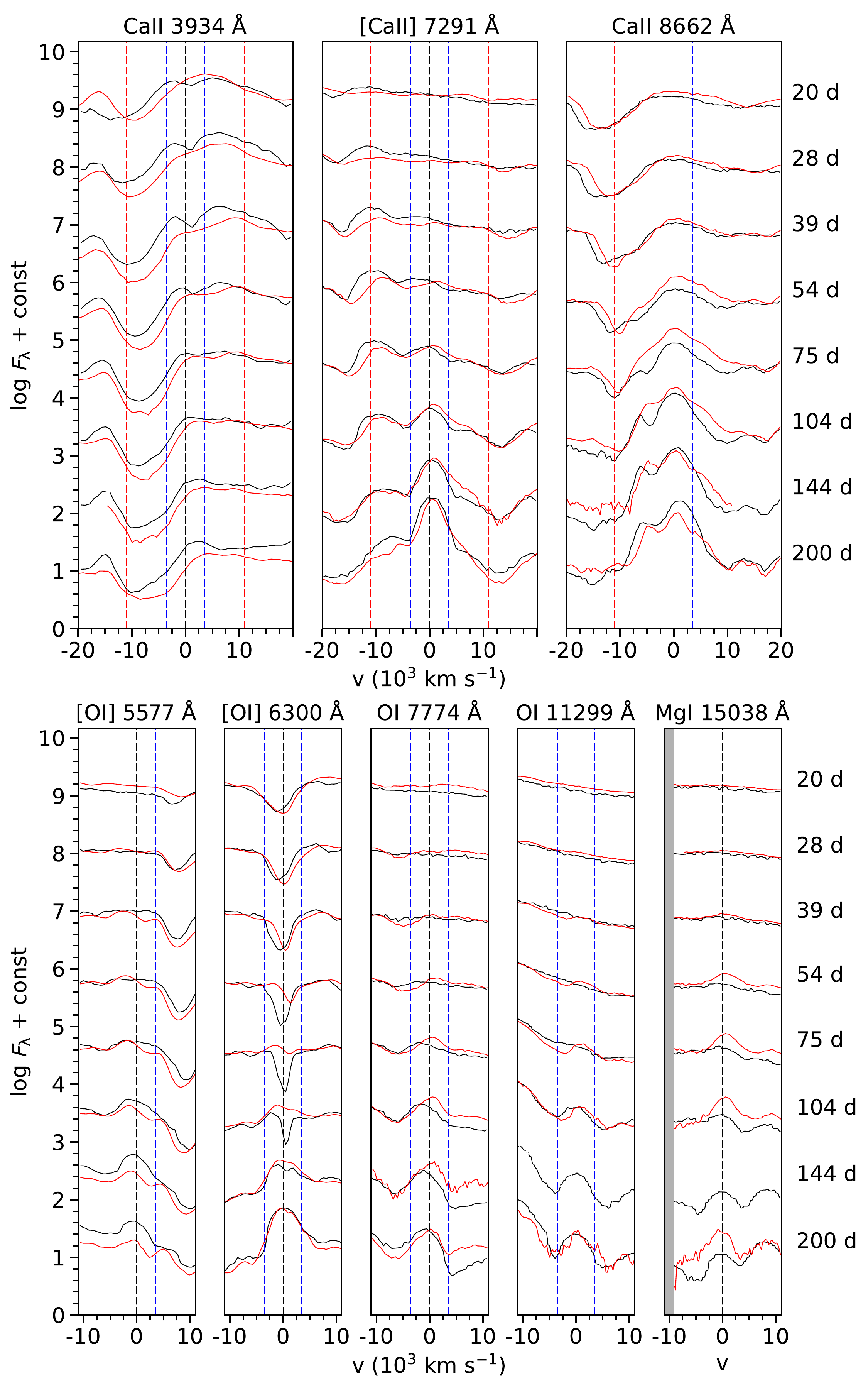}
\caption{Evolution of calcium, oxygen and magnesium lines for the standard model (black) compared to the observations of SN 2011dh (red). Spectra from  eight logarithmically spaced epochs between 20 and 200 days are shown, where the model and observed spectra at each epoch have been shifted by the same, but otherwise arbitrary amount. Otherwise as in Fig.~\ref{f_dh_spec_evo_H}.}
\label{f_dh_spec_evo_Ca_O_Mg}
\end{figure}

The line contribution from elements in the carbon to calcium range is shown in Fig.~\ref{f_12C_spec_trans_evo}, and except for the calcium lines which are strong at all times, the contribution increases after $\sim$40 days when the core, rich in these elements becomes increasingly transparent (see Fig.~\ref{f_12C_spec_cell_evo}). The upper panel of Fig.~\ref{f_dh_spec_evo_Ca_O_Mg} shows the evolution of the \ion{Ca}{ii} 3934, 3968 \AA, \ion{Ca}{ii} 8498, 8542, 8662 \AA~(hereafter \ion{Ca}{ii} triplet) and [\ion{Ca}{ii}] 7291, 7323 \AA~lines compared to SN 2011dh. The [\ion{Ca}{ii}] 7291, 7323 \AA~lines mainly originate from the Si/S clumps in the core, whereas the \ion{Ca}{ii} 3934, 3968 \AA~lines and the \ion{Ca}{ii} NIR triplet mainly originate from the hydrogen- and helium-rich material in the envelope. After $\sim$100 days an increasing contributing to the latter two comes from the Ni/He clumps in the helium envelope and the core. After $\sim$15 days the evolution of these lines is well reproduced by the model. However, at earlier times absorption in the \ion{Ca}{ii} 3934, 3968 \AA~lines and the \ion{Ca}{ii} triplet extends further out in the model (not shown). Again, this is possibly an effect of the choice of initial conditions, where we neglect the initial cooling of the thermal explosion energy.

The lower panel of Fig.~\ref{f_dh_spec_evo_Ca_O_Mg} shows the evolution of the [\ion{O}{i}] 5577 \AA, [\ion{O}{i}] 6300, 6364 \AA, \ion{O}{i} 7774 \AA, \ion{O}{i} 1.129, 1.130 $\mu$m and \ion{Mg}{i} 1.504 $\mu$m lines compared to SN 2011dh. All of these lines mainly originate from the oxygen-rich clumps in the core. The overall evolution of the oxygen lines is fairly well reproduced, but the [\ion{O}{i}] 5577 \AA, \ion{O}{i} 7774 \AA~and \ion{O}{i} 1.129, 1.130 $\mu$m lines appear later and are initially weaker than observed for SN 2011dh. The \ion{O}{i} 9263 \AA~line is also present in the model, but is blended with the [\ion{Co}{ii}] 9336, 9343 \AA~lines (see Sect.~\ref{s_spec_evo_Fe_Ni}). 

Similar to the oxygen lines, the \ion{Mg}{i} 1.504 $\mu$m line appears later and is weaker than observed for SN 2011dh, although it remains weaker also in the nebular phase (see \citetalias{Jer15} for a discussion of the strengths of the \ion{Mg}{i} lines in the nebular phase). The \ion{Mg}{i}] 4571 \AA~line does not emerge until $\sim$250 days in the model spectra but have begun to rise in the observed spectra at 200 days. We note, that whereas the radioactive Ni/He material is mixed into the helium envelope in the model, the oxygen and magnesium rich material is not, which may explain the early suppression of the oxygen and magnesium lines compared to observations. 

As mentioned in Sect.~\ref{s_spec_evo_He}, the \ion{Na}{i} 5890, 5896 \AA~lines dominate the \ion{He}{i} 5876 \AA~line after $\sim$40 days, and as seen in Fig.~\ref{f_dh_spec_evo_He_opt_NIR} the subsequent evolution is well reproduced. In the model, the feature emerging at $\sim$1.18 $\mu$m after $\sim$60 days is mainly caused by the \ion{C}{i} 1.176 $\mu$m line, and towards $\sim$150 days the [\ion{C}{i}] 8727 \AA~line, originating mainly from the O/C and He/C clumps in the core, begins to contribute significantly to the blend with the \ion{Ca}{ii} triplet.

\subsubsection{Scandium to manganese}
\label{s_spec_evo_Sc_Mn}

The line contribution from elements in the scandium to manganese range is shown in Fig.~\ref{f_12C_spec_trans_evo}, and is dominating in the 3000-4000 \AA~region and important in the 4000-5000 \AA~region at all times. After $\sim$40 days it also contributes significantly to the optical emission redwards 5000 \AA. The emission almost exclusively originates from singly ionized ions, and the strongest contribution comes from \ion{Ti}{ii}. It is the result of scattering and fluorescence in numerous transitions, and individual lines are hard to distinguish. Line-blocking by elements in the scandium to manganese range is important for the suppression of the emission bluewards $\sim$5000 \AA, and in particular in the 3000-4000 \AA~region, corresponding roughly to the $U$-band. 

\subsubsection{Iron to nickel}
\label{s_spec_evo_Fe_Ni}

The line contribution from elements in the iron to nickel range is shown in Fig.~\ref{f_12C_spec_trans_evo}, and this contribution is strong at all times, in particular in the 4000-5500 \AA~range. After $\sim$40 days the contribution increases also at other wavelengths, likely due to emission from the increasingly transparent core (see Fig.~\ref{f_12C_spec_cell_evo}), where about half of the Ni/He material resides. Except for the $U$-band (see Sect.~\ref{s_spec_evo_Sc_Mn}), blocking through scattering and fluorescence in numerous \ion{Fe}{ii} lines is the main cause for the suppression of the emission bluewards $\sim$5500 \AA, so important in shaping the spectra of SE SNe. However, with a few exceptions, like the \ion{Fe}{ii} 5169 \AA~line, individual iron lines are strongly blended and hard to distinguish. The contribution from nickel is insignificant at all times, but after $\sim$50 days cobalt begins to contribute to the spectrum with several distinct lines. 

\begin{figure}[tbp!]
\includegraphics[width=0.49\textwidth,angle=0]{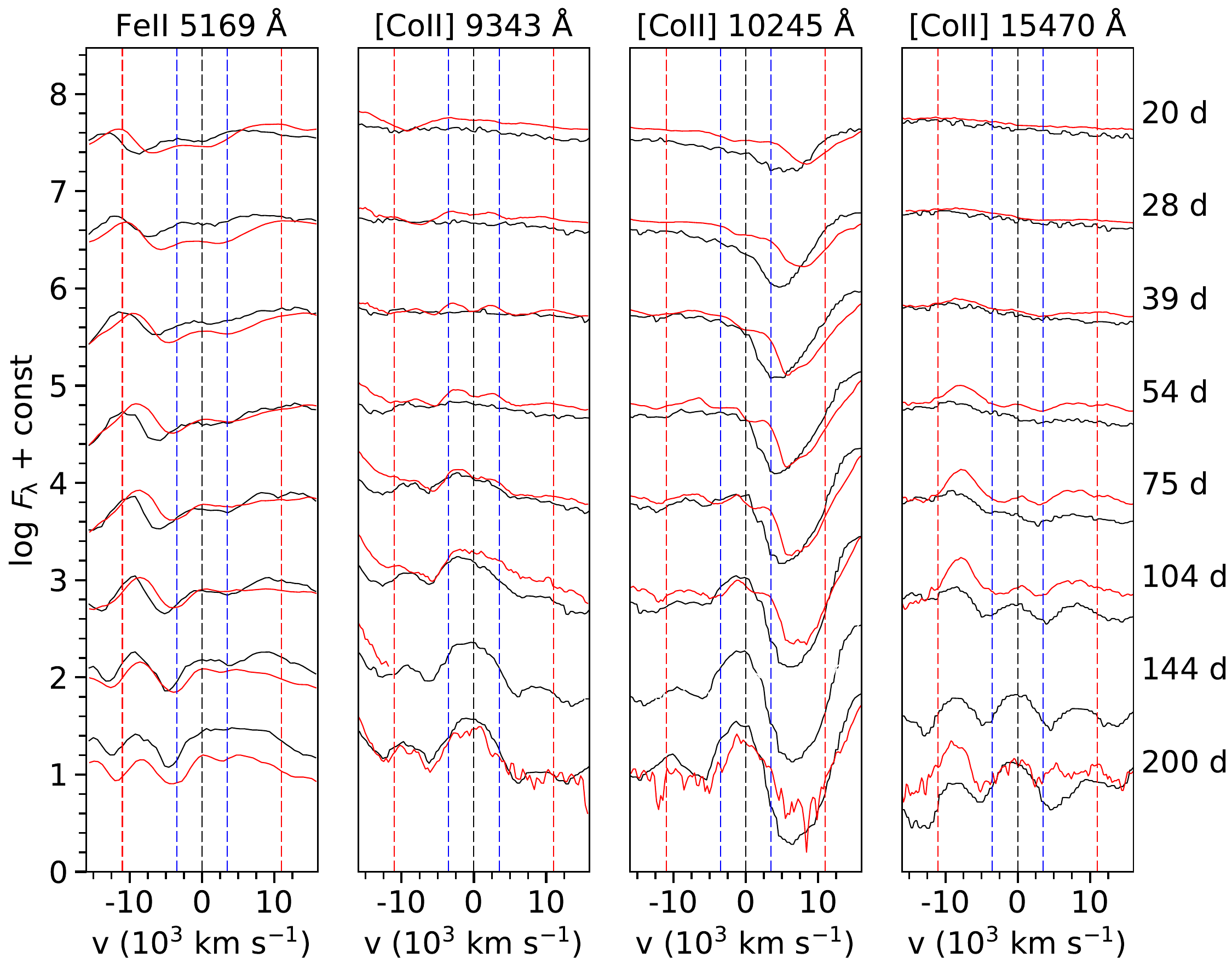}
\caption{Evolution of iron and cobalt lines for the standard model (black) compared to the observations of SN 2011dh (red). Otherwise as in Fig.~\ref{f_dh_spec_evo_Ca_O_Mg}. }
\label{f_dh_spec_evo_Fe_Co}
\end{figure}

Figure~\ref{f_dh_spec_evo_Fe_Co} shows the evolution of the \ion{Fe}{ii} 5169 \AA~and [\ion{Co}{ii}] 9336, 9343 \AA, [\ion{Co}{ii}] 1.019, 1.024, 1.028 $\mu$m and [\ion{Co}{ii}] 1.547 $\mu$m lines compared to SN 2011dh. Initially, the \ion{Fe}{ii} 5169 \AA~line originates from the hydrogen and helium-rich material in the envelope, but after $\sim$40 days it mainly originates from the Ni/He clumps in the helium envelope and the core. The \ion{Co}{ii} lines mainly originate from the Ni/He and Si/S clumps in the core, but with some contribution from the Ni/He clumps in the helium envelope. The \ion{Fe}{ii} 5169 \AA~line is reasonably well reproduced, but initially absorption extends to higher velocities than observed, a discrepancy that disappears towards 150 days. The overall evolution of the cobalt lines is fairly well reproduced, but note that the [\ion{Co}{ii}] 9336, 9343 \AA~lines are blended with the \ion{O}{i} 9263 \AA~line. The iron and cobalt lines are interesting as they are directly linked to the distribution of the Ni/He material in the ejecta, and the width of the cobalt lines seems to suggest that this distribution is similar in SN 2011dh as in the model. On the other hand, as the emission originates mainly from the core it is unclear how strong this constraint is.

\subsubsection{The infrared spectrum}
\label{sec_IR}

The NIR and mid-IR (MIR) spectrum is of special interest because of the relatively isolated lines especially in the nebular phase. Because of their low excitation temperature these lines are also relatively insensitive to temperature and therefore especially suitable for abundance estimates. Several elements, like Ne and Ar, also have some of the few strong lines in this range. The usefulness of this range was demonstrated by the Kuiper Airborne Observatory (KAO) observations of SN 1987A \citep[e.g.,][]{Wood93} and similar observations of more distant SNe will be possible with the NIRSPEC and MIRI instruments on the James Webb Space telescope (JWST). Because of their obvious interest for JWST we will in this section discuss this in some detail, and deviate from the comparison to SN 2011dh.

\begin{figure*}[tbp!]
\includegraphics[width=1.0\textwidth,angle=0]{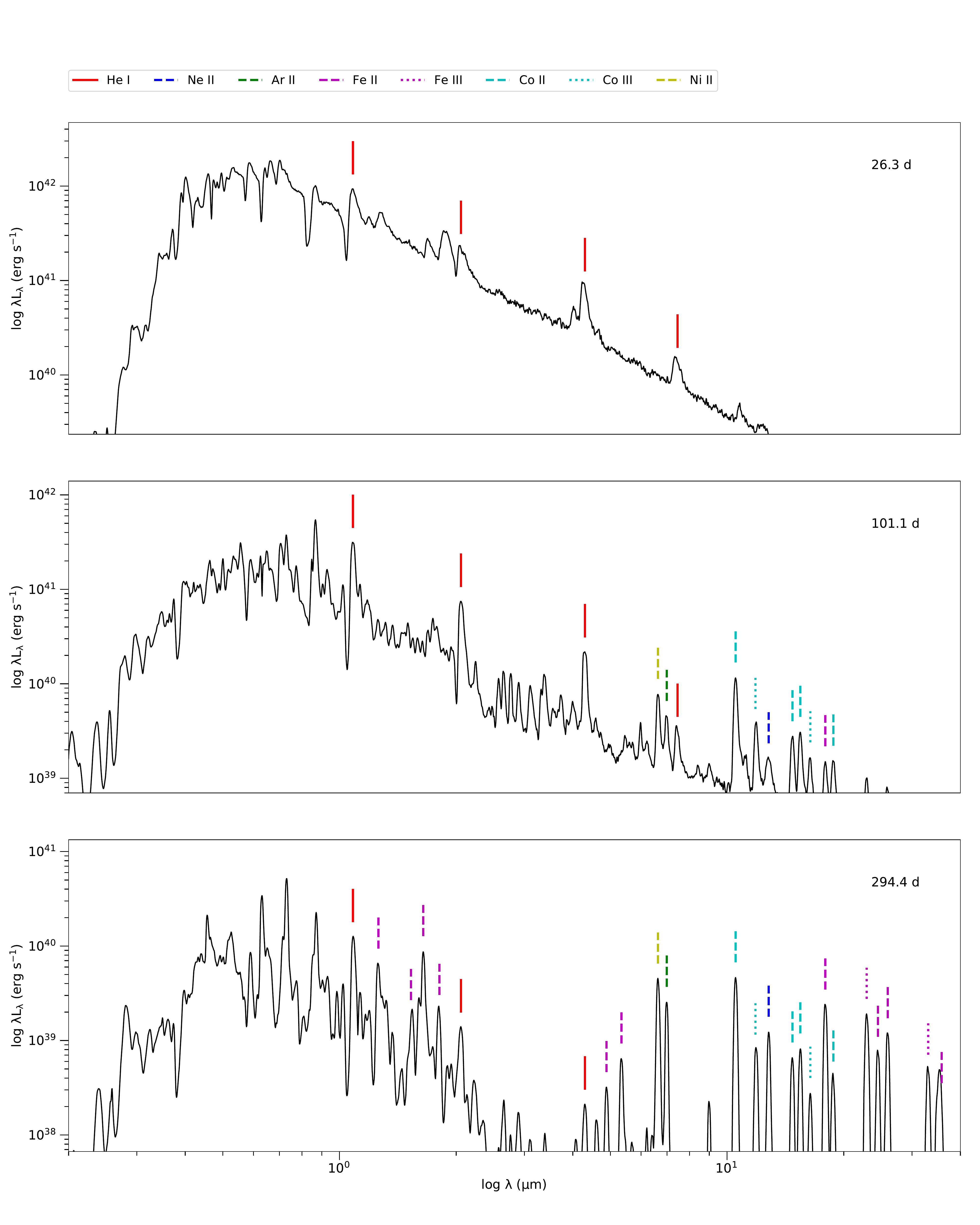}
\caption{Optical, NIR and MIR spectra for the standard model at 26 days (near peak), 101 days and at 294 days (nebular phase). The vertical lines show the strongest IR lines from the ions given in the legend at the top.}
\label{fig_NIR_MIR}
\end{figure*}

In Fig.~\ref{fig_NIR_MIR} we show the full spectra at three representative epochs. The IR spectrum at 26 days, close to maximum, is mainly an extension to the optical spectrum, with a few strong He I lines at $1.083, 2.058, 4.296$ and  $7.456~\mu$m and a strong continuum. While the $1.083~\mu$m line has a strong P-Cygni absorption the other lines are mainly in emission. At 100 days the IR spectrum is becoming more nebular with strong emission lines from [\ion{Ni}{ii}] $6.636~\mu$m, [\ion{Ar}{ii}] $6.985~\mu$m, [\ion{Co}{ii}] $10.52~\mu$m, and [\ion{Co}{iii}] $11.88~\mu$m. 

The last spectrum at 294 days is fully nebular above $\sim 2.5~\mu$m. There is at this phase only a weak continuum. In the NIR lines of [\ion{Fe}{ii}] $1.257, 1.534, 1.644$ and $1.810~\mu$m are strong. The line at $\sim$1.64 $\mu$m is, however, a blend of [\ion{Si}{I}] $1.646~\mu$m and [\ion{Fe}{ii}] $ 1.644~\mu$m.  The MIR lines have luminosities comparable to the optical lines due to the decreasing temperature. In addition to the [\ion{Ni}{ii}] $6.636~\mu$m,  [\ion{Ar}{ii}] $6.985~\mu$m, [\ion{Co}{ii}] $10.52~\mu$m, and [\ion{Co}{iii}] $11.88~\mu$m, which are strong at 100 days, also [\ion{Ne}{ii}] $12.814~\mu$m, [\ion{Fe}{ii}] $17.94, 24.52, 25.99, 35.77~\mu$m and [\ion{Fe}{iii}] $22.92, 33.04~\mu$m are now strong. Most of the lines correspond to fine-structure transitions in the ground state multiplets, which are close to be in LTE.  They are therefore  sensitive to the ionization fractions of the ions and elemental abundances, but less so to the temperature (as $h\nu/kT$ is usually small). 

Note, that we have not included any molecular or dust contribution in our models. In particular, both lines of CO and SiO from the fundamental and overtone bands may be important, as is seen in for example SN 1987A \citep[e.g.][and references therein]{Wood93}. As discussed in \citetalias{Jer15} and \citetalias{Erg15}, CO first overtone emission as well as a strong continuum excess in the IR attributed to dust were observed in SN 2011dh, which is visible in Fig.~\ref{f_dh_spec_evo_comp} as a growing discrepancy redwards $\sim$2 $\mu$m between our standard model and the observations of SN 2011dh. However, due to lack of IR data, it is unclear if this is a common feature in Type IIb SNe or not.

\subsubsection{Line velocities}
\label{s_spec_evo_line_vel}

\begin{figure}[tbp!]
\includegraphics[width=0.49\textwidth,angle=0]{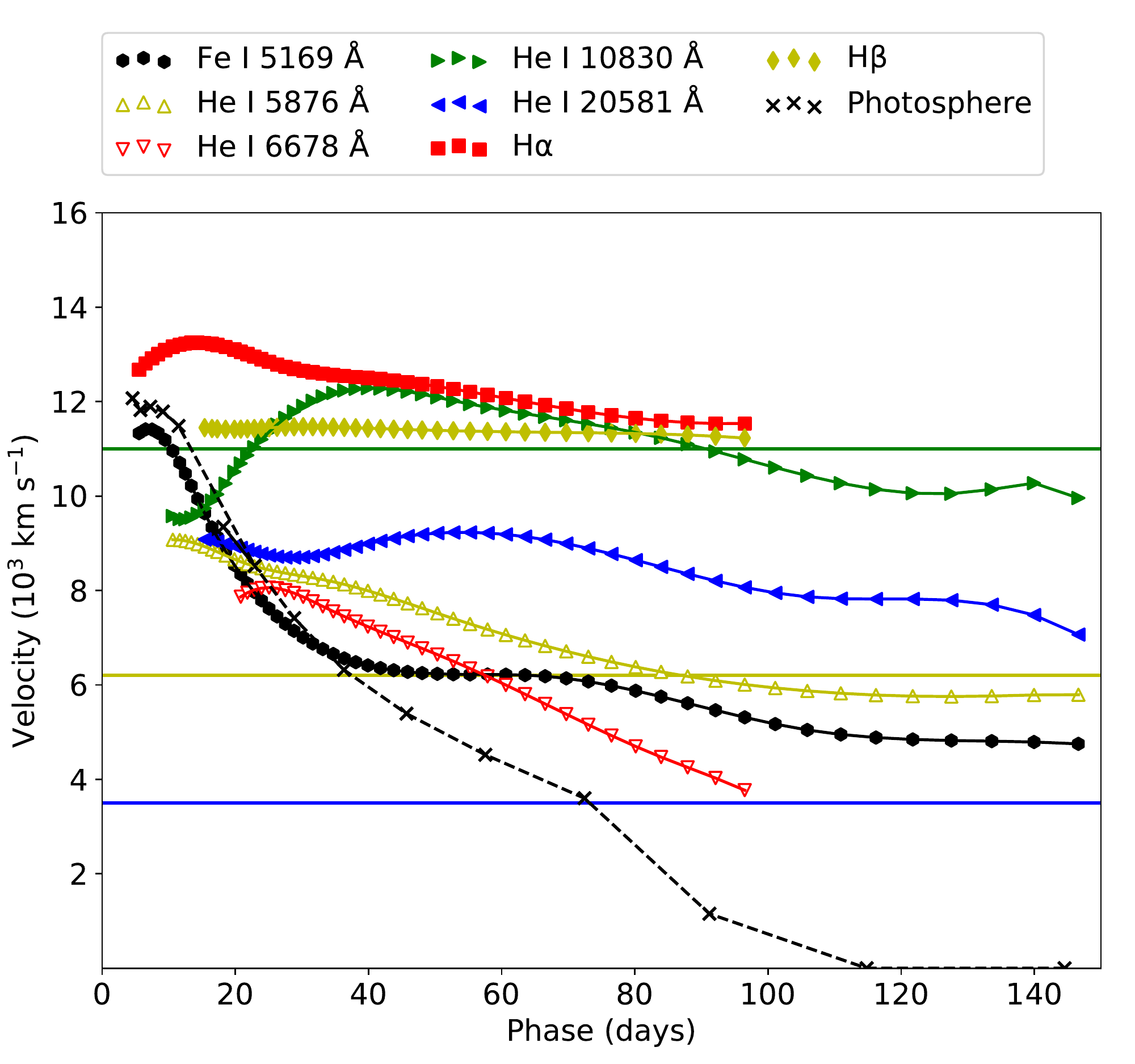}
\caption{Velocity evolution of the absorption maxima of the H$\alpha$ (red squares), H$\beta$ (yellow diamonds), \ion{He}{i} 5876 \AA~(yellow upward triangles), \ion{He}{i} 6678 \AA~(red downward triangles), \ion{He}{i} 10830 \AA~(green rightward triangles), \ion{He}{i} 20581 \AA~(blue leftward triangles) and \ion{Fe}{ii} 5169 \AA~(black circles) lines for the standard model. The black crosses show the velocity evolution of the (Rosseland mean) continuum photosphere, whereas the horizontal lines mark the outer borders of the carbon-oxygen core (blue) and inner/outer (green/yellow) helium envelope.}
\label{f_12C_vel_evo}
\end{figure}

Figure~\ref{f_12C_vel_evo} shows the velocity evolution of the absorption maxima of the H$\alpha$, H$\beta$, \ion{He}{i} 5876 \AA, \ion{He}{i} 6678 \AA, \ion{He}{i} 10830 \AA, \ion{He}{i} 20581 \AA~and \ion{Fe}{ii} 5169 \AA~lines, as well as the (Rosseland mean) continuum photosphere (compare; \citetalias{Erg15}: Fig.~14). The hydrogen lines show the highest velocities, have a flat evolution, and as discussed (Sect.~\ref{s_spec_evo_H}), they approach the velocity of the interface between the helium and hydrogen envelopes. The evolution is mostly in agreement with observations, but the high H$\alpha$ velocities observed before $\sim$10 days for SN 2011dh are not reproduced by the model, which could again be related to our choice of initial conditions. 

The helium lines appear between 10 and 15 days near the photosphere and then evolve quite differently, where the \ion{He}{i} 10830 \AA~velocity increases towards that of the interface between the helium and hydrogen envelope, the \ion{He}{i} 20581 \AA~velocity stays almost flat, and the \ion{He}{i} 5876 \AA~and 6678 \AA~velocities decline. The evolution of the \ion{He}{i} 10830 \AA~and 20581 \AA~velocities is in quite good agreement with observations, whereas the \ion{He}{i} 5876 \AA~and 6678 \AA~velocities differ more.

The evolution of the \ion{Fe}{ii} 5169 \AA~velocity follows that of the (Rosseland mean) continuum photosphere until $\sim$30 days, confirming the common assumption \citepalias[e.g.][]{Erg14} that this line is a good tracer of the photosphere during the diffusion phase. However, as mentioned before, this velocity is higher than observed for SN 2011dh, although the discrepancy disappears towards 150 days (see Fig.~\ref{f_dh_spec_evo_Fe_Co}). This seems to be a general tendency, and most lines as well as the photosphere have a bit higher velocities in the model during the diffusion phase, whereas this discrepancy disappears towards 150 days.
 
\subsection{Photometric evolution}
\label{s_phot_evo}

\begin{figure}[tbp!]
\includegraphics[width=0.49\textwidth,angle=0]{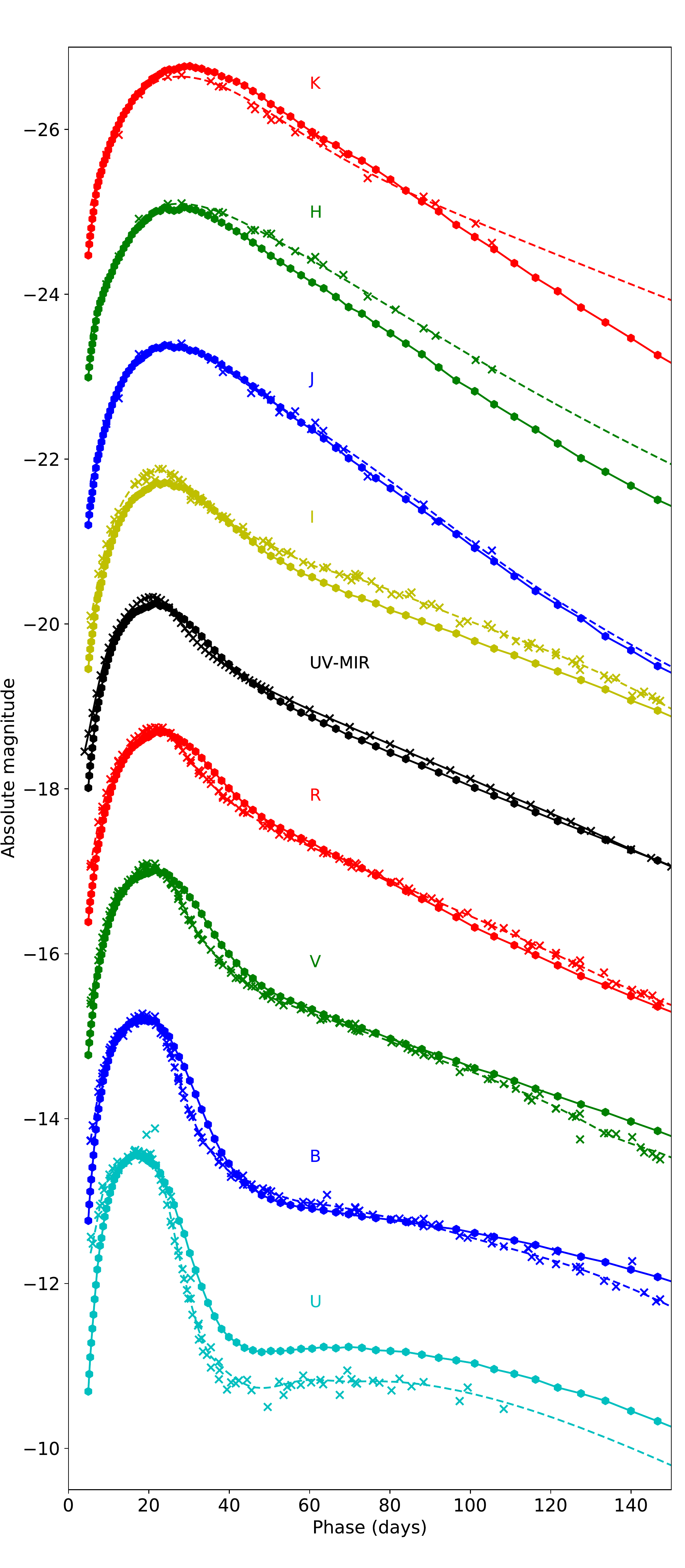}
\caption{Broad-band and bolometric lightcurves for the standard model (solid lines and circles) compared to observations of SN 2011dh (dashed lines and crosses). From bottom to top we show the U (cyan), B (blue), V (green), R (red), UV to MIR pseudo-bolometric (black), I (yellow), J (blue), H (green) and K (red) lightcurves, which for clarity have been shifted with 2.6, 1.2, 0.0, -1.4, -3.5, -4.4, -5.8, -7.5, -8.9 mags, respectively.}
\label{f_dh_lightcurve}
\end{figure}

Figure \ref{f_dh_lightcurve} shows the broad-band lightcurves for the standard model between 3 and 150 days compared to observations of SN 2011dh. In agreement with observations, the maximum occurs at increasingly later times for redder bands and the drop onto the tail is more pronounced (deeper and faster) for bluer bands. Also in agreement with observations, the early (before 100 days) tail decline rates are generally higher for redder bands, with the $J$-band lightcurve having the steepest slope and the $U$-band lightcurve being almost flat. As has been noted in several sample studies \citep[e.g.][]{Tad15,Tad17}, the aforementioned behaviour of the maximum and the subsequent decline is shared not only by SN 2011dh, but by SE SNe in general. It is also shared by the SE SNe NLTE models presented by \citet{Des15,Des16}, and the lightcurves of their Type IIb model 3p65Ax1 are qualitatively similar to those of the standard model. 

Although the model broad-band lightcurves agree quite well with the observations of SN 2011d, there are some differences worth noting. Most notable are the differences in the $U$, $H$ and $K$-bands and the evolution between 25 and 50 days, which is slower in $R$ and bluer bands than observed for SN 2011dh. The growing excess in the $K$-band could be related to dust and CO overtone emission \citepalias{Erg15}, Note, however, that there is no NIR observation between $\sim$100 and $\sim$200 days, so the evolution in this period is uncertain. The discrepancy in the $U$-band could stem from several sources, like the extinction, the photometric calibration and the metallicity adopted in the ejecta models. 

\subsection{Colour evolution}
\label{s_colour_evo}

\begin{figure}[tbp!]
\includegraphics[width=0.49\textwidth,angle=0]{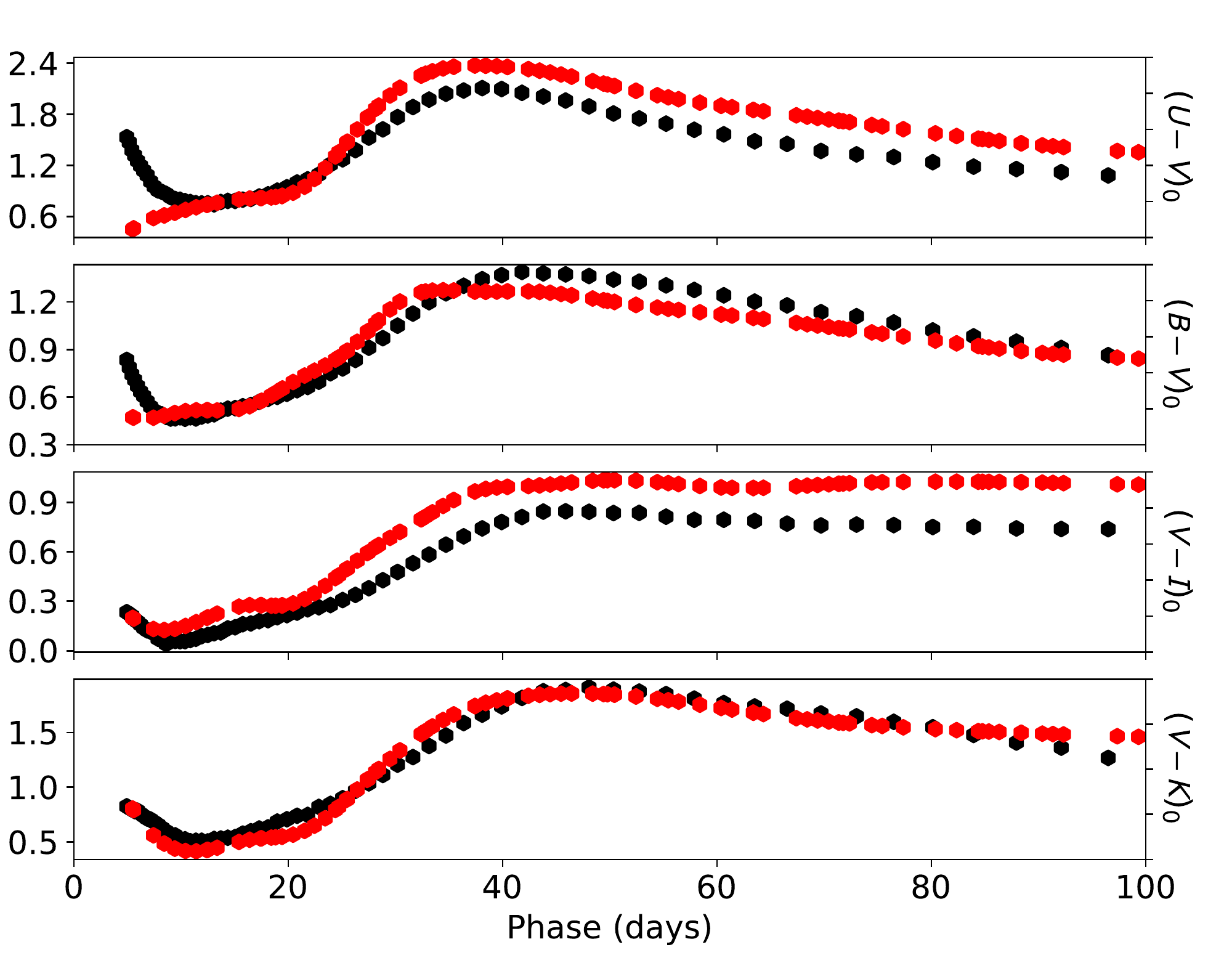}
\caption{$U-V$, $B-V$, $V-I$ and $V-K$ intrinsic colour evolution for the standard model (black) compared to observations of SN 2011dh (red).}
\label{f_dh_colour_evo}
\end{figure}

Figure \ref{f_dh_colour_evo} shows the intrinsic $U-V$, $B-V$, $V-I$ and $V-K$ colour evolution for the standard model between 3 and 100 days compared to observations of SN 2011dh. Initially, we see a blueward trend in all colours reaching a minimum at $\sim$10 days. Subsequently all colours redden and reach a maximum at $\sim$40 days, in turn followed by a slow blueward trend for all colours, although the $V-I$ colour stays almost constant. This behaviour is in agreement with observations, although SN 2011dh does not show an initial blueward trend in the $U-V$ and $B-V$ colours, likely due to the influence of an initial cooling tail, not present in the model due to our choice of initial conditions. 

As has been noted in several sample studies \citep[e.g.][]{Str17}, the properties of the colour evolution discussed here are shared not only by SN 2011dh, but by SE SNe in general. They are also shared by the SE SNe NLTE models presented by \citet{Des15,Des16}, and the colour evolution of their Type IIb model 3p65Ax1 is qualitatively similar to that of our standard model. As was noted for the lightcurve, the model evolution after $\sim$20 days is a bit slower than observed for SN 2011dh. The model $V-I$ and $U-V$ colours are bluer than observed for SN 2011dh, reflecting differences in the $I$- and $U$-band lightcurves.

\subsection{Bolometric evolution and summary}
\label{s_bol_evo}

Figure~\ref{f_dh_lightcurve} shows the pseudo-bolometric UV to MIR lightcurve for the standard model between 3 and 150 days compared to observations of SN 2011dh. Similar to the broad-band lightcurves, the agreement is good, but the diffusion peak in the model is slightly broader than the observed one, and also has a slightly different shape. They also differ before $\sim$3 days (not shown) due to the absence of a cooling tail in our model.

Overall, the agreement between the standard model and the observations of 2011dh is good, in particular since no fine tuning has been done. However, there are also differences. Some of these have straightforward explanations, as the stronger hydrogen lines (mass of hydrogen envelope) and the excess in the K-band (dust and CO). Other are harder to nail down, but we may speculate that the differences in the helium, oxygen and magnesium lines at early times are related to the macroscopic mixing. The slightly broader diffusion peak and higher photospheric velocities may indicate a lower ejecta mass, but as these quantities also depend on the macroscopic mixing this conclusion is uncertain. 

To better understand the differences a more extensive parameter study is needed, where the parameters of the macroscopic mixing, as well as the initial mass and explosion energy are explored. The limited set of models in this paper merely serves the purpose to demonstrate and study the effects of the clumping geometry on our results. Nevertheless, the overall good agreement in the comparison strengthens the conclusion in earlier works (\citealt{Mau11}; \citealt{Ber11}; \citetalias{Erg15}; \citetalias{Jer15}) that SN 2011dh originated from a star with initial mass of $\sim$12 M$_\odot$ that had lost all but a tiny fraction (<0.1 M$_\odot$) of its hydrogen envelope, strongly suggesting a binary origin.

Note, however, that the results based on hydrodynamical lightcurve modelling in \citealt{Ber11} and \citetalias{Erg15} neglect both macroscopic mixing and NLTE. As both of this may have a considerable effect on the bolometric lightcurve, the reader may wonder why the best-fit hydrodynamical models from these works are similar to model 12C. A possible explanation is that in a model like 12C, the effects of macroscopic mixing and NLTE on the bolometric lightcurve counteract and may cancel each other. That said, it remains to be seen if this is a general tendency or not.

\section{The effects of macroscopic mixing}
\label{s_effect_macro}

Mixing of the SN ejecta occurs in the explosion due to hydrodynamical instabilities \citep[e.g.][]{Mul91}, and is thought to take place on macroscopic scales only \citep[e.g.][]{Fry91}. The resulting 3-D structure is further altered by the expansion of clumps containing radioactive material due to heating from radioactive decays \citep[e.g.][]{Her91}. 

In JEKYLL, the macroscopic mixing of the ejecta is simulated through use of the Virtual Grid method \citepalias{Jer11}, where the compositional layers are assumed to be fragmented into spherical clumps, and mixed with each-other. As JEKYLL is the only spectral-synthesis code that combines time-dependence, NLTE and a self-consistent treatment of the macroscopic mixing, it is of great interest to investigate the effects of the macroscopic mixing on our results. To achieve this we compare the macroscopically mixed standard model presented in Sect.~\ref{s_application} with a microscopically mixed version, where the composition and density have been averaged over the compositional zones. To investigate the effects of the clumping geometry, we also explore a set of macroscopically mixed models differing in the expansion of the Ni/He and Si/S clumps as well as the sizes of the clumps. The full set of models and their differences are described in Sect.~\ref{s_ejecta_model}. 

As we will see, the macroscopic mixing has several effects, both on the state of matter and the radiative transfer, and in the latter case a geometrical aspect enters the problem. These effects are not limited to the Type IIb case explored here, and our results have implications for SE SNe as well as CC SNe in general. Below we discuss the effects on the state of matter, the radiative transfer, the radioactive energy deposition, and the observed spectra and lightcurves. 

\subsection{State of matter}
\label{s_effect_macro_matter}

The different composition and density in the clumps compared to microscopically mixed ejecta, give rise to a different state of matter, that is different populations of ionized and bound states and a different temperature. In addition, the state of matter depends on the radiation field and the radioactive energy deposition, which also differ in a macroscopically mixed model.

\begin{figure*}[tbp!]
\includegraphics[width=1.0\textwidth,angle=0]{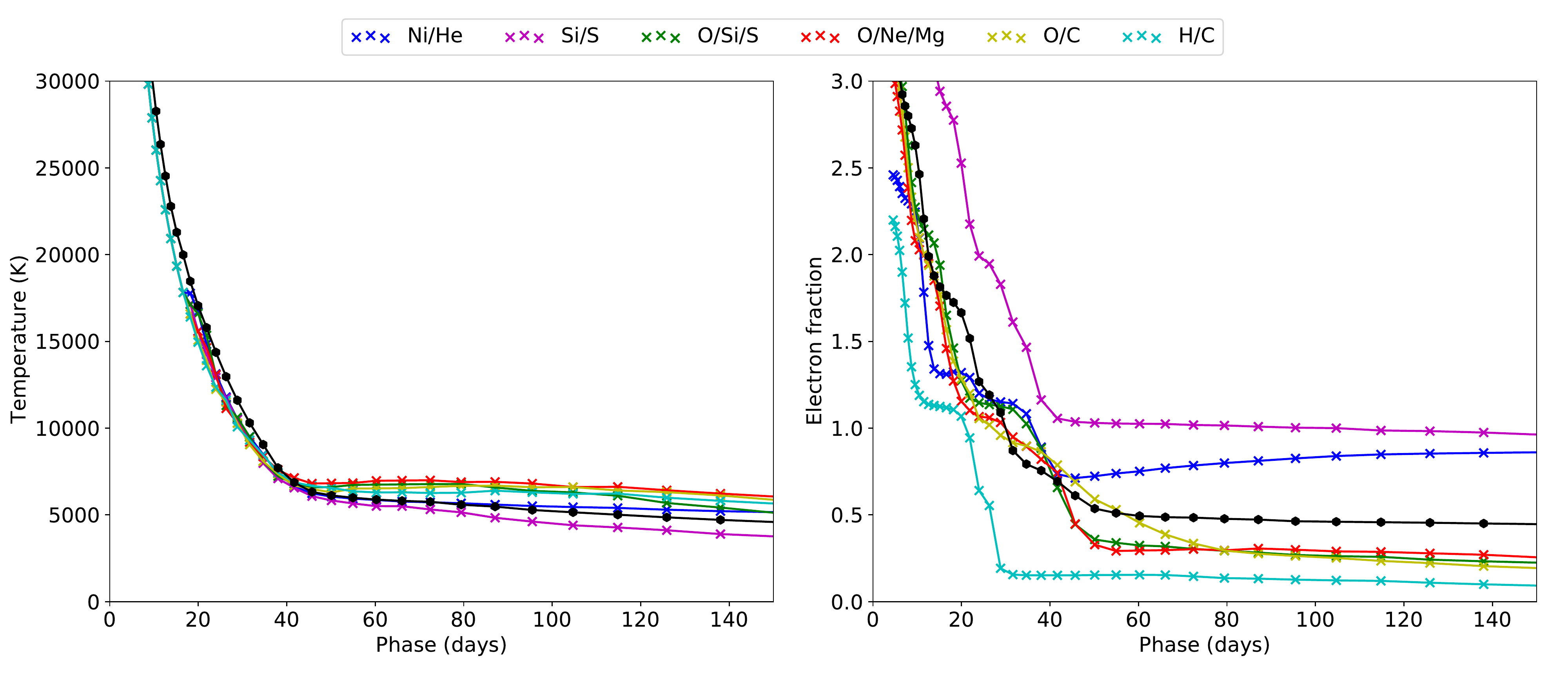}
\caption{Evolution of the temperature (left panel) and the electron fraction (right panel) in the Ni/He (blue), Si/S (magenta), O/Si/S (green), O/Ne/Mg (red), O/C (yellow) and He/C (cyan) clumps in the core (v=2900 km s$^{-1}$) of the macroscopically mixed standard model and the core of the microscopically mixed model (black).}
\label{f_12C_comp_matter_evo}
\end{figure*}

Figure~\ref{f_12C_comp_matter_evo}  shows the temperature and electron fraction in the different clump types in the core (v=2900 km s$^{-1}$) of the macroscopically mixed standard model and the core of the microscopically mixed model. In the beginning of the simulation the core is handled by the diffusion solver, which assumes the temperature to be the same in all clumps. The electron fraction differs, however, as it depends on the composition and density (through the Saha equation). At $\sim$15 days the core is handed over to the NLTE solver, but the difference in temperature remains small until $\sim$50 days. Thereafter we see an increasing difference, and the highest temperatures are achieved in the O/Ne/Mg and He/C clumps, and the lowest in the Si/S clumps. This is likely explained by a more efficient cooling in the calcium-rich Si/S clumps. The difference in electron fraction remains large at all times, and the highest electron fraction is achieved in the Si/S clumps and the lowest in the He/C clumps. This is likely explained by the lower ionization potential and density in the Si/S clumps. The results are consistent with those in \citetalias{Jer15} and \citet{Koz98a}, which discuss the evolution of the temperature and ionization in the nebular phase.

\begin{figure}[tbp!]
\includegraphics[width=0.49\textwidth,angle=0]{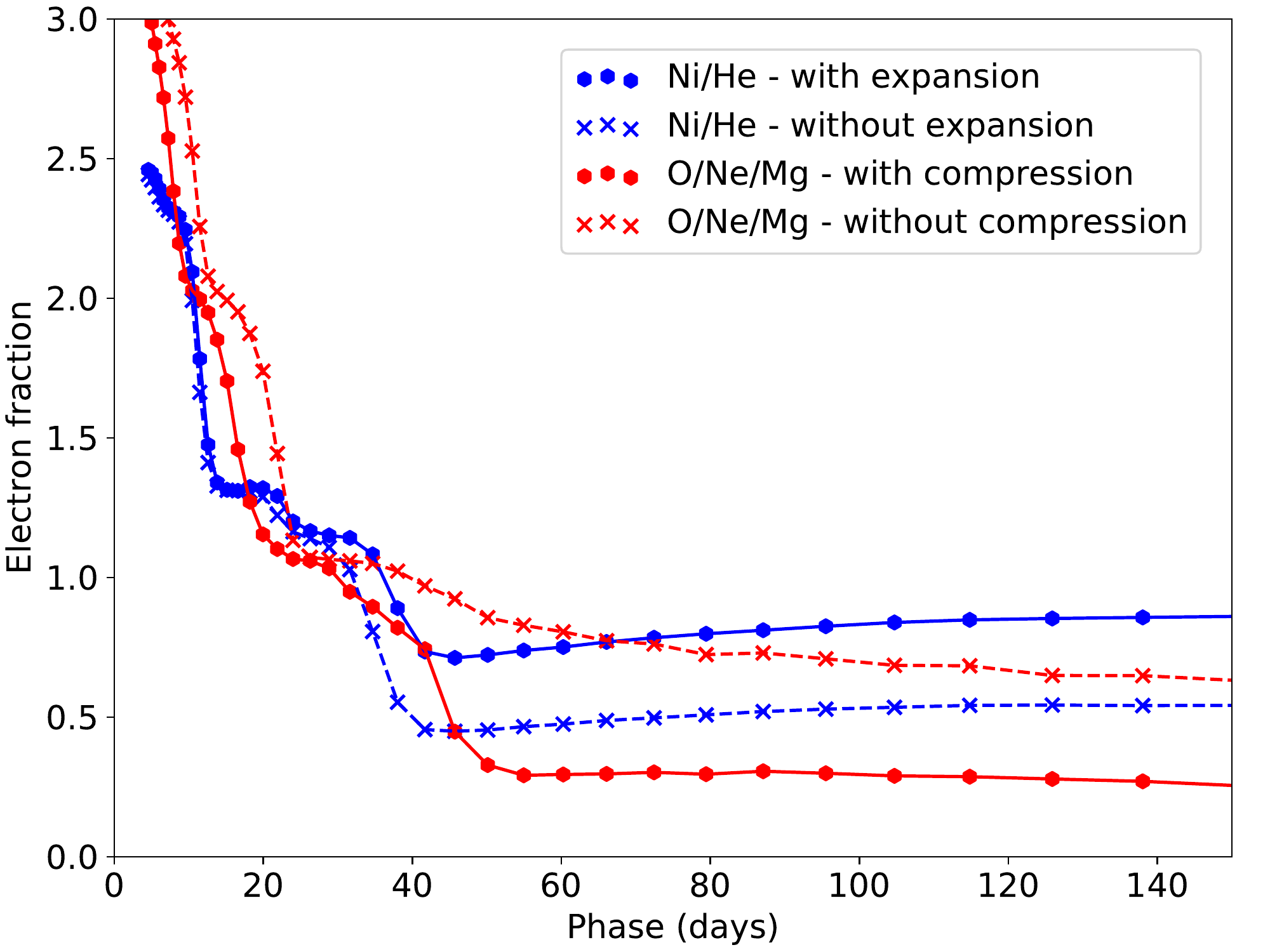}
\caption{Evolution of the electron fraction in the Ni/He (blue) and O/Ne/Mg (red) clumps in the core (v=2900 km s$^{-1}$) for the macroscopically mixed models with (circles and solid lines) and without (crosses and dashed lines) expansion/compression of these clumps.}
\label{f_12C_comp_xe_evo}
\end{figure}

An important effect of the difference in density between the models is on the degree of ionization. Due to the inverse density dependence of the ionization/recombination rates, the degree of ionization tends to decrease if the material is compressed, and vice versa. This is illustrated by Fig.~\ref{f_12C_comp_xe_evo}, which shows the electron fraction in the Ni/He and O/Ne/Mg clumps in the core (v=2900 km s$^{-1}$) for the macroscopically mixed models with and without expansion/compression of these clumps. As seen in the figure, the electron fraction is considerably higher in the expanded Ni/He clumps and considerably lower in the compressed O/Ne/Mg clumps. A similar effect can be seen in the helium envelope, in particular for the model with strong expansion of the Ni/He clumps. Among other things, the degree of ionization affects the electron scattering opacity in the clumps, and we will return to this issue in the next section.

\begin{figure}[tbp!]
\includegraphics[width=0.49\textwidth,angle=0]{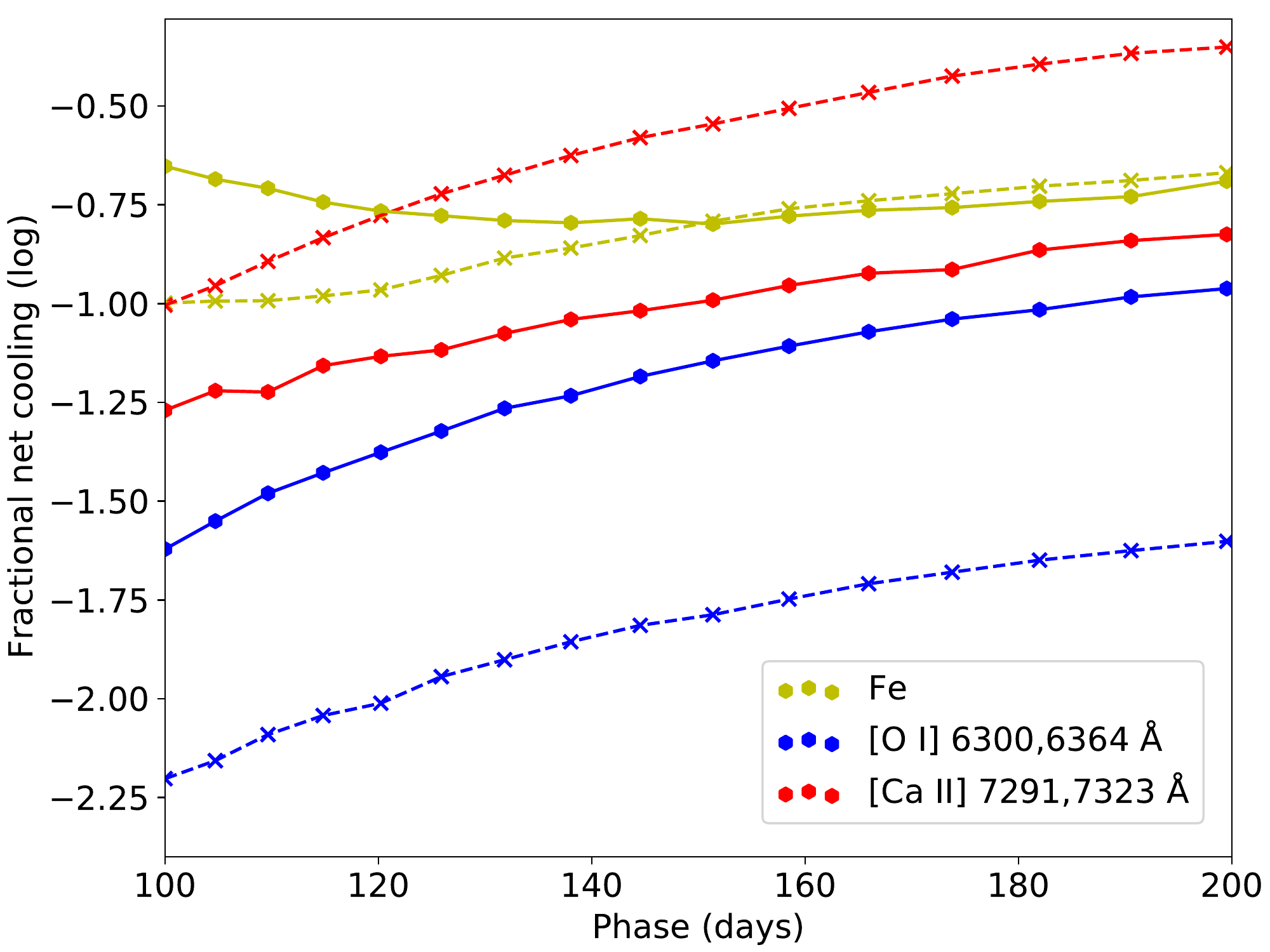}
\caption{Evolution of the (fractional) net collisional cooling in the nebular phase for the Fe lines (yellow) and the [\ion{O}{i}] 6300, 6364 \AA~(blue) and [\ion{Ca}{ii}] 7291, 7323 \AA~(red) lines in the core (v=2900 km s$^{-1}$) for the macroscopically mixed standard model (circles and solid lines) and the microscopically mixed model (crosses and dashed lines).}
\label{f_12C_comp_line_cooling_evo}
\end{figure}

An important effect of the difference in composition between the models is on the cooling rates. In microscopically mixed ejecta strong coolants as calcium are distributed uniformly, and may overtake the cooling from other elements \citep{Fra89}. This is especially important in the nebular phase, and influences lines driven by collisional cooling as the [\ion{O}{i}] 6300, 6364 \AA~and [\ion{Ca}{ii}] 7291, 7323 \AA~lines. This is illustrated by Fig.~\ref{f_12C_comp_line_cooling_evo}, where we show the net collisional cooling\footnote{Difference between collisional cooling and heating.} for these lines, as well as the total net collisional cooling for iron lines, in the core (v=2900 km s$^{-1}$) for the macroscopically mixed standard model and the microscopically mixed model. As seen in the figure, there is a strong difference between these models, and the net collisional cooling for the [\ion{Ca}{ii}] 7291, 7323 \AA~lines is much higher in the microscopically mixed model where calcium has been uniformly distributed. At the same time, the net collisional cooling for the [\ion{O}{i}] 6300, 6364 \AA~lines is much smaller in the microscopically mixed model, whereas the iron lines are less affected.

The effects of the macroscopic mixing on the state of matter due to the differences in density and composition have also been investigated by \citet{Des18b}, \citet{Des20} and \citet{Des21} using CMFGEN, who reach similar conclusions as discussed here. However, as the state of matter depends on the radiation field and the radioactive energy deposition, which are also affected by the macroscopic mixing (see Sects.~\ref{s_effect_macro_rad} and \ref{s_effect_macro_energy}), results from JEKYLL and CMFGEN may not be directly comparable.

\subsection{Radiative transfer}
\label{s_effect_macro_rad}

Except for the different state of matter in the clumps, which affects the opacities and emissivities, the geometrical arrangement of these  affect the radiative transfer. In particular, it affects the diffusion time, which is governed by the mean free path. In a macroscopically mixed medium that is spherically symmetric on average (which is assumed in JEKYLL), and in the limit of many clumps, the mean free path can be expressed as an effective mean free path $\skew{3.0}\hat{\lambda}_{\nu}(r)$, which is an average over all random arrangements of the clumps. 

In Appendix \ref{a_effective_opacity} we discuss the effective mean free path in more detail, and derive the effective diffusion approximation for such a medium in the case of a smooth and spherically symmetric energy density. This diffusion approximation differs from the ordinary one only in that the Rosseland mean opacity is replaced by an effective Rosseland mean opacity based on the effective mean free path. As discussed in Appendix \ref{a_effective_opacity}, the assumption of a smooth and spherically symmetric energy density is not strictly true, but as we demonstrate by full radiative transfer calculations on actual 3D-grids, it is well justified in the parameter space of interest for SNe.

As shown in Appendix \ref{a_effective_opacity}, there are a few important limiting cases in which the expression for the effective mean free path simplifies. First, in the limit of optically thin clumps, the effective opacity ($\hat{\kappa}_{\nu}=(\skew{3.0}\hat{\lambda}_{\nu}~\rho)^{-1}$) is given by a \textit{mass-average} of the material\footnote{Here and elsewhere the prefix material is used to distinguish the ordinary opacity from the effective opacity.} opacity in the different clump types, and is independent\footnote{Except implicitly through the dependence of the material opacity on the density.} of the clumping geometry (i.e.~the filling factors and the sizes of the clumps). Second, in the limit of optically thick clumps, the effective mean free path is given by a \textit{volume-average} of the mean free path in the different clump types, and depends on both the clumping geometry and the material opacity. Finally, in the limit of two distinct classes of clumps, one optically thick and one optically thin, the effective opacity depends \textit{only} on the clumping geometry, and is independent of the material opacity. This geometrical limit is discussed further in Appendix \ref{a_effective_opacity}.

\begin{figure}[tbp!]
\includegraphics[width=0.49\textwidth,angle=0]{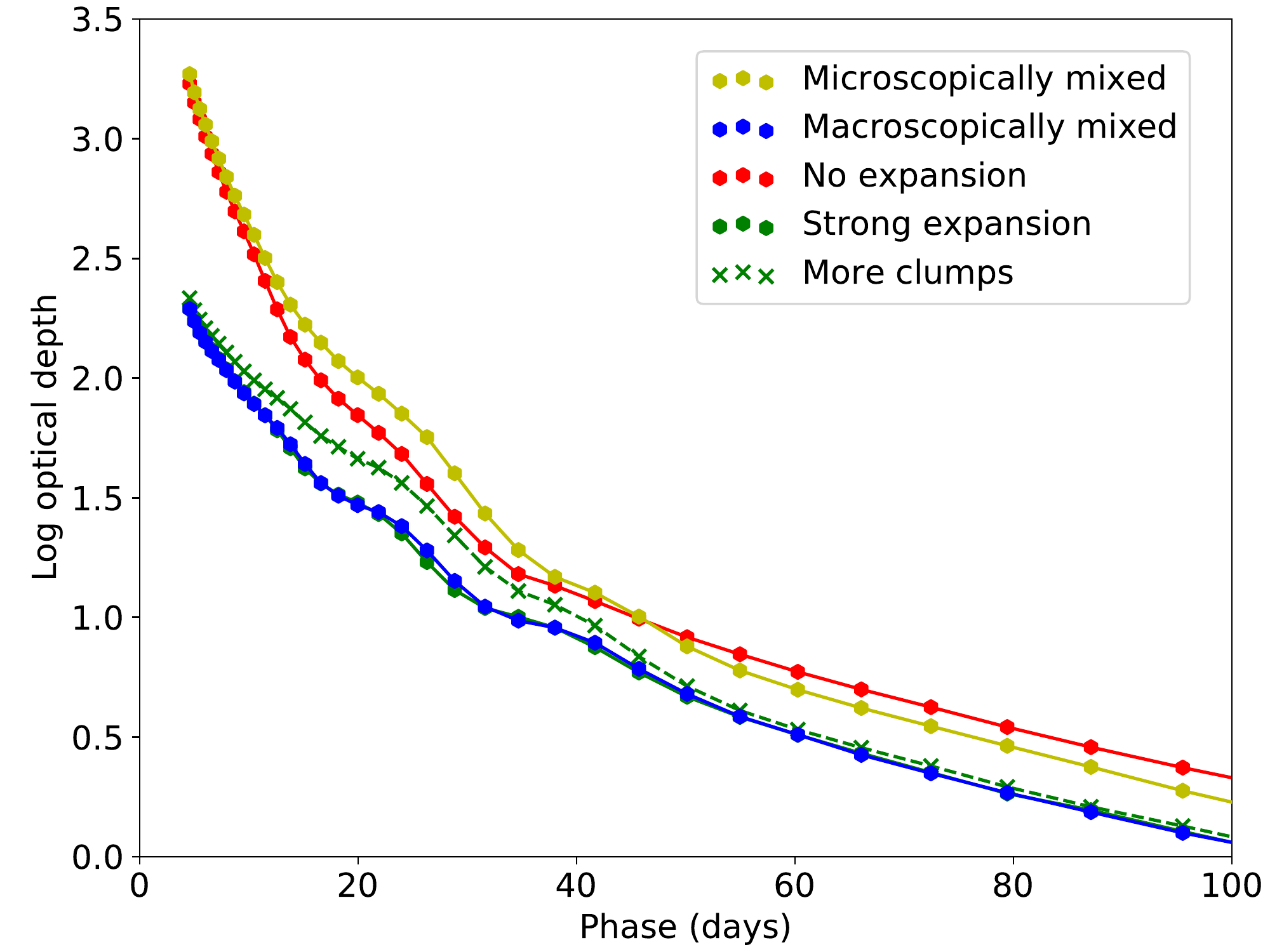}
\caption{Evolution of the Rosseland mean optical depth in the core for the macroscopically mixed standard model (blue circles) compared to the microscopically mixed model (red circles). In addition, we show the macroscopically mixed models with no expansion (yellow circles), strong expansion (green circles), and more clumps (green crosses).}
\label{f_12C_comp_tau_core_evo}
\end{figure}

Armed with this knowledge we now turn to our models. Figure~\ref{f_12C_comp_tau_core_evo} shows the effective Rosseland mean optical depth in the core for the full set of models. As seen in the figure, the effect of the macroscopic mixing is quite dramatic during the first $\sim$40 days and is mainly caused by the expansion of the clumps containing radioactive material. While the effective Rosseland mean opacity in the model without expansion of these clumps is similar to that in the microscopically mixed model, it is almost ten times lower in the standard model where these clumps have been expanded. This can be understood, because at early epochs the clumps are in the optically thick regime, and in this limit the effective mean free path is given by a volume average over the different clump types. In the standard model, the volume is dominated by the Ni/He clumps which due to expansion have a much longer mean free path. In Appendix \ref{a_effective_opacity} we show that if the opacity is uniform and unaffected by the expansion, and one type of clumps have been expanded by a factor $f_{\mathrm{E}}$ to achieve a filling factor of $\Phi_{\mathrm{E}}$, the decrease of the effective opacity in the optically thick limit is given by
\begin{equation}
  \label{eq_19b}
  R=\frac{(1-\Phi_{\mathrm{E}})^{2}} {(1-\Phi_{\mathrm{E}}/f_{\mathrm{E}})} + \Phi_{\mathrm{E}}~f_{\mathrm{E}} \geq 1,
\end{equation}
which gives $R \approx 8.5$ for the standard model. If $R \gg 1$,  $R \approx \Phi_{\mathrm{E}}~f_{\mathrm{E}}$, which shows that a strong decrease can be achieved if the clumps have been expanded by a large factor to fill a large fraction of the volume. After $\sim$40 days, the clumps gradually move into the optically thin regime, and the geometrical effect on the effective Rosseland mean opacity disappears. However, as seen in the figure, the effective Rosseland mean opacity is still lower in the standard model. This can be understood, because in the optically thin limit the effective opacity is given by a mass-average over the different clump types. In the standard model, the mass is dominated by the oxygen-rich clumps, which due to compression have a lower degree of ionization (Sect.~\ref{s_effect_macro_matter}), and therefore a lower electron scattering opacity. This recombination effect, which occurs in the optically thin regime, is similar to the findings by \citet{Des18}, although in their work the compressed clumps were surrounded by a void medium.

It is also warranted to investigate the effect of the clump size on the effective Rosseland mean opacity. The clump size is important because if these are small enough we move into the optically thin regime. As seen in Fig.~\ref{f_12C_comp_tau_core_evo}, if the number of clumps is increased by a factor of 100 (corresponding to 10$^{3/2}$ times smaller clumps), the effective Rosseland mean opacity becomes considerably higher than in the standard model between $\sim$10 and $\sim$50 days. Before this, the clumps in both models are still in the optically thick regime, and after this, the clumps in both models have entered the optically thin regime.

\begin{figure}[tbp!]
\includegraphics[width=0.49\textwidth,angle=0]{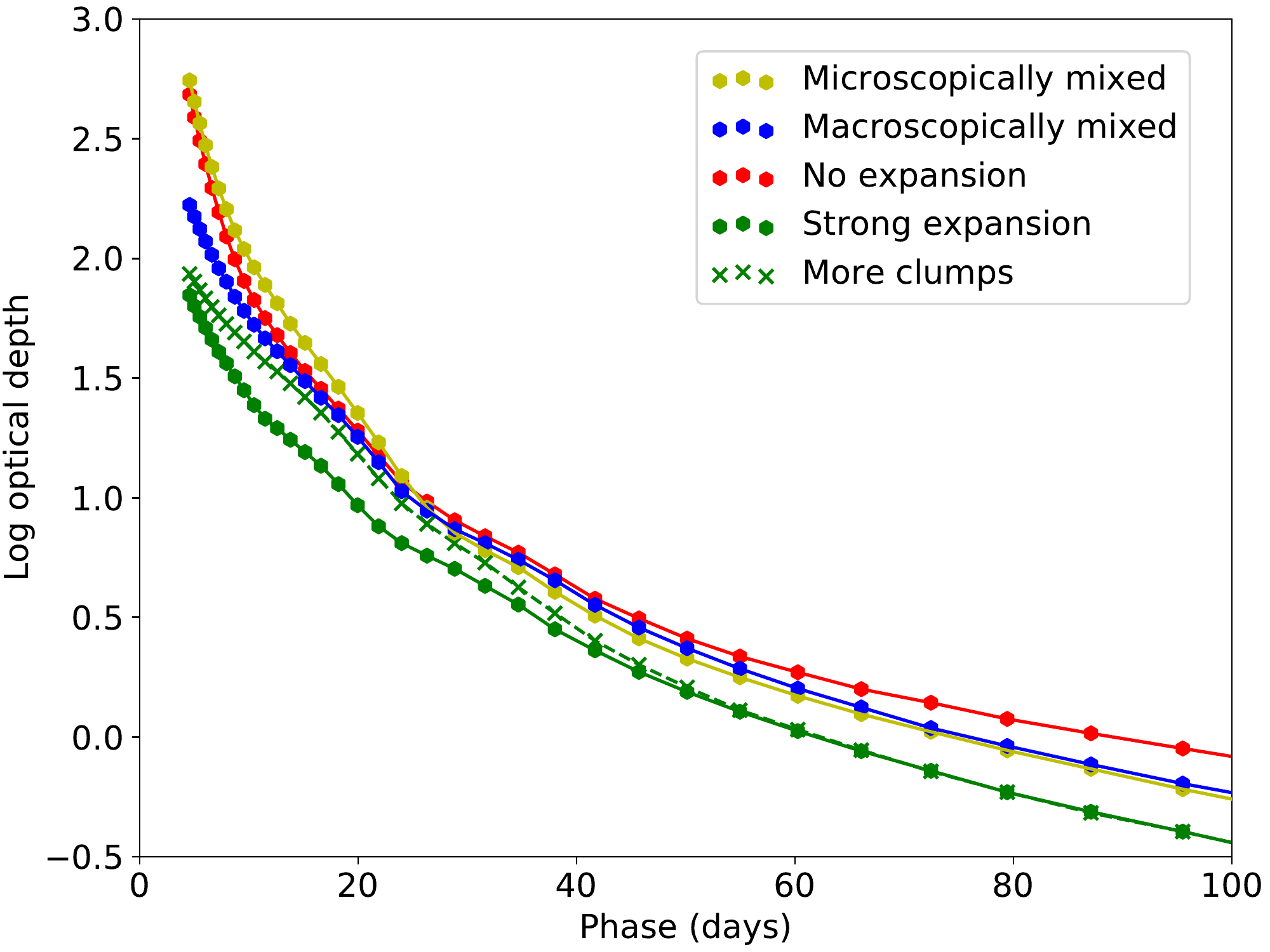}
\caption{Evolution of the Rosseland mean optical depth in the inner helium envelope for the macroscopically mixed standard model (blue circles) compared to the microscopically mixed model (red circles). In addition, we show the macroscopically mixed models with no expansion (yellow circles), strong expansion (green circles), and more clumps (green crosses).}
\label{f_12C_comp_tau_He_env_inner_evo}
\end{figure}

\begin{figure}[tbp!]
\includegraphics[width=0.49\textwidth,angle=0]{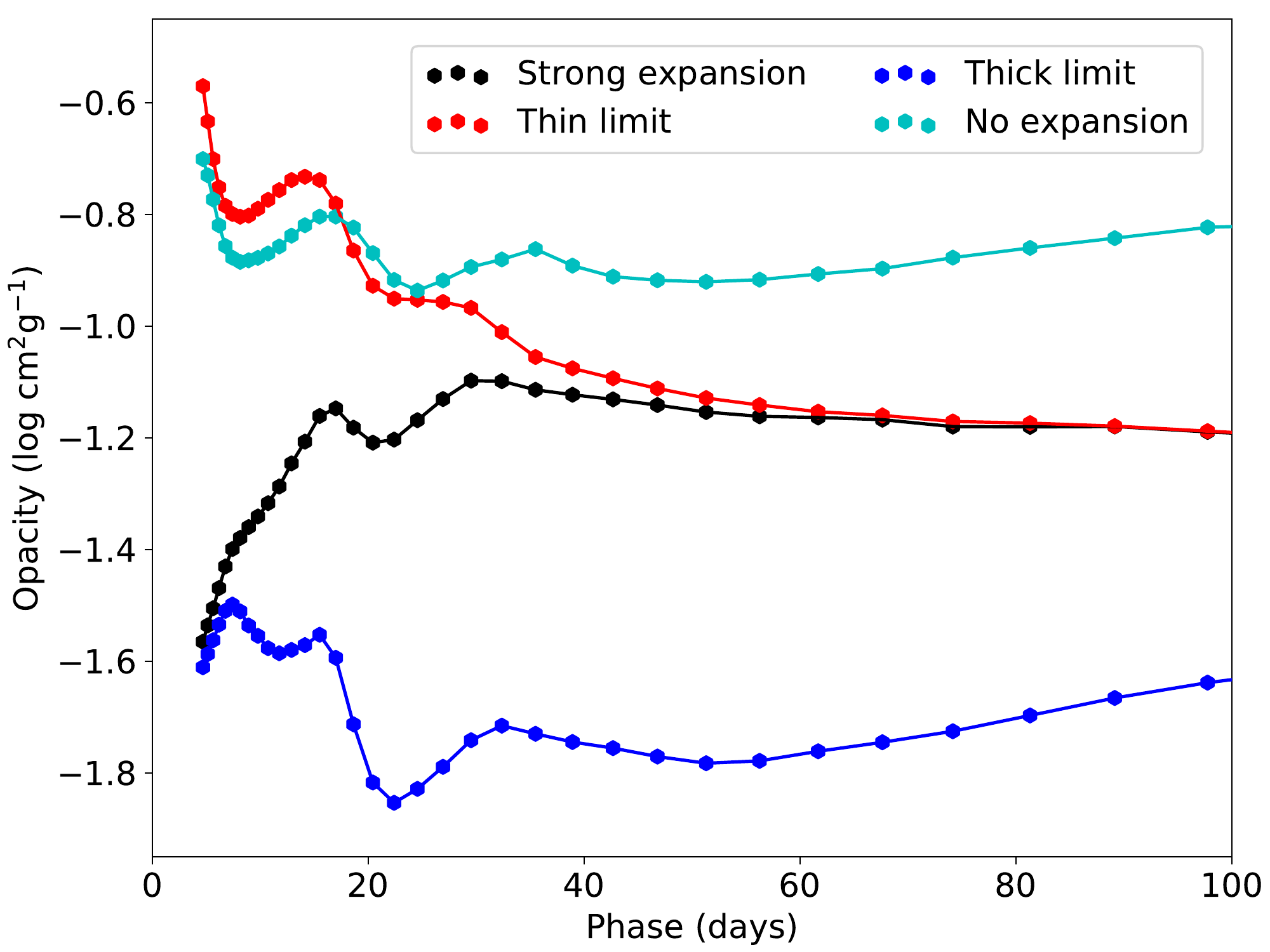}
\caption{Evolution of the Rosseland mean opacity in the inner helium envelope (v=4700 km s$^{-1}$) for the macroscopically mixed model with strong expansion (black) and the corresponding Rosseland mean opacity in the limits of optically thick (blue) and thin (red) clumps. In addition, we show the evolution of the Rosseland mean opacity for the macroscopically mixed model without expansion (cyan).}
\label{f_12C_comp_kappa_evo}
\end{figure}

As shown in Fig.~\ref{f_12C_comp_tau_He_env_inner_evo}, the effect of the macroscopic mixing is similar in the inner helium envelope, although it is weaker in the standard model with less expansion of the Ni/He clumps. The transition from the optically thick to the optically thin regime is illustrated by Fig.~\ref{f_12C_comp_kappa_evo}, where we show the effective Rosseland mean opacity in the inner helium envelope (v=4700 km s$^{-1}$) for the model with strong expansion of the Ni/He clumps together with the corresponding optically thick and thin limits. As seen in the figure, the effective Rosseland mean opacity initially follows the optically thick limit, begins to deviate at $\sim$5 days, and then gradually approaches and finally reaches the optically thin limit at $\sim$60 days. Note, that the difference between the optically thick and thin limit is modulated by the opacity in the clumps, which initially is higher in the Ni/He clumps due to a higher bound-free opacity, and then becomes lower in the helium-rich clumps due a lower electron scattering opacity.

In Fig.~\ref{f_12C_comp_kappa_evo} we also show the effective Rosseland mean opacity for the model without expansion of the Ni/He and Si/S clumps. This comparison shows that the decrease of the effective Rosseland mean opacity in the model where these clumps are expanded is dominated by the geometrical effect until $\sim$35 days, after which it is dominated by the decrease in material opacity in the helium-rich clumps due to the lower electron scattering opacity. These effects are complementary, and broadly speaking, the geometrical effect operates in the optically thick regime, whereas the recombination effect operates in the optically thin regime. The geometrical effect is, however, potentially much stronger, and in our models it dominates.

\subsection{Radioactive energy deposition}
\label{s_effect_macro_energy}

In a macroscopically mixed model the radioactive decays occur solely in the Ni/He and Si/S clumps. The leptons emitted in the decays are assumed to deposit their energy locally, whereas the deposition of $\gamma$-ray energy depends on the radiative transfer. Early on, the Ni/He and Si/S clumps are optically thick to the $\gamma$-rays and all the radioactive decay energy is deposited locally. When the Ni/He and Si/S clumps approach the optically thin regime, the $\gamma$-rays begin to leak out and may deposit their energy in other clumps. As discussed in Sect.~\ref{s_effect_macro_rad}, the radiative transfer in a macroscopically mixed model depends on the clumping geometry, but once the optical depth of all clumps approaches zero this dependence is lost and the radiative transfer is governed by the mass-averaged $\gamma$-ray opacity, which is the same as in a microscopically mixed model.

\begin{figure}[tbp!]
\includegraphics[width=0.49\textwidth,angle=0]{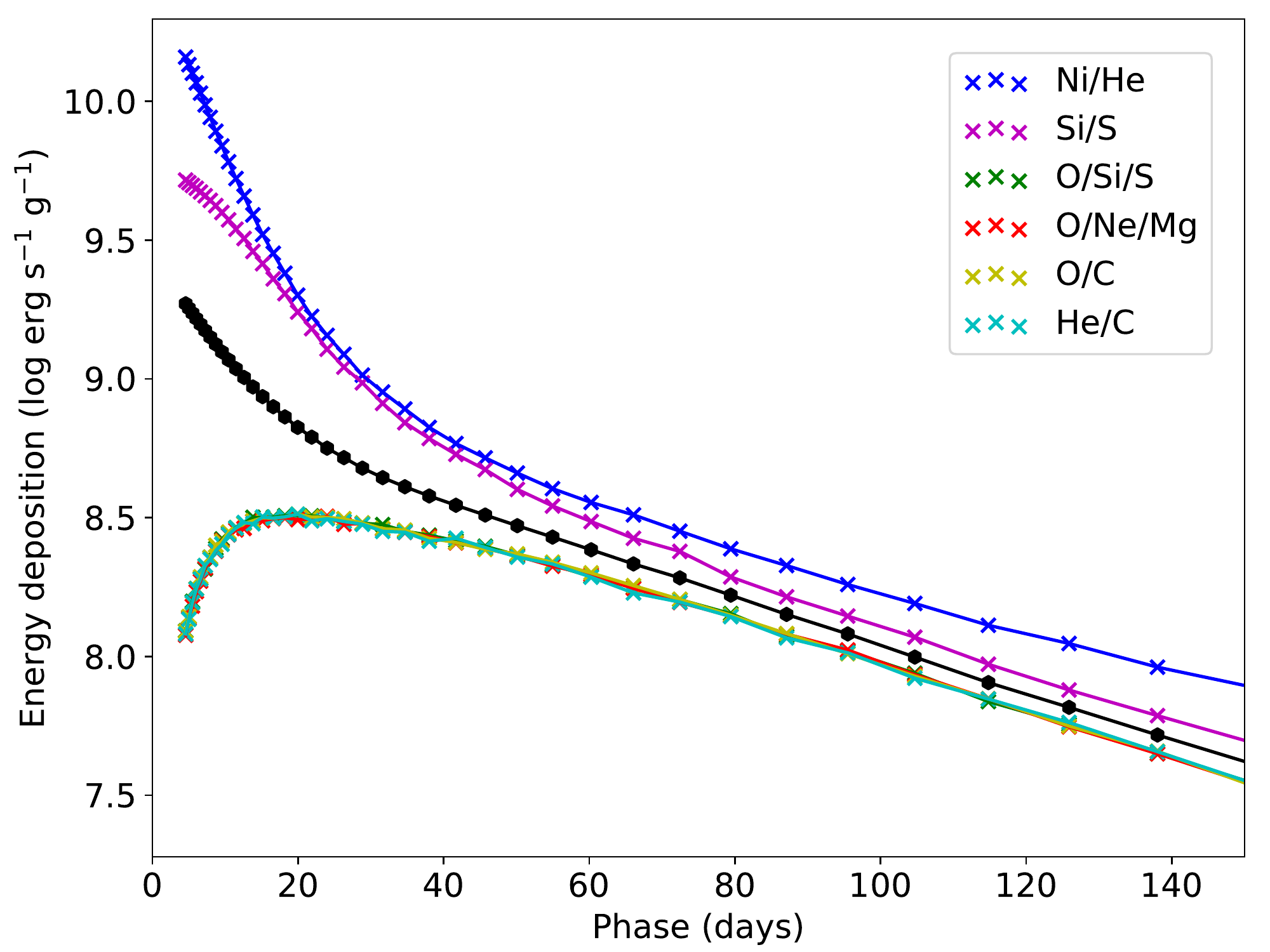}
\caption{Evolution of the radioactive energy deposition in the Ni/He (blue), Si/S (magenta), O/Si/S (green), O/Ne/Mg (red), O/C (yellow) and He/C (cyan) clumps in the core (v=2900 km s$^{-1}$) of the macroscopically mixed standard model and the core of the microscopically mixed model (black). Note that the energy deposition is similar in the oxygen- and helium-rich clumps, so these curves overlap.}
\label{f_12C_comp_eRadio_evo}
\end{figure}

Figure~\ref{f_12C_comp_eRadio_evo} shows the energy deposition in the different clump types in the core (2900 km s$^{-1}$) of the macroscopically mixed standard model and the core of the microscopically mixed model. As expected, we see that the radioactive decay energy is almost solely deposited in the Ni/He and Si/S clumps at early times. At $\sim$10 days, the $\gamma$-rays begin to leak out from the Ni/He and Si/S clumps and the energy deposition rises drastically in the other clumps. As time goes the $\gamma$-ray energy deposition in the clumps approaches a common value. However, as the leptons still deposit their energy locally in the Ni/He and Si/S clumps, the total radioactive energy deposition is always higher in these. This difference increases with time as an increasing fraction of the $\gamma$-rays escapes from the ejecta.

Interestingly enough, the radial deposition profile is almost the same in all models, and the total radioactive energy deposition differs only by a few percent. This might sound surprising, but to deposit their energy non-locally the $\gamma$-rays have to traverse a large number of clumps, and in that limit the absorption distribution functions are independent of the clumping geometry. The main effect of the macroscopic mixing on the radioactive energy deposition is therefore on the local distribution of the energy over the compositional zones. This might differ in the case of large-scale asymmetries, but as explained, such effects are not explored in this paper.

\subsection{Observed spectra and lightcurves}
\label{s_effect_macro_obs}

We now turn to the observed lightcurves and spectra, and analyse what effect the macroscopic mixing has on these. We begin with the bolometric lightcurve, which is mainly affected by the effects on the radiative transfer discussed in Sect.~\ref{s_effect_macro_rad}, and then turn to the spectra, which are mainly affected by the effects on the state of matter discussed in Sect.~\ref{s_effect_macro_matter}.

\subsubsection{Bolometric lightcurve}

\begin{figure}[tbp!]
\includegraphics[width=0.49\textwidth,angle=0]{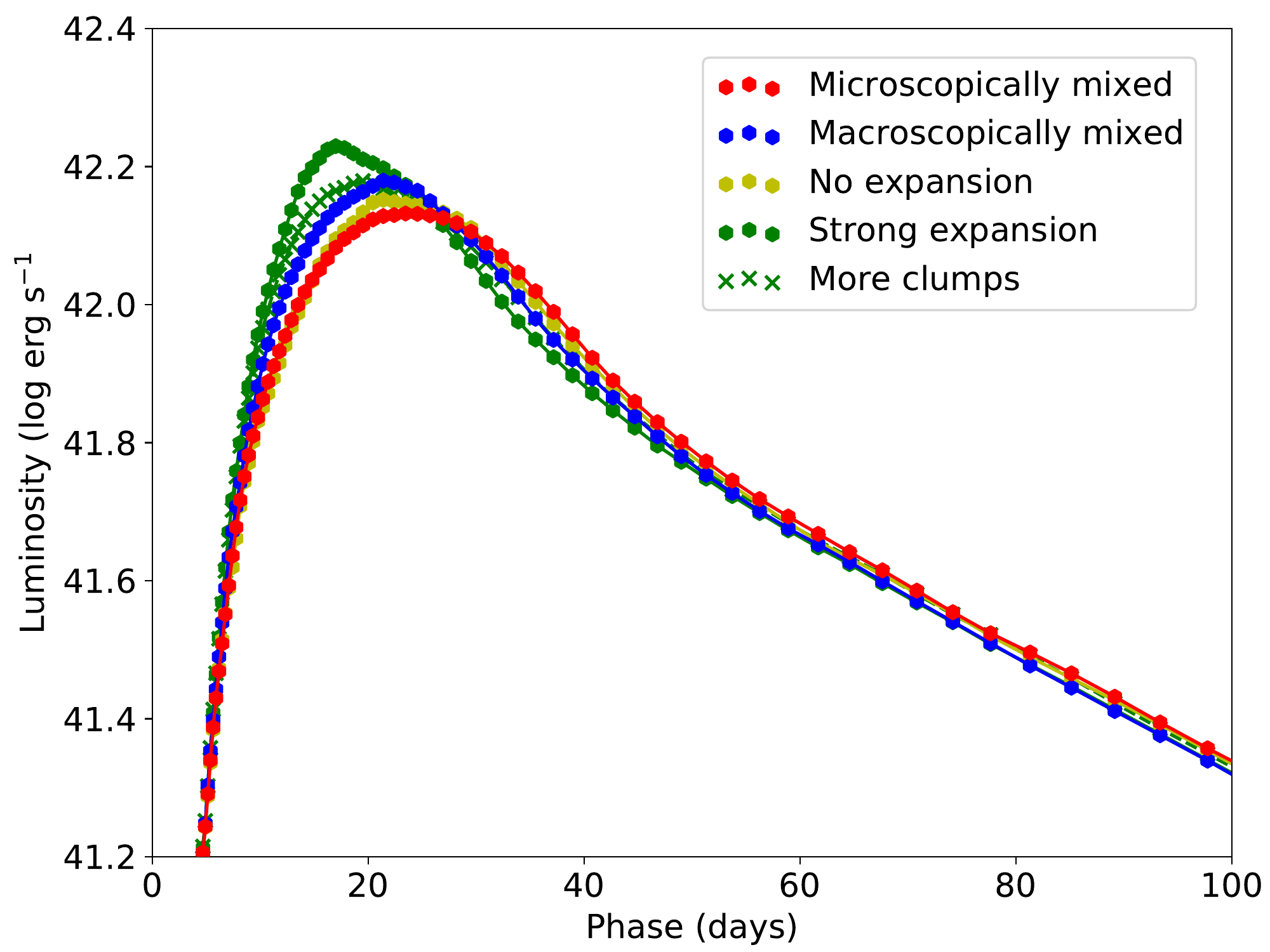}
\caption{Bolometric lightcurve for the macroscopically mixed standard model (blue circles) compared to the microscopically mixed model (red circles). In addition, we show the macroscopically mixed models with no expansion (yellow circles), strong expansion (green circles), and more clumps (green crosses).}
\label{f_12C_vc_comp_lc_evo}
\end{figure}

Figure \ref{f_12C_vc_comp_lc_evo} shows the bolometric lightcurves for the full set of models. Compared to the microscopically mixed model, the peak of the bolometric lightcurve is brighter, narrower and occurs 3 days earlier in the macroscopically mixed standard model, which is what we would expect from the lower effective Rosseland mean opacity in this model. Once on the tail, the difference disappears as the opacity difference decreases and the luminosity is not determined by the diffusion time anymore. 
The bolometric lightcurve for the macroscopically mixed model \emph{without} expansion of the Ni/He and Si/S clumps, which has similar effective Rosseland mean opacity as the microscopically mixed model (Sect.~\ref{s_effect_macro_rad}), is almost identical to that of the microscopically mixed model. Clearly, the expansion of the Ni/He and Si/S clumps may have a strong effect on the bolometric lightcurve through its influence on the effective Rosseland mean opacity. For the macroscopically mixed model with strong expansion of the Ni/He clumps in the helium envelope the effect is quite dramatic, and the peak occurs seven days before that of the microscopically mixed model and is 0.25 mag brighter. 

However, the effect also depends on the size of the clumps, and the bolometric lightcurve for same model with 100 times more clumps (which are therefore 10$^{3/2}$ times smaller) is more similar to that of the macroscopically mixed standard model. As discussed in Sect.~\ref{s_effect_macro_rad}, the reason for this is that the geometrical effect on the effective Rosseland mean opacity disappears when the clumps becomes optically thin. Unfortunately, no strong constraints exist on the size of the clumps and the expansion of the clumps containing radioactive material in the helium envelope, neither observational nor theoretical, and further work on this issue is highly warranted.

The effect of macroscopic mixing on the lightcurves is not restricted to the case explored here, and will affect the lightcurves of SE SNe in general. It may be stronger or weaker, depending on the expansion of the clumps containing radioactive material, the amount and distribution of such material, and the size of these and other clumps. However, if the clumps containing radioactive material are expanded by a large factor to fill a large fraction of the volume, it is potentially dramatic (Sect.~\ref{s_effect_macro_rad}). This is exemplified by the Type IIb models explored here, for which the product of the expansion and filling factors is 8.5 in the core and 2.5 (standard model) and 7.2 (strong expansion) in the helium envelope. The effect could lead to systematic underestimates of the ejecta masses of SE SNe, as it is not included in codes commonly used to estimate these from the bolometric lightcurve. We note, however, that these codes do not include NLTE either, which at least for the Type IIb models explored here works in the opposite direction \citepalias{Erg18}. 

Recently, \citet{Des19} found a similar effect on the lightcurves of Type II SNe using multi-D simulations with a simplified treatment of the matter. In these simulations, the ejecta were assumed to consist of high density clumps surrounded by a lower density medium, but the density contrast was not directly linked to the expansion of radioactive material as in our models. The authors also introduce the concept of optically thick "macroclumps", and their qualitative discussion of the effect on the radiative transfer is in line with our discussion of the effective opacity in the limit of optically thick clumps. We note, however, that the concept of effective opacity presented here (Sect.~\ref{s_effect_macro_rad} and Appendix~\ref{a_effective_opacity}) is quite fruitful, as it allows us to quantify the effect of the macroscopic mixing on the radiative transfer. The MC method used in JEKYLL to calculate it may be incorporated also in traditional codes, for example in diffusion-based hydrodynamical codes, commonly used to calculate the bolometric lightcurves of SNe. Remember, though, that the treatment of the macroscopic mixing in JEKYLL is still simplified (see Sect. \ref{s_method_macro}), and large-scale asymmetries and other deviations from our idealized parametrization might affect the results.

\subsubsection{Spectral evolution}

\begin{figure}[tbp!]
\includegraphics[width=0.49\textwidth,angle=0]{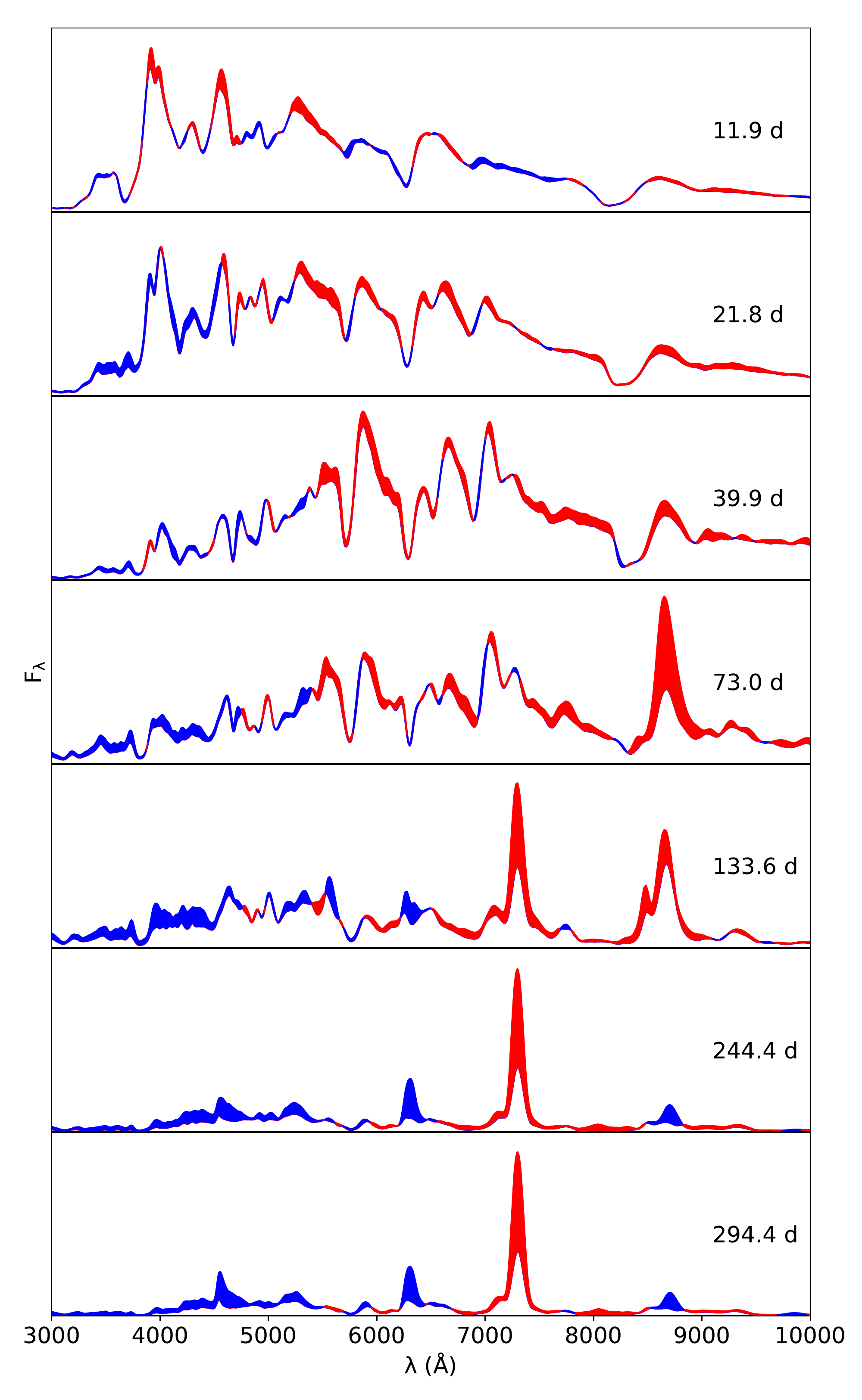}
\caption{Optical spectral evolution for the macroscopically mixed standard model compared to the microscopically mixed model. The difference has been highlighted in blue/red wherever the flux in the standard model is stronger/weaker.}
\label{f_12C_vc_comp_spec_evo_1}
\end{figure}

Figure \ref{f_12C_vc_comp_spec_evo_1} shows the optical spectral evolution for the macroscopically mixed standard model compared to the microscopically mixed model. Initially the spectral evolution is very similar in the two models, but after $\sim$40 days when the photosphere recedes into the macroscopically mixed layers (see Fig~\ref{f_12C_matter_evo}) they become increasingly different. In particular, the [\ion{Ca}{ii}] 7291, 7323 \AA~lines are much stronger, whereas the [\ion{O}{i}] 6300, 6364 \AA~lines are much weaker in the microscopically mixed model. Other differences emerge in \ion{Ca}{ii} triplet around $\sim$100 days and rewards of $\sim$5000 \AA, where the emission becomes much weaker in the microscopically mixed model after $\sim$100 days. The latter effect is caused by a redistribution in wavelength of the Fe emission from the core due to differences in temperature and ionization between the models (see Sect.~\ref{s_effect_macro_matter}).

The [\ion{Ca}{ii}] 7291, 7323 \AA~and [\ion{O}{i}] 6300, 6364 \AA~lines are mainly driven by collisional cooling, and as discussed in Sect.~\ref{s_effect_macro_matter}, the net collisional cooling rate of the [\ion{Ca}{ii}] 7291, 7323 \AA~lines is increased in the microscopically mixed model, whereas the net collisional cooling rate of the [\ion{O}{i}] 6300, 6364 \AA~lines is decreased. As was previously pointed out by \citet{Fra89}, this is because calcium is a very effective coolant by a factor of $\sim$10$^3$, and when calcium-rich material is mixed with other compositional zones it overtakes the cooling from other less efficient coolants. The difference in the [\ion{Ca}{ii}] 7291, 7323 \AA~and [\ion{O}{i}] 6300, 6364 \AA~lines become very strong after $\sim$100 days, and it is clear that macroscopic mixing needs to be taken into account in the nebular phase for the model to be realistic. This is particularly true for the [\ion{O}{i}] 6300, 6364 \AA~lines, given their importance for constraining the initial mass of the progenitor star \citepalias[see e.g.][]{Jer15}. Note that recent 3-D simulations of convection in stars show that some microscopic mixing of calcium- and oxygen-rich material may actually occur in the late stages of stellar evolution \citep[e.g.][]{Yad20}. Therefore a correct modelling of \emph{both} the [\ion{Ca}{ii}] 7291, 7323 \AA~and [\ion{O}{i}] 6300, 6364 \AA~lines is important for a reliable mass determination.

\begin{figure}[tbp!]
\includegraphics[width=0.49\textwidth,angle=0]{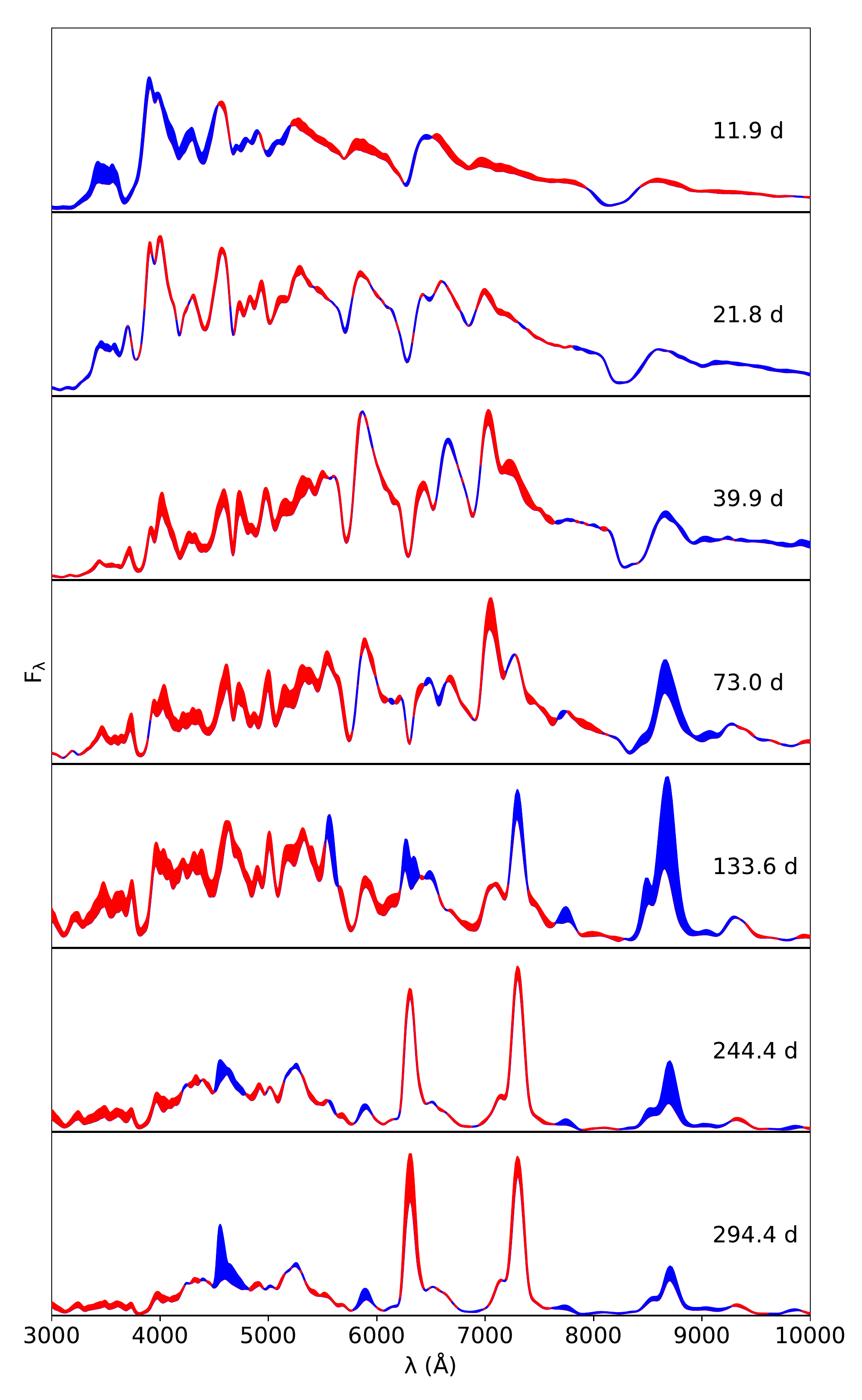}
\caption{Optical spectral evolution for the macroscopically mixed standard model compared to the model without expansion. The difference has been highlighted in blue/red wherever the flux in the standard model is stronger/weaker.}
\label{f_12C_vc_comp_spec_evo_2}
\end{figure}

It is also of interest to compare models with different clumping geometries. Figure \ref{f_12C_vc_comp_spec_evo_2} shows the optical spectral evolution for the macroscopically mixed standard model compared to the model without expansion of the Ni/He and Si/S clumps. Initially, the spectral evolution is very similar, but after $\sim$40 days several differences emerge. The most distinct differences are seen in the \ion{Ca}{ii} triplet after $\sim$100 days, and the \ion{Mg}{i}] 4571 \AA~line after $\sim$250 days. The difference in the \ion{Ca}{ii} triplet, which is actually a blend with the [\ion{C}{i}] 8727 \AA~line (Sect.~\ref{s_spec_evo_C_Ca}), originates from various zones in the core and the inner helium envelope, whereas the difference in the \ion{Mg}{i}] 4571 \AA~line can be traced back to emission from the oxygen-rich clumps in the core (see \citetalias{Jer15} for a further discussion on this). In both cases the observations of SN 2011dh are in better agreement with the model with expanded Ni/He and Si/S clumps. This is in line with constraints based on small-scale variations in the [\ion{O}{i}] 6300, 6364 \AA~and \ion{Mg}{i}] 4571 \AA~line profiles (\citetalias{Erg15}), which suggest strong compression of the oxygen-rich clumps in the core of SN 2011dh. 

The compression of the oxygen-rich material has also been investigated by \citet{Des21} for a set of steady-state SE SN models at an epoch of 200 days, and a similar effect on the \ion{Mg}{i}] 4571 \AA~line and the \ion{Ca}{ii} triplet blend can be seen, although it varies among the models and a considerable lower compression factor (5 instead of 21) was used. Note, that the [\ion{O}{i}] 6300, 6364 \AA~lines are only weakly affected by the compression in our model, although the fraction of \ion{O}{i} is increased by a factor of $\sim$2. Most of the models in \citet{Des21} show a similar effect on the ionization of O, whereas the effect on the [\ion{O}{i}] 6300, 6364 \AA~lines varies among the models. This underscores the complexity of how the compression affects the strength of different lines.

\begin{figure*}[tbp!]
\includegraphics[width=1\textwidth,angle=0]{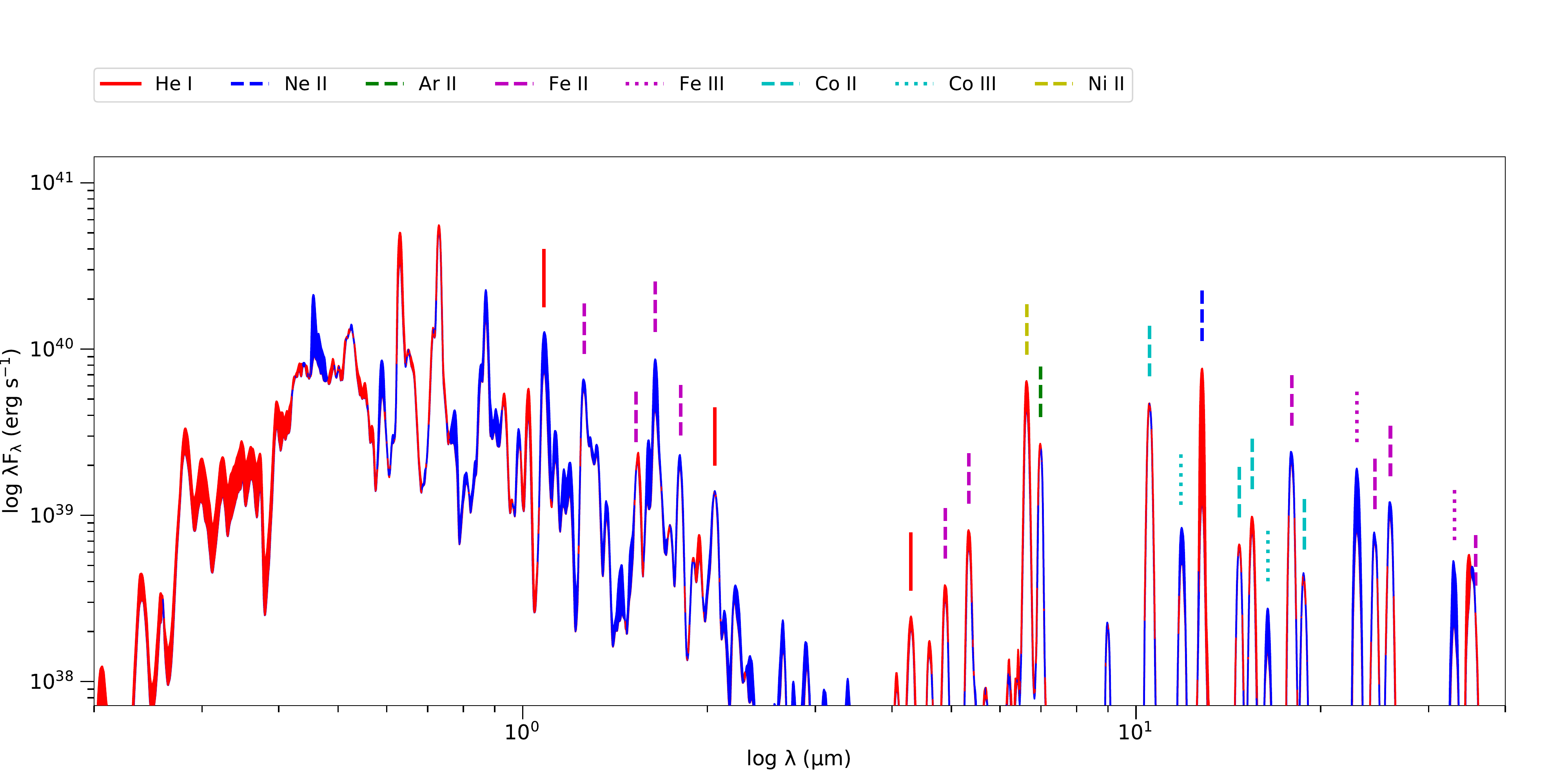}
\caption{Optical, NIR and MIR spectrum at 294 days (nebular phase) for the macroscopically mixed standard model compared to the model without expansion. The difference has been highlighted in blue/red wherever the flux in the standard model is stronger/weaker. The vertical lines show the strongest IR lines from the ions given in the legend at the top.}
\label{fig_NIR_MIR_308}
\end{figure*}

Further and more robust constraints on the expansion/compression of the clumps can be obtained from IR observations. In Fig. \ref{fig_NIR_MIR_308} we show the full spectrum at 294 days for the macroscopically mixed standard model and the model without expansion of the Ni/He and Si/S clumps. The most important difference is a change of ionization, as can be seen for especially the [Ne II], [Fe III] and [Co III] lines. The [Fe III] and [Co III] lines originate from the Ni/He clumps in the core and the inner helium envelope, and are directly linked to the expansion of these. There are also [\ion{Fe}{ii}] and [\ion{Co}{ii}] lines available, so the (weak) temperature dependence in the [\ion{Fe}{iii}] and [\ion{Co}{iii}] line-strengths can be removed by using line-ratios between singly and double ionized lines. Furthermore, the [\ion{Ne}{ii}] 12.8 $\mu$m line, which differs strongly, originates mainly from the O/Ne/Mg clumps in the core, and is directly linked to the compression of these. As mentioned, dust and molecule emission may contribute in the MIR. However as demonstrated by for example SN 1987A \citep[e.g][and references therein]{Wood93}, the contrast between the comparatively narrow lines and the dust continuum should make the lines easily observable. In order to better constrain the expansion/compression of the clumps, MIR observations of Type IIb SNe would therefore be very useful, something that may now be possible with the JWST.

\section{Conclusions}
\label{s_conclusions}

In \citetalias{Erg18} we presented the lightcurve and spectral synthesis code JEKYLL, which has the unique capability to self-consistently model the evolution of SNe taking into account both NLTE and the macroscopic mixing of the ejecta.

In this paper we use JEKYLL to evolve a macroscopically mixed Type IIb model through the photospheric and nebular phase. The model belongs to a set of models presented in \citetalias{Jer15}, corresponds to a star with an initial mass of 12 M$_\odot$, and was shown to compare well with nebular observations of SN 2011dh in \citetalias{Jer15} and \citetalias{Erg15}. Here we provide a detailed comparison with SN 2011dh, and show that this model also reproduces the photospheric spectra and broad-band lightcurves well. Some quantitative differences exist, however, and to find a model that improves the agreement is a challenge to be addressed in future work. This would require a larger grid of SN models, ideally based on self-consistent 3D explosion models and binary evolutionary stellar models. Our work complements and advances earlier work on SN 2011dh, and strengthens the conclusion that this SN originated from a star with an initial mass of $\sim$12 M$_\odot$ that had lost all but a tiny fraction (<0.1 M$_\odot$) of its hydrogen envelope, strongly suggesting a binary origin.

Self-consistent modelling of a SN through the \emph{both} photospheric and nebular phase taking into account both NLTE and macroscopic mixing has not been presented before, and the results show that our understanding of Type IIb SNe has reached a rather mature level. Most important observational aspects of Type IIb SNe, many of which are shared by SE SNe in general, are well reproduced, and the modelling may serve as a prototype for future work on these SNe. The spectral range is not limited to the optical and NIR, and with JWST in mind we present detailed MIR spectra, demonstrating their rich content of information in especially the nebular phase. Note, however, that we do not include emission from dust and molecules in these spectra.

We also present a thorough investigation of the effects of the macroscopic mixing on the SN evolution, partly based on a new method to calculate the effective Rosseland mean opacity. Although the models are tailored for SN 2011dh, the results have implications for Type IIb SNe, SE SNe and CC SNe in general. The effects are investigated by comparing macroscopically and microscopically mixed models, and by varying the clumping geometry in the macroscopically mixed models. In our treatment the clumping geometry is parametrized by the size of the clumps and their filling factors, where the latter are determined by the expansion of the Ni/He and Si/S clumps. 

In the photospheric phase, we find strong effects on the effective Rosseland mean opacity in the macroscopically mixed regions, which in turn affects the observed lightcurves. The diffusion peak of the bolometric lightcurve is considerably narrower (rise time decreased by 11 percent) in in the macroscopically mixed case, and differs strongly (rise time decreased by 29 percent) if the Ni/He clumps in the helium envelope are assumed to expand more than in our standard model. The effect is mainly geometrical, and is driven by the expansion of clumps containing radioactive material. This tends to decrease the effective opacity, and in the limit of optically thick clumps, the decrease is roughly given by the product of the expansion factor and the filling factor. In our models, this product has a value of 8.5 in the core, and 2.5 (standard model) or 7.2 (strong expansion) in the helium envelope. The effect also depends on the typical clumps size, as it disappears when the clumps becomes optically thin. In addition to the geometrical effect, there is also a recombination effect, which tends to decrease the effective opacity in the limit of optically thin clumps. These effects are complementary and operate in different regimes, but the geometrical effect is potentially stronger, and in our models it dominates.

The effect of the macroscopic mixing on the effective opacity has implications for lightcurve modelling of SE SNe in general, as it is not taken into account in hydrodynamical codes commonly used. If the clumps are expanded by a large factor to fill a large fraction of the volume the effect is potentially dramatic, and even a more modest effect could lead to systematic underestimates of the ejecta masses of SE SNe. One path forward could be to incorporate our method for the effective Rosseland mean opacity into hydrodynamical codes, but as these codes do not include NLTE either, the opacities would still be uncertain. In any case, the effect depends strongly on the clumping geometry, which is not well constrained.

In the nebular phase, we find strong effects on the collisional cooling rates in the macroscopically mixed regions, which affect the strength of lines driven by collisional cooling. In particular, the [\ion{Ca}{ii}] 7291, 7323 \AA~lines are much weaker and the [\ion{O}{i}] 6300, 6364 \AA~lines much stronger in the macroscopically mixed case. The reason for this is that calcium, which is a very efficient coolant, overtakes the cooling from oxygen if the calcium- and oxygen-rich material is mixed together. This has implications for modelling of CC SNe in general, as the strength of the [\ion{O}{i}] 6300, 6364 \AA~lines is often used as a tracer for the initial mass. Note, that only a small amount of microscopic mixing of the calcium-rich material, as may actually occur during the late progenitor evolution, may be sufficient for this effect to kick in.

Contrary to the photospheric phase, when the macroscopically mixed regions are mainly hidden below the photosphere, we find clear effects of the clumping geometry on the spectra in the nebular phase. In particular, lines from \ion{Mg}{I}], \ion{Ca}{II}, [\ion{Ne}{II}], [\ion{Co}{III}], and [\ion{Fe}{III}] are sensitive to the expansion/compression of the clumps, mostly due to its impact on the ionization. Such lines therefore provide a path forward to better constrain the expansion/compression of the clumps in Type IIb and SE SNe. The most robust constraints can be derived from observations of fine-structure lines in the MIR, as may now be possible with the JWST.

The effects of the macroscopic mixing on the NLTE state of matter discussed here are in qualitative agreement with results obtained by \citet{Fra89}, \citetalias{Jer15}, \citet{Des20} and \citet{Des21}, whereas the effect of the macroscopic mixing on the radiative transfer has similarities with the effect of clumping on the radiative transfer in Type IIP SNe found by \citet{Des19}. However, the radiative transfer and the NLTE state of matter are coupled, so these effects depend on each other, and a self-consistent treatment of NLTE and macroscopic mixing is the unique feature that JEKYLL offers. In \citetalias{Erg18} we investigated the effect of NLTE on the lightcurve and spectra, and from \citetalias{Erg18} and this work it is clear that \textit{both} NLTE and macroscopic mixing are necessary ingredients to accurately model the lightcurves and spectra of Type IIb SNe throughout their evolution.

\begin{acknowledgements} 
This work has been supported by grants from the Swedish Research Council and the Swedish National Space Board, and the computations were performed with resources provided by the Swedish National Infrastructure for Computing (SNIC) at Parallelldator-centrum (PDC). We are grateful to Peter Lundqvist for support and discussions, and to Anders Jersktrand for providing the 12C ejecta model and parts of the atomic data used in the simulations. We also thank the anonymous referee for constructive comments and suggestions that helped us to improve the paper.
\end{acknowledgements}

\appendix

\section{Configuration}
\label{a_configuration}

JEKYLL was configured to run in time-dependent mode (with respect to the radiative transfer), and to use a full NLTE solution including the following; radiative bound-bound, bound-free and free-free processes, collisional bound-bound and bound-free processes, non-thermal excitation (He only), ionization and heating, as well as two-photon processes. Charge-transfer was not included. We use the diffusion solver above an optical depth of 50, and a recombination correction based on the total recombination rates  while still enforcing detailed balance. In addition, packet control \citepalias{Erg18} was turned on to assure good sampling of the radiation field in all frequency regions. The number of $\Lambda$-iterations per time-step was set to 4. As discussed in \citetalias{Erg18}, this gives a well converged solution, which has also been verified for the models used in this paper.

\section{Atomic data}
\label{a_atomic_data}

The atomic data used is the default choice described in \citetalias{Erg18}, but has been extended with more levels and a full NLTE solution for ionization stages V and VI. This makes only a small difference for the observed lightcurves and spectra, and tests show that the simulations are not sensitive to further changes in the number of levels and ionization stages. Using online data provided by NIST\footnote{\url{www.nist.gov}} and R. Kurucz\footnote{\url{www.cfa.harvard.edu/amp/ampdata/kurucz23/sekur.html}}, these ions where updated to include 100 levels (or as many as were available) for elements lighter than Scandium and 300 levels (or as many as were available) for heavier elements. Total recombination rates for these ions were adopted from the online table provided by S. Nahar\footnote{\url{www.astronomy.ohio-state.edu/~nahar/_naharradiativeatomicdata/}} whenever available, and otherwise from \citet{Shu82}.

\section{Radiative diffusion in a macroscopically mixed medium}
\label{a_effective_opacity}

Radiative diffusion in a macroscopically mixed medium differs from that in a smooth medium due to the varying mean free path in the clumps. As we will see, this may lead to strong deviations in the effective opacity that the photons experiences compared to a smooth medium. In this appendix we discuss diffusion in a macroscopically mixed medium that is smooth and spherically symmetric on average, that is we restrict ourselves to small-scale deviations from spherical symmetry. We also assume that the medium is isotropic on average, that is these deviations have no preferred orientation. We begin by deriving the effective diffusion approximation for such a medium.

\subsection{The effective diffusion approximation}
\label{a_diffusion_approximation}

First, consider a general 3-D medium, and let us derive an expression for the radial flux in terms of the photon energy density $u$ and the distribution function for absorption $f$. Based on this we will re-derive the diffusion approximation for a smooth and spherically symmetric medium, and then generalize to a macroscopically mixed medium that is smooth, isotropic and spherically symmetric on average.

For simplicity we assume a time-independent radiation field and a static medium. As we will demonstrate in Sect.~\ref{a_example_calculations} using full radiative transfer calculations, this assumption does not affect the results as long as the diffusion time in the clumps is small compared to other relevant time-scales. In terms of the intensity $I_{\nu}$ for photons with frequency $\nu$, propagating in direction $\mathbf{d}$ with polar angle $\theta$, the radial flux at $\mathbf{r}$ is then given by
\begin{equation}
\label{eq_a_1}
F_{\nu}(\mathbf{r}) = \int I_{\nu}(\mathbf{r},\mathbf{d})~\cos{\theta}~d\Omega
\end{equation}

Writing this in terms of the emissivity $j_{\nu}$ for photons emitted into the ray at $\mathbf{r'}$ and distance $l$, and the cumulative distribution function for absorption along the ray $\mathcal{F}_{\nu}$ we get
\begin{equation}
\label{eq_a_2}
F_{\nu}(\mathbf{r}) = \int \int_{0}^{\infty}~j_{\nu}(\mathbf{r'},\mathbf{d})~[1-\mathcal{F}_{\nu}(l,\mathbf{r'},\mathbf{d})]~\cos{\theta}~dl~d\Omega
\end{equation}

Assuming that $j_{\nu}=(c/4\pi)~\kappa_{\nu}~\rho~u_{\nu}$ (which applies in the optically thick limit), and noting that $\kappa_{\nu}(\mathbf{r'}) \rho(\mathbf{r'})~[1-\mathcal{F}_{\nu}(l,\mathbf{r'},\mathbf{d})]=f_{\nu}(l,\mathbf{r},\mathbf{-d})$, we then arrive at
\begin{equation}
\label{eq_a_3}
F_{\nu}(\mathbf{r})= \frac{c}{4\pi}~\int \int_{0}^{\infty} f_{\nu}(l,\mathbf{r},\mathbf{-d})~u_{\nu}(\mathbf{r'})~\cos{\theta}~dl~d\Omega
\end{equation}

Now, consider a smooth and spherically symmetric medium. In this case, Eq.~\ref{eq_a_3} can be written
\begin{equation}
\label{eq_a_4}
F_{\nu} = \frac{c}{2}~\int_{0}^{\pi} \int_{0}^{\infty} f_{\nu}(l)~u_{\nu}(r-l~\cos{\theta})~\cos{\theta}~\sin{\theta}~dl~d\theta
\end{equation}

To obtain the diffusion approximation we make a Taylor expansion of $u_{\nu}$ with respect to $r$, and noting that $\int_{0}^{\pi} \cos{\theta}~\sin{\theta}~d\theta=0$ this gives
\begin{equation}
\label{eq_a_5}
F_{\nu} \approx -\frac{c}{2}~\int_{0}^{\pi}\int_{0}^{\infty} f_{\nu}(l)~\frac{du_{\nu}}{dr}~l~\cos^{2}{\theta}~\sin{\theta}~dl~d\theta=-\frac{\lambda_{\nu}c}{3}~\frac{du_{\nu}}{dr},
\end{equation}
where $\lambda_{\nu}$ is the mean free path given by $\int_{0}^{\infty} f_{\nu}(l)~l~dl$. Integrating over frequency and assuming a blackbody radiation field we arrive at 
\begin{equation}
  \label{eq_a_6}
  F=-\frac{4 a c T^{3}}{3\kappa_{R}\rho}~\frac{dT}{dr},
\end{equation}
where
\begin{equation}
\label{eq_a_7}
\kappa_{R}= \frac{\displaystyle \int_{0}^{\infty} \frac{dB_{\nu}}{dT} d\nu}{\displaystyle \int_{0}^{\infty}  \frac{1}{\kappa_{\nu}} \frac{dB_{\nu}}{dT} d\nu} 
\end{equation}
is the Rosseland mean opacity.

Next, consider a macroscopically mixed medium consisting of $n$ types of randomly distributed clumps with filling factor $\Phi_{i}$(r) and density $\rho_{i}(r)$, which is smooth, isotropic and spherically symmetric on average. In such a medium it is natural to chose a formulation in terms of angular averages, which in the limit of many clumps are equivalent to averages over all random arrangements of the clumps. Taking such an average of Eq.~\ref{eq_a_3} we get
\begin{equation}
\label{eq_a_8}
\left<F_{\nu}(\mathbf{r})\right>= \frac{c}{4\pi}~\int \int_{0}^{\infty} \left<f_{\nu}(l,\mathbf{r},\mathbf{-d})~u_{\nu}(\mathbf{r'})\right>~\cos{\theta}~dl~d\Omega
\end{equation}

To proceed we make the critical assumption that the photon energy density is smooth, spherically symmetric and independent of the random arrangement of the clumps, that is $u_{\nu}(\mathbf{r})=u_{\nu}(r)$. This is not strictly true, and in Sect.~\ref{a_example_calculations} we will compare our results to those of full MC calculations where we relax this assumption. Noting that all average quantities are smooth, isotropic and spherically symmetric we get
\begin{equation}
\label{eq_a_9}
\left<F_{\nu}\right> = \frac{c}{2}~\int_{0}^{\pi} \int_{0}^{\infty} \left<f_{\nu}\right>(l)~u_{\nu}(r-l~\cos{\theta})~\cos{\theta}~\sin{\theta}~dl~d\theta,
\end{equation}
which is analogous to Eq.~\ref{eq_a_4}. It is convenient to write $\left<f_{\nu}\right>$ as $\sum \Phi_{i}~\left<f_{\nu,i}\right>$, where $f_{\nu,i}$ is the distribution function for photons originating from clumps of type $i$. This relation follows from the fact that the probability for a point to lie within a clump of type $i$ is $\Phi_{i}$. In analogy with Eq.~\ref{eq_a_5} we then get the effective diffusion approximation in a macroscopically mixed medium
\begin{equation}
\label{eq_a_10}
\left<F_{\nu}\right>\approx-\frac{\skew{3.0}\hat{\lambda}_{\nu} c}{3}~\frac{du_{\nu}}{dr},
\end{equation}
where $\skew{3.0}\hat{\lambda}_{\nu}$ is the effective mean free path given by
\begin{equation}
  \label{eq_a_11}
  \skew{3.0}\hat{\lambda}_{\nu} = \sum_{i=0}^{n} \Phi_{i}~\skew{3.0}\hat{\lambda}_{\nu,i} = \sum_{i=0}^{n} \Phi_{i}\int_{0}^{\infty} \langle f_{\nu,i} \rangle~l~dl
\end{equation}

It is also convenient to define a corresponding effective opacity as $\hat{\kappa}_{\nu}=(\skew{3.0}\hat{\lambda}_{\nu}~\langle \rho \rangle)^{-1}$, where $\langle \rho \rangle= \sum \Phi_{i} \rho_{i}$ is the average density. In analogy with Eq.~\ref{eq_a_6} we then get 
\begin{equation}
  \label{eq_a_12}
  \left<F\right>=-\frac{4 a c T^{3}}{3\hat{\kappa}_{R} \langle \rho \rangle}~\frac{dT}{dr},\\
\end{equation}
where
\begin{equation}
  \label{eq_a_13}
  \hat{\kappa}_{R} = \frac{\displaystyle \int_{0}^{\infty} \frac{dB_{\nu}}{dT} d\nu~}{\displaystyle \int_{0}^{\infty} \frac{1}{\hat{\kappa}_{\nu}} \frac{dB_{\nu}}{dT} d\nu}
\end{equation}
is the effective Rosseland mean opacity in a macroscopically mixed medium. 

The effective Rosseland mean opacity depends on the effective mean free path, which in turn depends on $\langle f_{\nu,i} \rangle$. These averages of the absorption distribution functions can be written
\begin{equation}
    \label{eq_a_14}
    \langle f_{\nu,i} \rangle (l) = \langle \exp(-\kappa_{\nu,i}~\rho_{i}~l_{i})~\prod_{j=0}^{n} \exp(-\kappa_{\nu,j}~\rho_{j}~l_{j})~\kappa_{\nu}(l)~\rho(l)\rangle
\end{equation}
where $i$ and $j$ are the types of the originating and subsequent clumps, and $l_{i}$ and $l_{j}$ are the (total) pathlengths through them. In general, calculating them is complicated, and in JEKYLL we use a MC method described in \citetalias{Erg18}, where the MC packets are terminated once an interaction with the matter is drawn. This method is used to calculate the Rosseland mean opacity in the inner part of the ejecta, where the diffusion solver is used.

\subsection{Limiting cases}
\label{a_limiting_cases}

There are a few limits, in which case the averages of the absorption distribution functions can be simplified, and the effective mean free path easily calculated. In this section we discuss these limits, and derive expressions for the effective mean free path based on Eq.~\ref{eq_a_14}.

In the optically thin limit, the first exponential factor in Eq.~\ref{eq_a_14} approaches one, and as the number of clumps traversed before absorption approaches infinity, the $l_{j}$:s in the subsequent exponential factors approach their average values. In addition, the trailing opacity factor becomes independent and may be averaged separately. As the probability for a point to lie within a clump of type $j$ is $\Phi_{j}$, $\langle l_{j} \rangle=\Phi_{j} l$ and $\langle \kappa_{\nu}~\rho \rangle=\sum \Phi_{j}\kappa_{\nu,j}~\rho_{j}$, and we get
\begin{equation}
    \label{eq_a_15}
    \langle f_{\nu,i} \rangle (l) \approx \prod_{j=0}^{n} \exp(-\kappa_{\nu,j}~\rho_{j}~\langle l_{j} \rangle)~\langle \kappa_{\nu}~\rho \rangle = \exp(-\overline{\kappa}_{\nu}~\rho~l)~\overline{\kappa}_{\nu}~\langle \rho \rangle,
\end{equation}
where
\begin{equation}
    \label{eq_a_16}
    \overline{\kappa}_{\nu}=\sum_{j=0}^{n} \Phi_{j}~\kappa_{\nu,j}~\rho_{j}/\langle \rho \rangle
\end{equation}
is a mass average of the opacity over the different clump types. Therefore, in this limit, the effective mean free path is given by
\begin{equation}
  \label{eq_a_17}
  \skew{3.0}\hat{\lambda}_{\nu} = \frac{1}{\overline{\kappa}_{\nu}~\langle \rho \rangle},
\end{equation}
As $\overline{\kappa}_{\nu}$ is a mass-average, the effective mean free path in the optically thin limit is $\emph{independent}$ of the clumping geometry.

In the optically thick limit, all photons are absorbed in the originating clump, and we get
\begin{equation}
  \label{eq_a_18}
  \langle f_{\nu,i} \rangle (l) \approx \exp(-\kappa_{\nu,i}~\rho_{i}~l)~\kappa_{\nu,i}~\rho_{i}
\end{equation}
Therefore, in this limit the effective mean free path is given by
\begin{equation}
  \label{eq_a_19}
  \skew{3.0}\hat{\lambda}_{\nu} = \sum_{i=0}^{n} \Phi_{i}~\lambda_{\nu,i},
\end{equation}
which is a volume average of the mean free path over the different clump types. Contrary to the thin limit, the effective mean free path depends on \emph{both} the material opacity and the clumping geometry, although only through the filling factors.

Using Eqs.~\ref{eq_a_17} and \ref{eq_a_19}, the ratio between the effective mean free path in the optically thick and thin limit (or equivalently, the ratio between the effective opacity in the optically thin and thick limit) can be expressed as
\begin{equation}
  \label{eq_a_20}
  R_{\nu}= \sum_{i=0}^{n} \Phi_{i} \skew{3.0}\hat{\lambda}_{\nu,i} \sum_{j=0}^{n} \Phi_{j} \frac{1}{\skew{3.0}\hat{\lambda}_{\nu,j}} = \sum_{i=0}^{n} \sum_{j=0}^{n} \Phi_{i} \Phi_{j} \frac{\skew{3.0}\hat{\lambda}_{\nu,i}}{\skew{3.0}\hat{\lambda}_{\nu,j}} \geq 1,
\end{equation}
where the last condition follows from the fact that $\skew{3.0}\hat{\lambda}_{\nu,i}/\skew{3.0}\hat{\lambda}_{\nu,j}+\skew{3.0}\hat{\lambda}_{\nu,j}/\skew{3.0}\hat{\lambda}_{\nu,i} \geq 2$. Therefore, the effective mean free path is the same in both limits only if the mean free path is the same in all clump types, and is otherwise longer in the optically thick limit. Note, that for a medium with equal opacity in all clumps, Eq.~\ref{eq_a_20} can be expressed as
\begin{equation}
  \label{eq_a_21}
  R_{\nu}= \sum_{i=0}^{n} \sum_{j=0}^{n} \Phi_{i} \Phi_{j} \frac{\rho_{i}}{\rho_{j}} \geq 1
\end{equation}
Therefore, in this case, the effective mean free path is the same in both limits only if the density is the same for all clump types, and is otherwise longer in the optically thick limit. In other words, the effective mean free path is increased and the effective opacity decreased by density variations.

In the transitional regime between the optically thick and thin limit, the dependence of the effective mean free path on the clumping geometry is more complicated, and also more profound. In fact, in the limit where one class of clumps is fully transparent and another fully opaque, the effective mean free path depends \emph{only} on the clumping geometry and is \emph{independent} of the material opacity. In this limit, $\langle f_{\nu,i} \rangle$ for photons originating in the transparent clumps becomes the distribution function of hitting the opaque clumps, and the effective mean free path may be written as
\begin{equation}
    \label{eq_a_22}
    \skew{3.0}\hat{\lambda}_{\nu} = \Phi~\lambda_{G}
\end{equation}
where the geometrical mean free path $\lambda_{G}$ is the average distance to the opaque clumps, and $\Phi$ the filling factor for the transparent clumps. If the filling factor for the opaque clumps is small we may use an analogy with the material opacity and write $\langle f_{\nu,i} \rangle (l)=\exp(-\chi_{G}~l)~\chi_{G}$, where $\chi_{G}$ is the geometrical cross-section per volume produced by the opaque clumps. Therefore, in this case we have $\lambda_{G}=1/\chi_{G}$.

The concept of geometrical mean free path is also useful to analyse the behaviour in the transitional regime for less extreme cases, and we may define $\lambda_{G,i}$ as the average distance for photons originating in a region occupied by clumps of type $i$ to regions occupied by clumps of other types. It is also practical to define a geometrical optical depth as $\tau_{G,\nu,i}=\lambda_{G,i}~\kappa_{\nu,i}~\rho_{i}$, indicating to which extent photons originating in one region leak into other regions.

\subsection{Test case}
\label{a_example_calculations}

In order to further study the behaviour of the effective opacity, and to test the validity of the results, we explore a test case with a clumping geometry similar to the Type IIb SN models in the paper. For this case we calculate the effective opacity and the limits using our approximate theory developed in Sects.~\ref{a_diffusion_approximation} and \ref{a_limiting_cases}, and in Sect.~\ref{a_validation} we compare the results to full radiative transfer calculations on actual 3D-grids, where the assumptions upon which this theory relies are relaxed.

Consider a smooth medium with constant grey opacity which we divide into two types of "clumps" of equal mass and mass fractions 0.9 and 0.1. Imagine then that we expand the latter (while keeping the average density constant) by a factor of 9 to achieve a filling factor of 0.9. The clumping parameters of such a medium are similar to the macroscopically mixed standard model, where the expanded clumps correspond to the Ni/He clumps. Due to expansion, the material mean free path in these clumps is 81 times longer than in the compressed clumps, but as the filling factor for the expanded region is large, the geometrical optical depth is less different, and is 6 times lower in the expanded region than in the compressed region.

\begin{figure}[tbp!]
\includegraphics[width=0.49\textwidth,angle=0]{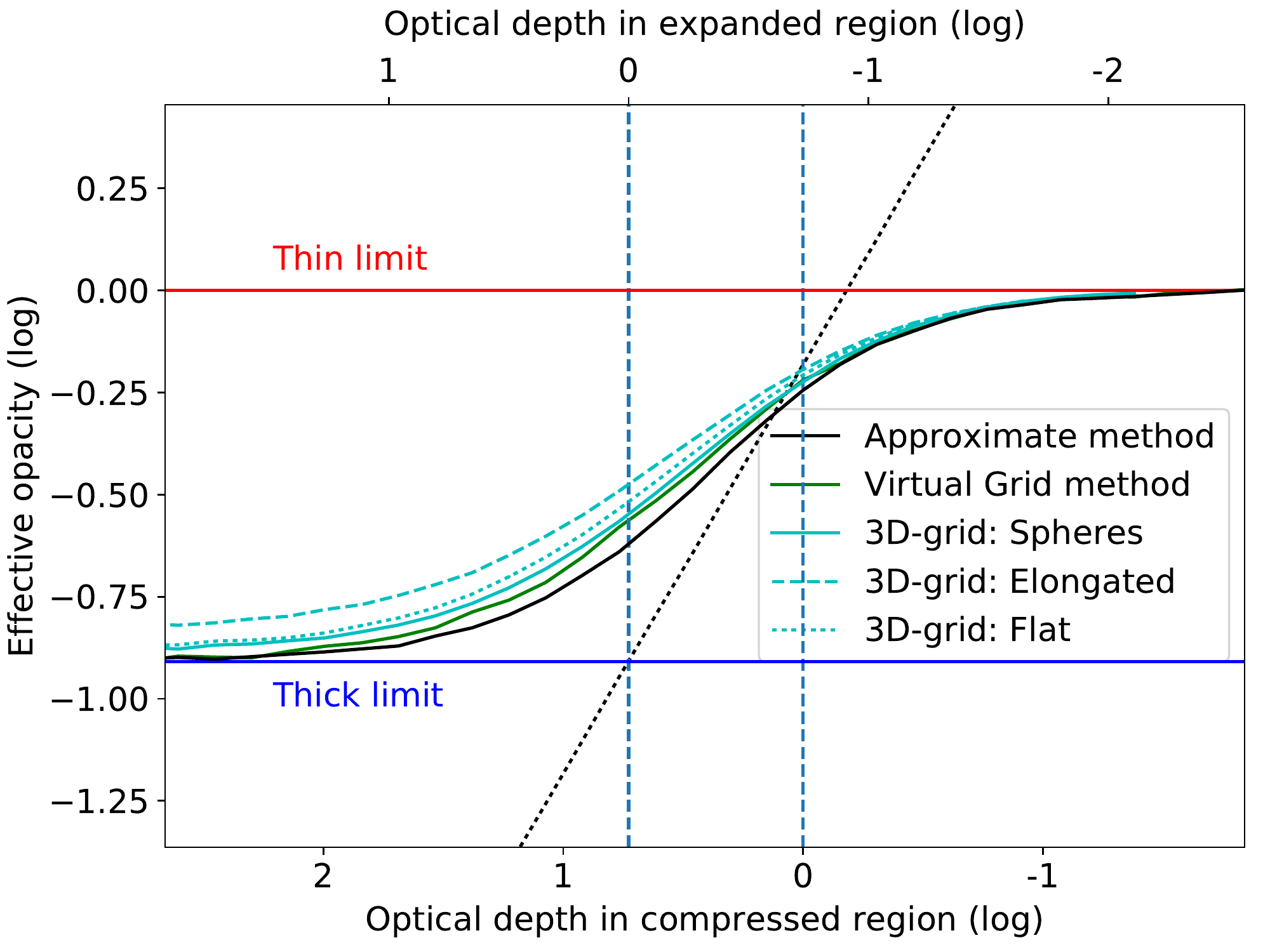}
\caption{Effective opacity (black) for a homologously expanding medium with clump properties similar to the core of the Type IIb SN models compared to full radiative transfer calculations using the Virtual Grid method (yellow) and actual 3D-grids containing spherical (solid cyan), elongated (dotted cyan) and flat (dashed cyan) clumps. We also show the optically thin (blue), optically thick (red) and geometrical (dotted black) limits. The opacity is given relative to the thin limit, and the geometrical optical depth in the compressed and expanded regions is shown at top and bottom, respectively.}
\label{f_effective_opacity_model_1}
\end{figure}

The calculated effective opacity and the corresponding limits for this medium as it is homologously expanded are shown in Fig.~\ref{f_effective_opacity_model_1} as a function of geometrical optical depth. Initially, when the geometrical optical depth is  high in all regions, the effective opacity follows the optically thick limit. Then, when the geometrical optical depth approaches one in the expanded region, photons begin to escape from this region and the effective opacity enters the transitional regime where it gradually increases. It remains there until the geometrical optical depth approaches one also in the compressed region, and then gradually settles on the optically thin limit, where the geometrical optical depth is low in all regions. The conditions are not extreme enough for the effective opacity to follow the geometrical limit, but it is worth noting that it does not diverge to much from this limit when the expanded region is thin and the compressed region is thick.

\begin{figure}[tbp!]
  \includegraphics[width=0.49\textwidth,angle=0]{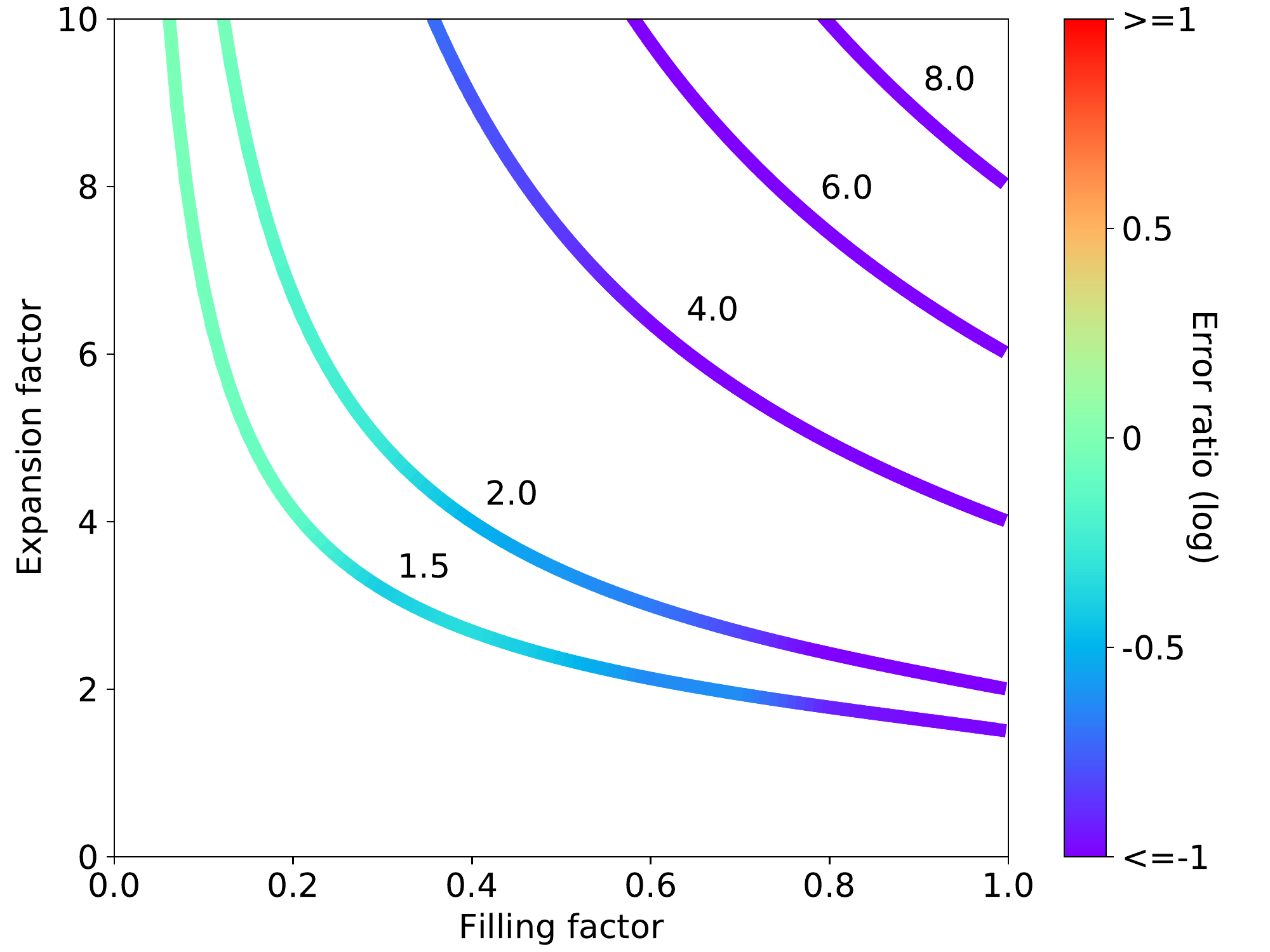}
  \caption{Effective opacity decrease in a medium with uniform grey opacity, where one type of clumps have been expanded by a factor $f_{\mathrm{E}}$ to reach a filling factor of $\Phi_{E}$. Contours are shown for a decrease of the effective opacity by a factor of 1.5, 2, 4, 6, and 8. The colour of the contours shows the ratio of the error from using our method to the error from not using it.}
  \label{f_opacity decrease}
\end{figure}

In the optically thin limit, the effective opacity is independent of the clumping geometry and equals that of a smooth medium, but from Eq.~\ref{eq_a_21} we see that in the optically thick limit it decreases by a factor
\begin{equation}
  \label{eq_a_23}
  R=\Phi_{\mathrm{C}}/f_{\mathrm{C}}+\Phi_{\mathrm{E}}~f_{\mathrm{E}}\frac{(1-\Phi_{\mathrm{E}})^{2}}{(1-\Phi_{\mathrm{E}}/f_{\mathrm{E}})} + \Phi_{\mathrm{E}}~f_{\mathrm{E}},
\end{equation}
where $f_{\mathrm{C}}$ is the compression factor for the high density clumps, $f_{\mathrm{E}}$ the expansion factor for the low density clumps, and we have used $\Phi_{\mathrm{C}}=1-\Phi_{\mathrm{E}}$ and $f_{\mathrm{C}}=(1-\Phi_{\mathrm{E}}/f_{\mathrm{E}})/(1-\Phi_{\mathrm{E}})$. This gives a decrease of the effective opacity by a factor of 9 for the numbers in this example. The general dependence of the opacity decrease in the optically thick limit on the filling factor and expansion factor for the low density clumps is illustrated in Fig.~\ref{f_opacity decrease}. From Eq.~\ref{eq_a_23} we see that the effective opacity can not decrease more than the clumps have been expanded. If $R \gg 1$, the opacity decrease is approximately given by $\Phi_{\mathrm{E}}~f_{\mathrm{E}}$.

\subsection{Validation}
\label{a_validation}

Now, let us return to the assumptions upon which our approximate theory relies. The validity of these assumptions can be tested by comparisons to full radiative transfer calculations on actual 3D-grids. It is also of interest to compare to the Virtual Grid method used for the radiative transfer in the outer part of the ejecta, as the virtual nature of the grid may cause differences compared to actual 3D-grids. Note, that the reason that we use the Virtual Grid method in the outer part and our approximate method in the inner part is the computational cost. Our approximate method is considerable cheaper than the Virtual Grid method, which in turn is considerable cheaper than a calculation on an actual 3D-grid.

Let us first relax the assumption of a spherically symmetric energy density using full radiative transfer calculations for the case discussed in Sect.~\ref{a_example_calculations}. The calculations were performed with the Virtual Grid method and on an actual 3D-grid containing spherical clumps. In the latter case, the compressed spherical clumps were placed in an ambient expanded medium as an actual 3D-grid can not be strictly divided into spherical clumps. A time-independent radiation field and a static medium were assumed and the energy was injected at an inner boundary. In Fig.~\ref{f_effective_opacity_model_1} we show the effective opacity obtained from the effective diffusion approximation if we force the average flux and energy density to comply with these calculations. 

Examining first the case when the radiative transfer is performed using the Virtual Grid method, the difference between this and our approximate method is small. In the optically thick and thin limits there is no difference at all, but in the transitional region the effective opacity is slightly underestimated by our approximate method. Turning to the case when the radiative transfer is performed on an actual 3D-grid containing spherical clumps, the result is similar to the Virtual Grid method except in the optically thick regime, where the effective opacity approaches the thick limit more slowly. This shows that the Virtual Grid method has its limitations, and can not catch all aspects of the radiative transfer on an actual 3D-grid. Nevertheless, overall the differences are small, and for the case explored the assumption of a spherically symmetric energy density seems justified.

We also want to explore the effect of the geometry of the clumps on the result, as both the methods used in JEKYLL assume that these are spherical. Therefore we have performed full radiative transfer calculations on actual 3D-grids containing flat (2D-like) and elongated (1D-like) clumps, both ellipsoids with an axis ratio of 10. The setup and the assumptions are the same as above, and the results as a function of geometrical optical depth are shown in Fig.~\ref{f_effective_opacity_model_1}. As seen there is a clear difference in the case of flat (2D-like) clumps, whereas elongated (1D-like) and spherical clumps behave similarly.  Although the difference in the flat (2D-like) case is modest it shows that the geometry of the clumps has an effect on the effective opacity. Investigations of the geometry of clumps in real SN ejecta or from 3D explosion models would be of great interest.

It is also of interest to explore a larger parameter space in terms of filling factors and expansion factors. To achieve this we have run a set of radiative transfer calculations on actual 3D-grids containing spherical clumps. These models lie on the contour lines of constant opacity decrease in the $\Phi_E$-$f_E$ plane shown in Fig.~\ref{f_opacity decrease}, which more or less spans the region of interest for SNe. We then calculated the difference in effective opacity between our method and the 3D-calculations near the point where the former settles on the thick limit. In the figure we show the ratio of the error from using our method to the error from not using it in colour coded logarithmic units. As seen in Fig.~\ref{f_opacity decrease} our method is well justified over most of the parameter space, and is otherwise still better than not using it. The relative error is everywhere less than $\sim$30 percent and $\sim$15 percent on average. A similar comparison between the Virtual Grid method and the calculations on actual 3D-grids gives a similar but slightly better agreement. It should be noted, however, that in both cases the agreement seems to degrade at small filling factors, in particular in combination with high expansion factors, and to be generally applicable, our methods may need to be modified.

Finally, let us relax the assumptions of a time-independent radiation-field and a static medium using full radiative transfer calculations with JEKYLL. In this case, the timescales of several physical processes becomes important, and we therefore explore a case analogues to that above, but with radioactive decay energy injected into the expanded clumps and expansion cooling taken into account. For the case when the diffusion time in the clumps was short compared to the decay and expansion timescales, the result from the full radiative transfer calculation was in good agreement with our approximate method, and was in fact very similar to that obtained with the Virtual Grid method in the time-independent/static case (see Fig.~\ref{f_effective_opacity_model_1}). A sufficient condition for our approximate method to apply is therefore that the diffusion time in the clumps is short compared to the decay and expansion timescales, and how well that condition is fulfilled in our SN models is discussed separately in Appendix \ref{a_tdiff_clump}.

\section{The diffusion time in the clumps}
\label{a_tdiff_clump}

If the diffusion time in the clumps is shorter than the expansion and decay time-scales, we do not expect any ongoing hydrodynamical expansion of the Ni/He and Si/S clumps (Sect.~\ref{s_method_macro}). As JEKYLL do not have hydrodynamical capabilities, it is important to investigate how well this condition is fulfilled in our models. The same condition also applies to the diffusion solver, where a common temperature/energy density in the clumps is assumed for several reasons (Sect~\ref{s_method_macro} and Appendix~\ref{a_validation}).

\begin{figure}[tbp!]
\includegraphics[width=0.49\textwidth,angle=0]{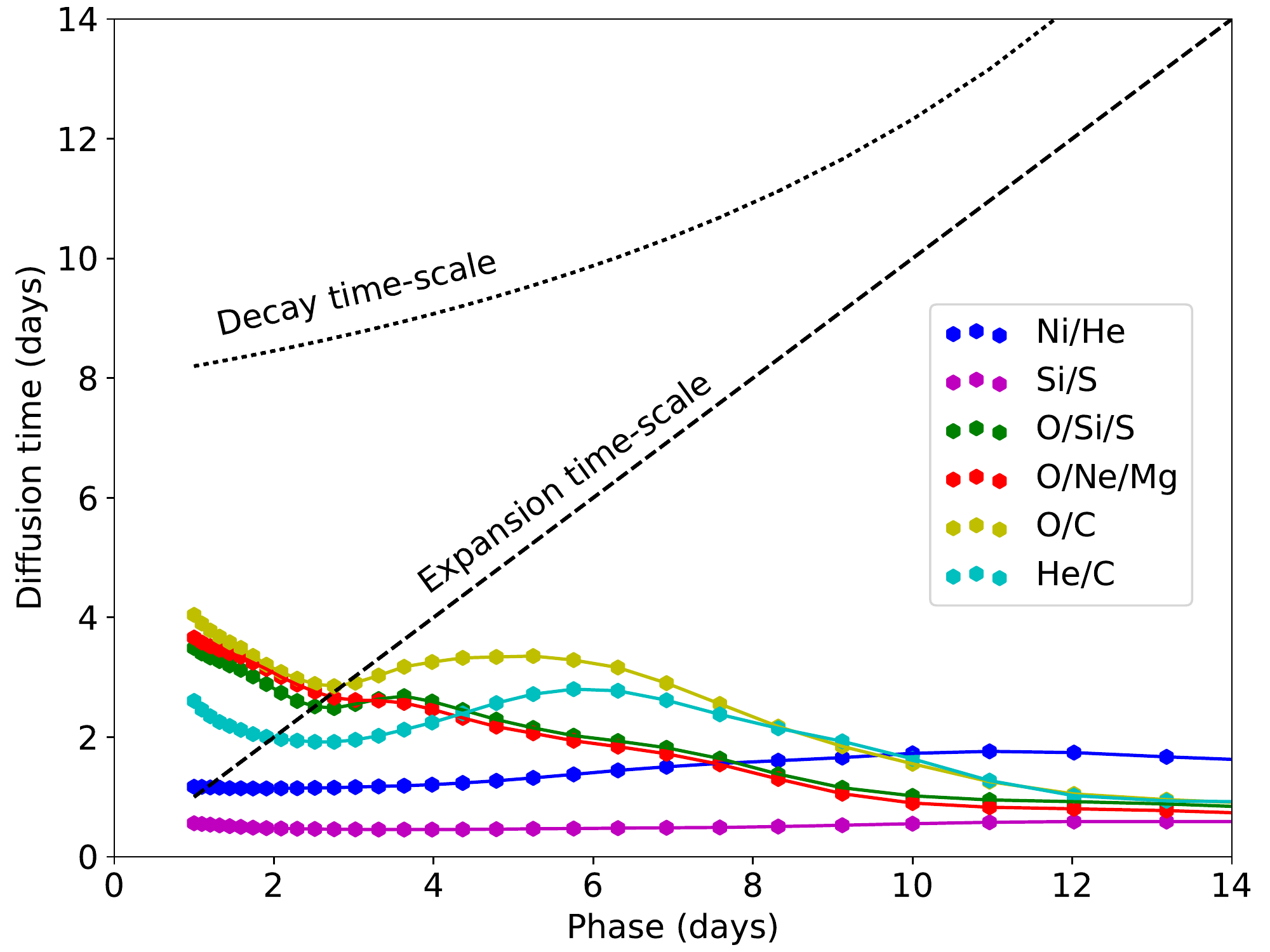}
\caption{Diffusion time in the Ni/He (blue), Si/S (magenta), O/Si/S (green), O/Ne/Mg (red), O/C (yellow) and He/C (cyan) clumps in the core of the macroscopically mixed standard model compared to the expansion (dashed black line) and decay (dotted black line) time-scales.}
\label{f_tdiff_clump}
\end{figure}

In Fig.~\ref{f_tdiff_clump} we show the diffusion time in the clumps in the core of the standard model compared to the expansion and decay time-scales. As clumps may join to form bigger regions if the filling factor is large, we have used the geometrical mean free path $\lambda_{G,i}$ (see Appendix \ref{a_diffusion_approximation}) instead of the individual clump size. Following \citet{Arn80,Arn82}, the diffusion time in an expanding medium has been calculated as
\begin{equation}
    \tau_{m,i}=\sqrt{2~\tau_{0,i}~\tau_h},
\end{equation}
where
\begin{equation}
    \tau_{0,i}=\frac{3}{\pi^{2} c} \lambda_{G,i}^{2}~\rho_{i}~\kappa_{i}
\end{equation}
is the diffusion time in a static medium, $\tau_h$ is the expansion timescale, and the $i$ refers to the clump type.
As seen in the figure, the condition is increasingly well fulfilled after a few days. The diffusion time in the helium envelope and in the model with more clumps is much lower, so this applies to all our models.

\section{Additional spectral figures}
\label{a_additional_figures}

\begin{figure*}[tbp!]
\includegraphics[width=1.0\textwidth,angle=0]{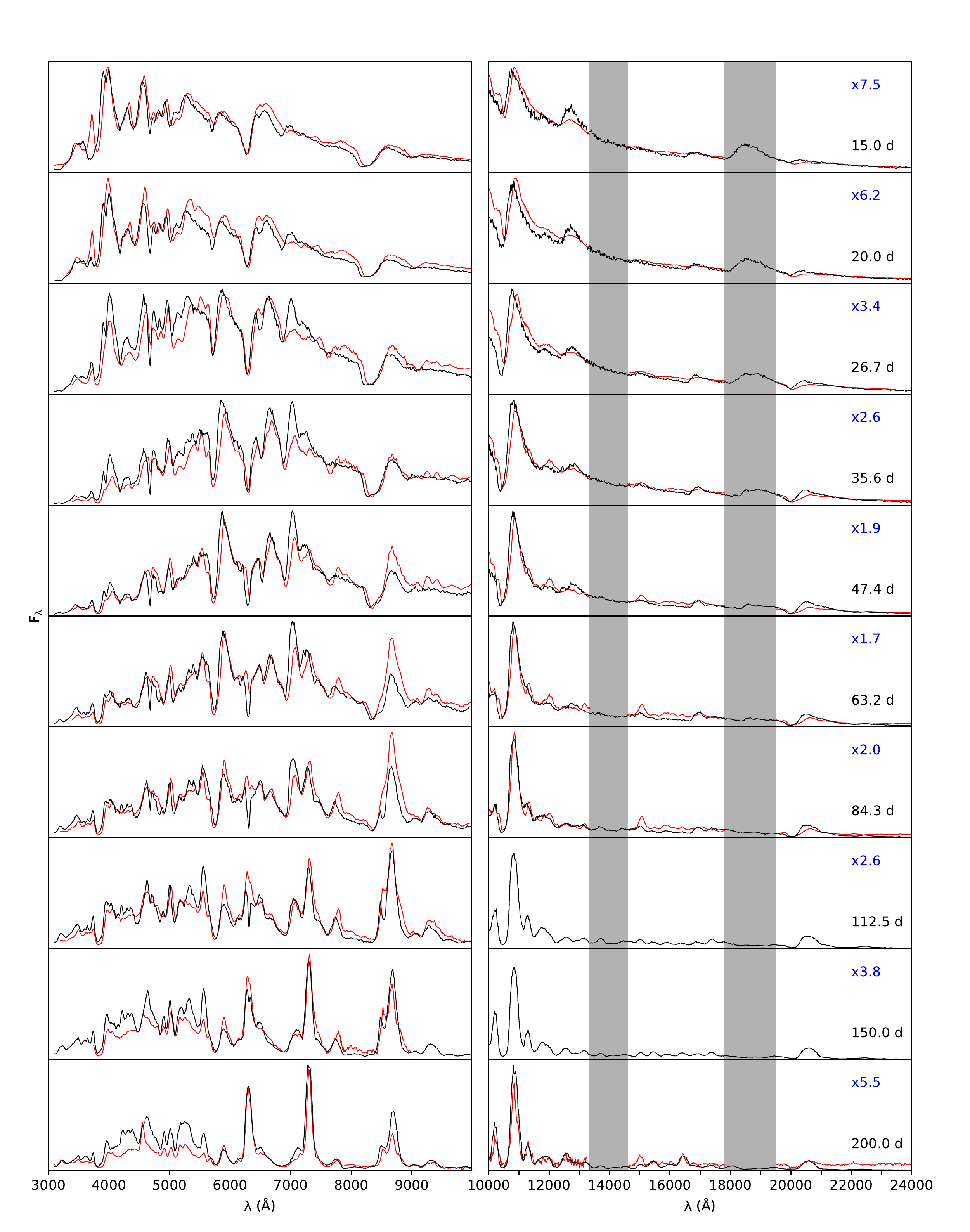}
\caption{Spectral evolution on a linear flux scale in the optical (left panel) and the NIR (right panel) for the standard model (black) compared to the observations of SN 2011dh (red). Otherwise as in Fig~\ref{f_dh_spec_evo_comp}.}
\label{f_dh_spec_evo_comp_linear}
\end{figure*}

For convenience, we provide a set of additional spectral figures. First, in Fig.~\ref{f_dh_spec_evo_comp_linear} we show the same comparison between the standard model and the observations of SN 2011dh as in Fig.~\ref{f_dh_spec_evo_comp}, but on a linear flux scale.

\begin{figure*}[tbp!]
\includegraphics[width=1.0\textwidth,angle=0]{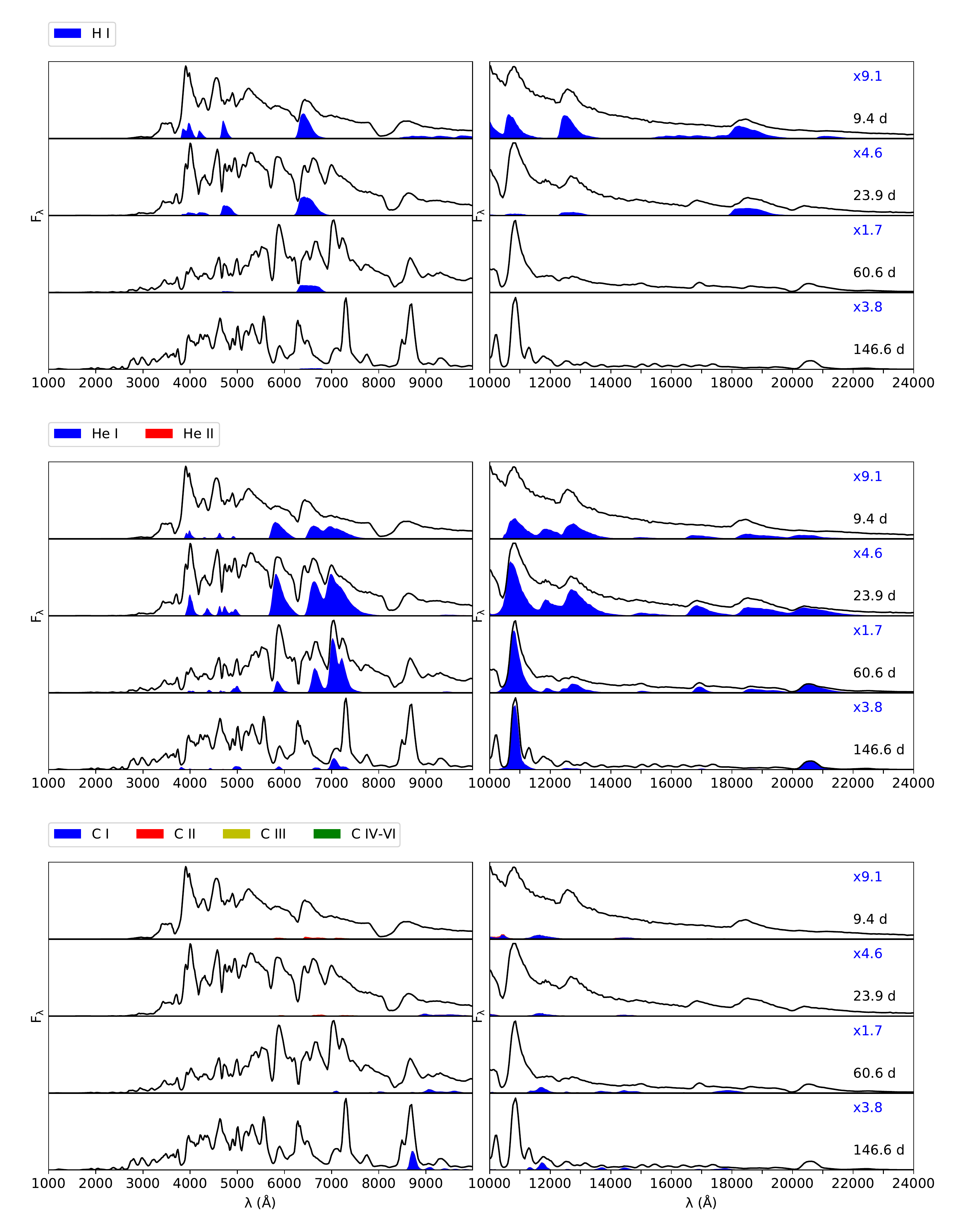}
\caption{Bound-bound contribution from ionisation stages I (blue), II (red), III (yellow) and higher (green) of hydrogen (upper panel), helium (middle panel) and carbon (lower panel) to the spectral evolution of the standard model.}
\label{f_12C_spec_trans_evo_ion_1}
\end{figure*}

\begin{figure*}[tbp!]
\includegraphics[width=1.0\textwidth,angle=0]{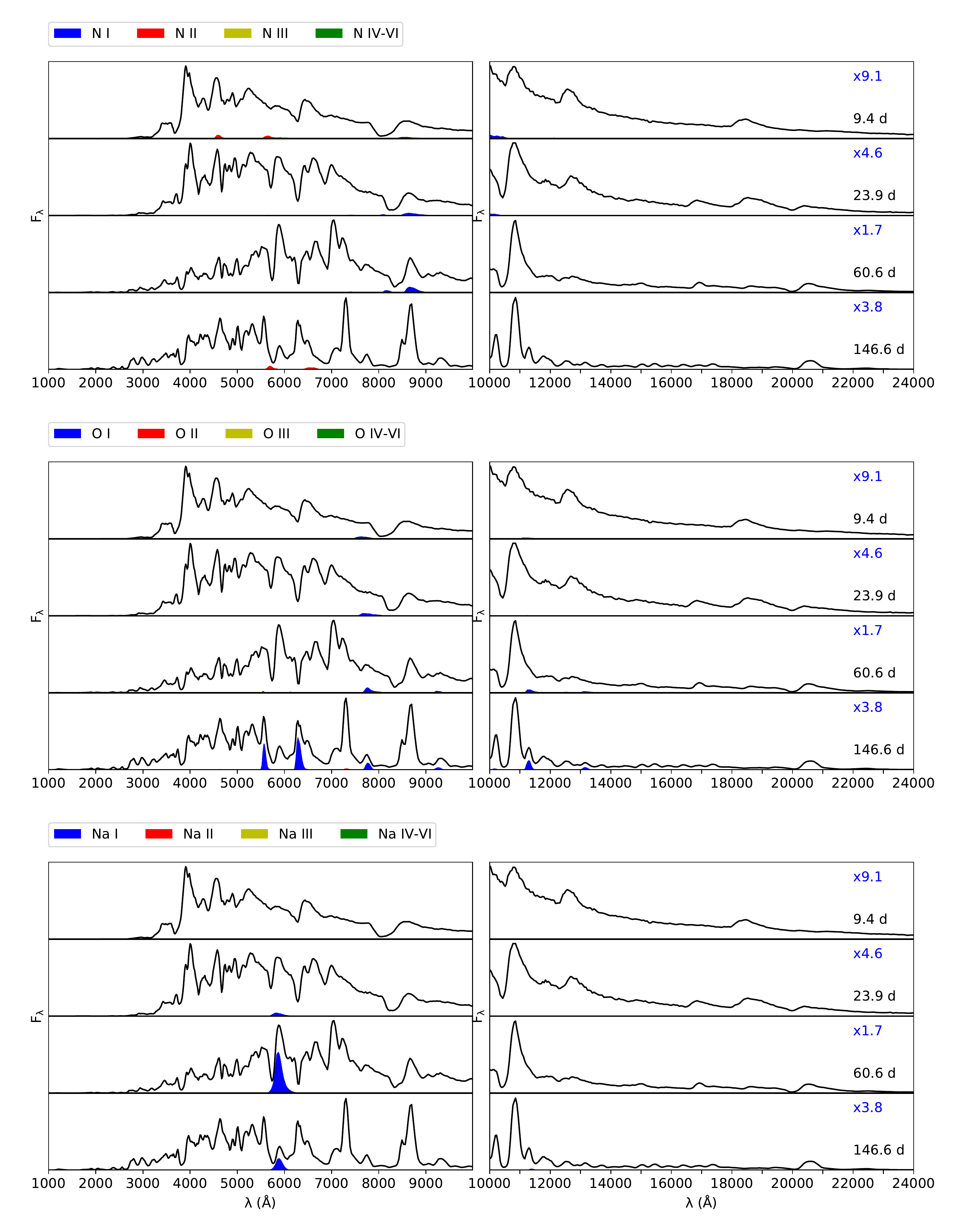}
\caption{Bound-bound contribution from ionisation stages I (blue), II (red), III (yellow) and higher (green) of nitrogen (upper panel), oxygen (middle panel) and sodium (lower panel) to the spectral evolution of the standard model.}
\label{f_12C_spec_trans_evo_ion_2}
\end{figure*}

\begin{figure*}[tbp!]
\includegraphics[width=1.0\textwidth,angle=0]{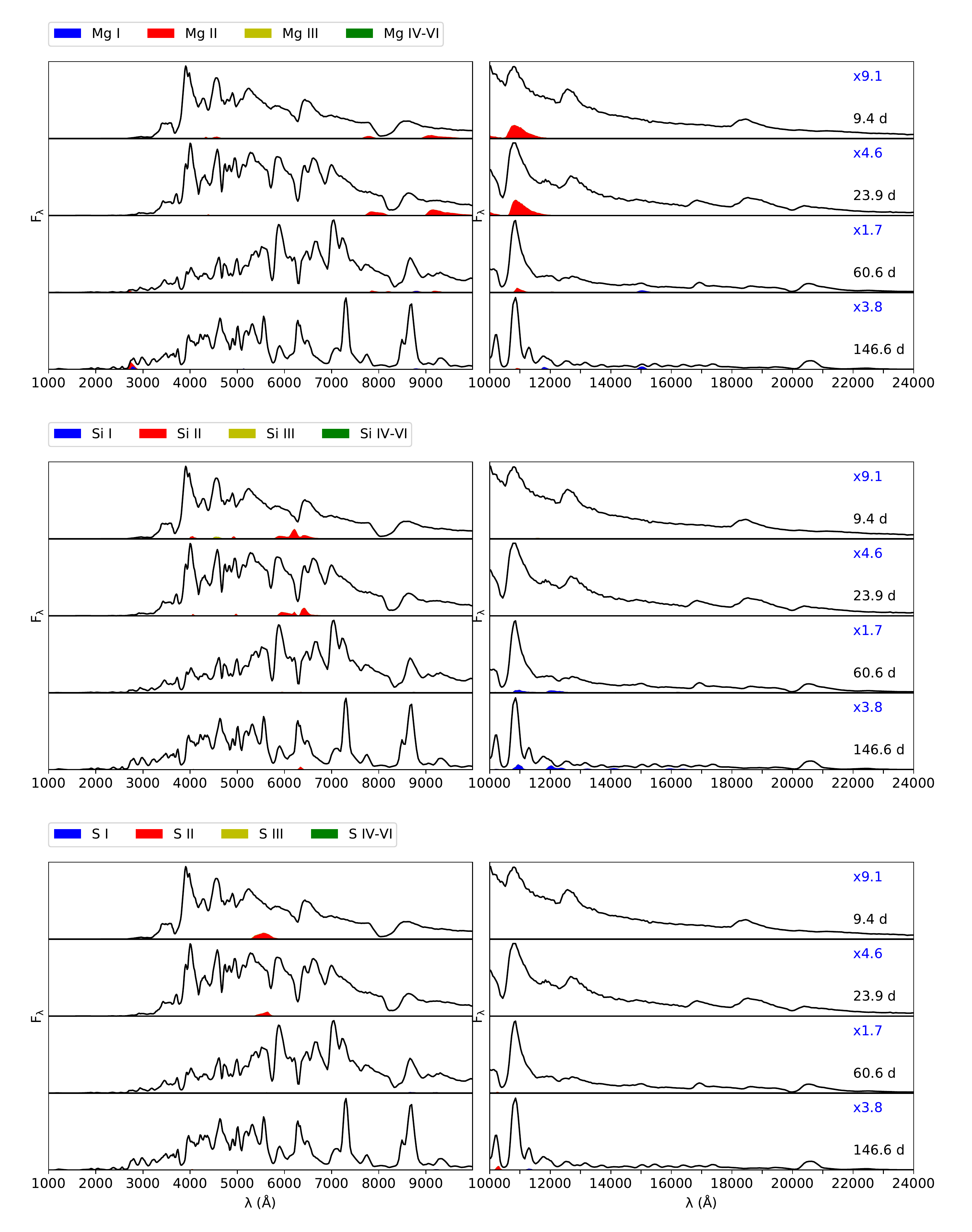}
\caption{Bound-bound contribution from ionisation stages I (blue), II (red), III (yellow) and higher (green) of magnesium (upper panel), silicon (middle panel) and sulphur (lower panel) to the spectral evolution of the standard model.}
\label{f_12C_spec_trans_evo_ion_3}
\end{figure*}

\begin{figure*}[tbp!]
\includegraphics[width=1.0\textwidth,angle=0]{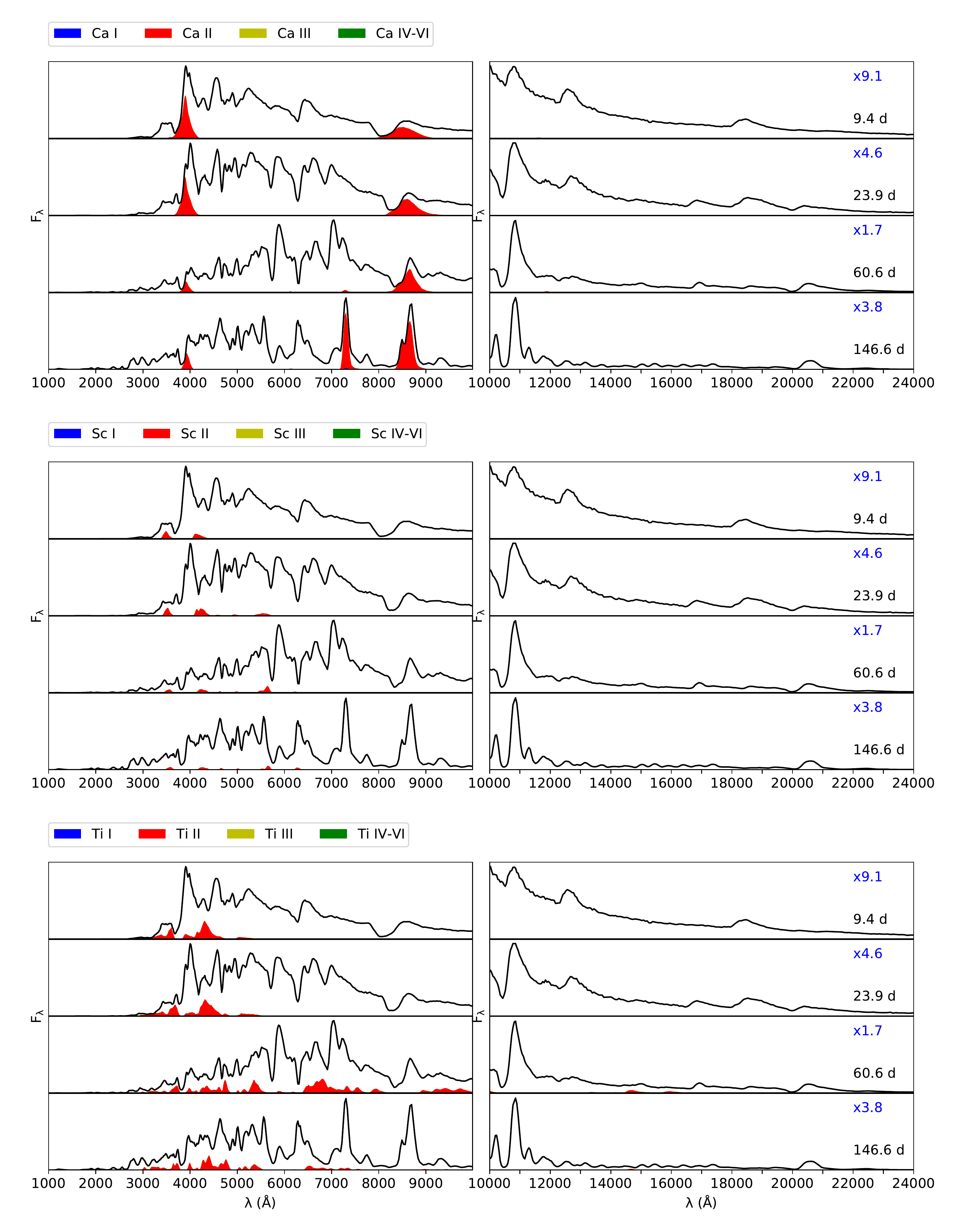}
\caption{Bound-bound contribution from ionisation stages I (blue), II (red), III (yellow) and higher (green) of calcium (upper panel), scandium (middle panel) and titanium (lower panel) to the spectral evolution of the standard model.}
\label{f_12C_spec_trans_evo_ion_4}
\end{figure*}

\begin{figure*}[tbp!]
\includegraphics[width=1.0\textwidth,angle=0]{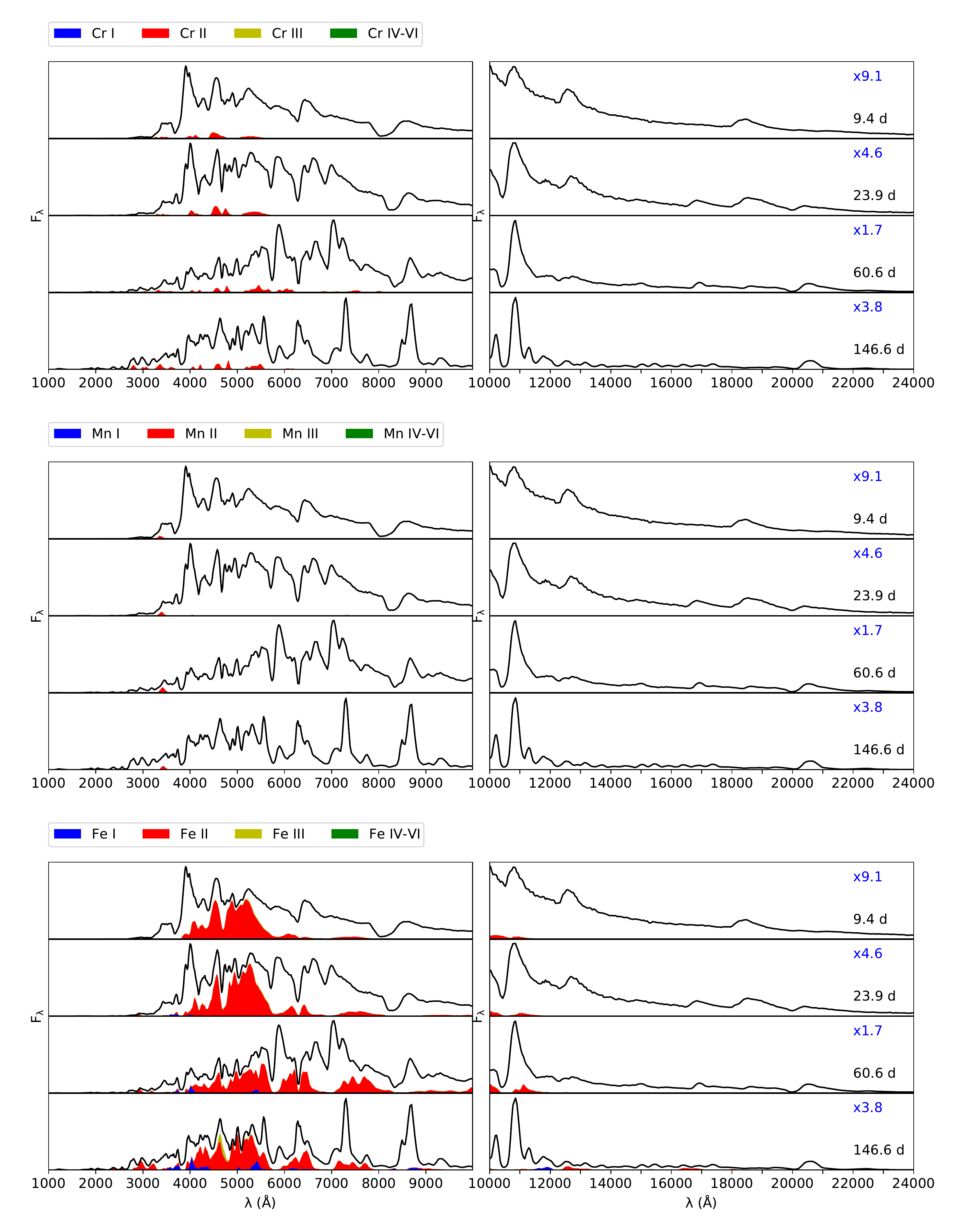}
\caption{Bound-bound contribution from ionisation stages I (blue), II (red), III (yellow) and higher (green) of chromium (upper panel), manganese (middle panel) and iron (lower panel) to the spectral evolution of the standard model.}
\label{f_12C_spec_trans_evo_ion_5}
\end{figure*}

\begin{figure*}[tbp!]
\includegraphics[width=1.0\textwidth,angle=0]{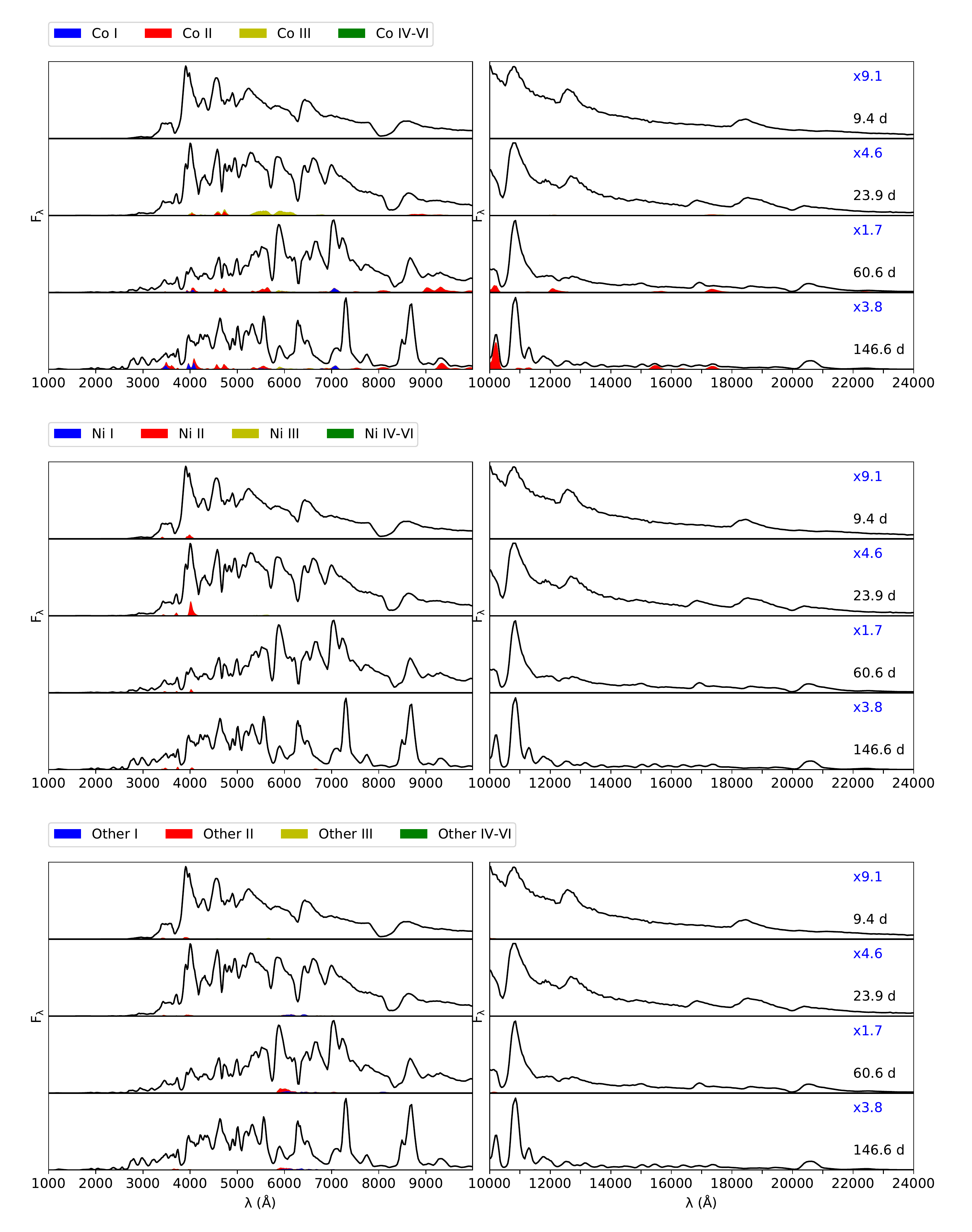}
\caption{Bound-bound contribution from ionisation stages I (blue), II (red), III (yellow) and higher (green) of cobalt (upper panel), nickel (middle panel) and other elements (lower panel) to the spectral evolution of the standard model.}
\label{f_12C_spec_trans_evo_ion_6}
\end{figure*}

Second, in Fig.~\ref{f_12C_spec_trans_evo_ion_1}-\ref{f_12C_spec_trans_evo_ion_6} we show the bound-bound contribution (last scattering/emission event excluding electron scattering) from ionization stages I, II, III and higher of hydrogen, helium, carbon, nitrogen, oxygen, sodium, magnesium, silicon, sulphur, calcium, scandium, titanium, chromium, manganese, iron, cobalt, nickel and other elements to the spectral evolution of the standard model.

\begin{figure*}[tbp!]
\includegraphics[width=1.0\textwidth,angle=0]{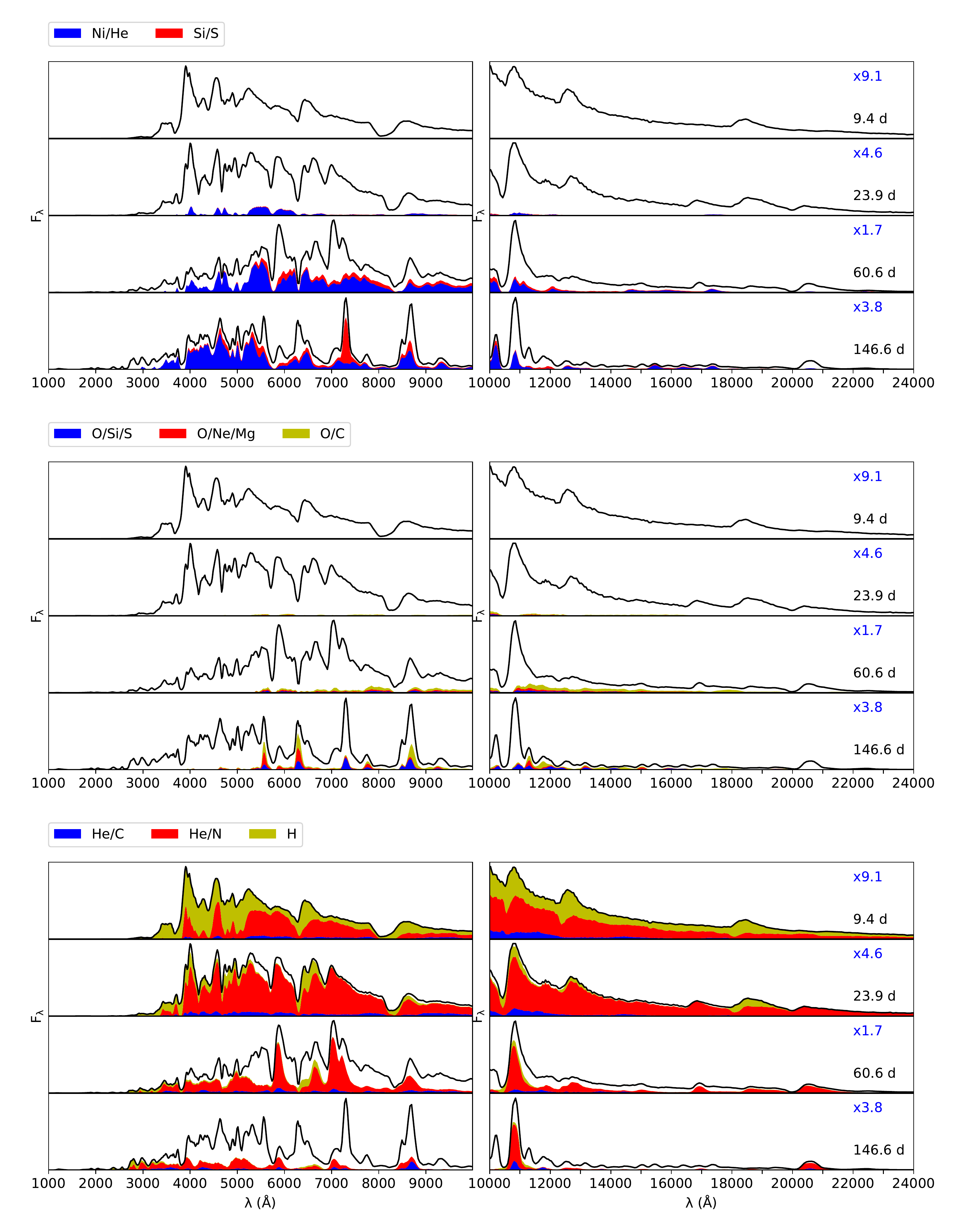}
\caption{Bound-bound contribution from the nickel-rich zones (Ni/He: blue, Si/S: red), the oxygen-rich zones (O/Si/S: blue, O/Ne/Mg: red, O/C: yellow) and the hydrogen- and helium-rich zones (H/C: blue, He/N: red, H: yellow) to the spectral evolution of the standard model.}
\label{f_12C_spec_cell_evo_zone}
\end{figure*}

Finally, in Fig.~\ref{f_12C_spec_cell_evo_zone} we show the bound-bound contribution (last scattering/emission event excluding electron scattering) from the nickel-rich (Ni/He, Si/S), oxygen-rich (O/Si/S, O/Ne/Mg, O/C) and hydrogen- and helium-rich (He/C, He/N, H) compositional zones to the spectral evolution of the standard model.

\bibliographystyle{aa}
\bibliography{spec-IIb-paper}

\label{lastpage}

\end{document}